\documentclass[twocolumn]{aastex61}

\received{}
\revised{}
\accepted{}
\submitjournal{ApJ}

\usepackage{natbib}
\shorttitle{HOT PLASMA IN A LIMB ACTIVE REGION}
\shortauthors{Parenti et al.}

\begin{document}


\title{Spectroscopy of very hot plasma in non-flaring parts of a solar limb active region: spatial and temporal properties}

\correspondingauthor{Susanna Parenti}
\email{susanna.parenti@ias.u-psud.fr}

\author{Susanna Parenti}
\affiliation{Institut d'Astrophysique Spatiale, \\
CNRS - Univ. Paris-Sud, Universit\'{e} Paris-Saclay\\
 Bat. 121, F-91405 Orsay, France}

\author{Giulio del Zanna}
\affiliation{DAMTP, Centre for Mathematical Sciences,\\
 Wilberforce road,\\
  Cambridge, UK}

\author{Antonino Petralia}
\affiliation{Dipartimento di Fisica \& Chimica, Universit\'{a} di Palermo, \\
Piazza del Parlamento 1, \\
90134 Palermo, Italy}

\author{Fabio Reale}
\affiliation{Dipartimento di Fisica \& Chimica, Universit\'{a} di Palermo,\\
 Piazza del Parlamento 1, \\
 90134 Palermo, Italy}
\affiliation{INAF-Osservatorio Astronomico di Palermo, Piazza del Parlamento 1, 90134 Palermo, Italy}

\author{Luca Teriaca}
\affiliation{Max-Planck-Institut f{\"u}r Sonnensystemforschung,\\
 Justus-von-Liebig-Weg 3, \\
 37077 G{\"o}ttingen, Germany}

\author{Paola Testa}
\affiliation{Harvard-Smithsonian Center for Astrophysics,\\
 60 Garden Street,\\
  MS 58, Cambridge MA 02138, USA}

\author{Helen E. Mason}
\affiliation{DAMTP, Centre for Mathematical Sciences,\\
 Wilberforce road Cambridge, UK}

\begin{abstract}
In this work we investigate the thermal structure of an off-limb active region in various non-flaring areas, as it provides key information on the way these structures are heated. In particular, we concentrate in the very hot component  ($>3~\mathrm{MK}$) as it is a crucial element to discriminate between different heating mechanisms. We present an analysis using Fe and Ca emission lines from both SOHO/SUMER and HINODE/EIS. A dataset covering all ionization stages from Fe\,{\footnotesize X} to Fe\,{\footnotesize XIX} has been used for the thermal analysis (both DEM and EM). Ca XIV is used for the SUMER-EIS radiometric cross-calibration.
 We show how the very hot plasma is present and persistent almost everywhere in the core of the limb AR.
The off-limb AR is
clearly structured in Fe\,{\footnotesize XVIII}.
Almost everywhere, the EM analysis reveals plasma at $10
~\mathrm{MK}$ (visible in Fe\,{\footnotesize XIX} emission) which is
down to $0.1\%$ of EM of the main $3 ~\mathrm{MK}$ plasma. We estimate the power law index of the hot tail of the EM to be between $-8.5$ and $-4.4$. However, we leave an open question on the possible existence of a small minor peak at around $10 ~\mathrm{MK}$. 
The absence in some part of the AR of Fe\,{\footnotesize XIX} and Fe\,{\footnotesize XXIII} lines (which fall into our spectral range) enables us to determine an upper limit on the EM at such temperatures.
Our results include a new Ca\,{\footnotesize XIV} 943.59 \AA~ atomic model.  
\end{abstract}

\keywords{Sun: activity --- Sun: corona --- Sun: UV radiation --- techniques: spectroscopic}

\section{Introduction} \label{sec:intro}

Decades of observations have unveiled the difficult task of 
understanding how and where the energy is deposited for 
the creation and maintenance  of the solar corona. 
Various physical mechanisms have been proposed \citep[see][for a review]{reale14} which 
may dominate depending on the local environment. 
Active regions (AR) are the most visible manifestation of the enduring corona. These are the hottest part of the non flaring corona, with strong UV and X-ray emission. ARs 
are generally composed of coronal loops which are classed in two main
thermal categories: the hot and low-lying loops (3MK), less dense than
equilibrium conditions,  which are concentrated in the core; the warm
loops (1MK), larger and more dense than equilibrium conditions, which are 
located above the AR core  \citep{reale14}. It is not yet clear if these different
properties are the result of different physical processes at work or 
are the manifestation of the same process observed in a different evolutionary state. 

In recent years new observational signatures of coronal heating have
been identified, which we believe, if completely understood, could be
essential to progressing our understanding of the corona: 
a small amount of very hot plasma ($>5 ~\mathrm{MK}$) has been
detected in several quiescent ARs (see e.g. \cite{reale14} for a
review). This detection is important as it confirms a prediction made 
several years earlier from the the modeling of coronal loops heated
impulsively over small (sub-resolution) scales
\citep[e.g.][]{parker88, cargill94, Klimchuk06,  klimchuk15}. 
Such models predict at larger spatial scales (at the spatial resolution
reached by the modern instruments) a multi-thermal plasma along the
line of sight. The very hot plasma provides evidence of the transitory 
(short-lifetime) state of the cooling plasma in these loops. 
The importance of observing such small amounts of very hot plasma
relies on the fact that it is unique to impulsive heating events. 
This is the signature, for an unresolved spatial scale, of thin loops (strands) 
each heated independently over a short time.

One of the various diagnostics techniques used to test coronal
heating models is the sampling of the  heating-cooling process
in loops, which shows different observational signatures depending 
on the way the heating happens. The differential emission measure
(DEM) distribution with the temperature samples this process and is 
a common way of investigating the heating of loops \citep{cargill14}. 
A typical AR DEM increases  as a power law of the temperature up to a 
maximum around $ 3 ~\mathrm{MK}$, with an index which may change
and probably depends on the level  of magnetic flux. Typically 
 $EM \sim T^{3-5}$ \citep{warren12, delzanna15b}.
This index constrains the frequency at which the energy is released in
the form of heating.  At temperatures above the peak, the DEM decreases drastically, 
but the difficulty of the measurements (mostly carried out with EUV
data) makes it hard to define the shape of this distribution above 
about $ 5 ~\mathrm{MK}$. In most of the cases it can be fitted with a power law function, but simulations of nanoflares heating reveal  that this is not always the case \citep{barnes16}.  This high temperature component of the DEM 
is the signature of the first phase of cooling which, for this reason, 
possibly conserves more information on the heating process and the amount 
of energy deposited \cite[e.g.][]{parenti06}. 
The existing EUV instruments have difficulty in detecting emission
above about $ 5 ~\mathrm{MK}$ as there is very little plasma at these 
temperatures \citep{winebarger12} and because only a few relatively
weak UV spectral lines form at these temperatures.  For these reasons the DEM is not well constrained at high temperatures.
X-ray spectra are more suitable, see for instance results from SMM \cite{delzanna14} and references therein. 
The upper limits in the DEM/EM imposed by the the measured fluxes that we
found in the literature using EUV spectra are given by Hinode/EIS Ca\,{\footnotesize
  XVII}, SOHO/SUMER Fe\,{\footnotesize XVIII} and SDO/AIA 94 channel (Fe\,{\footnotesize XVIII}).
 Estimations from these datasets of the power law index 
($EM \sim  T^{\alpha}$) change from -6 to about -14
(\cite{warren12} see Table \ref{tab:slopes}).  The sounding
rocket EUNIS-13 \citep{brosius14} enabled measurements of extended
Fe\,{\footnotesize XIX} emission in an on-disk AR. Without the possibility of performing a full DEM analysis, they provided a 
Fe\,{\footnotesize XII}/Fe\,{\footnotesize XIX} EM ratio of about 0.59 in the AR core. This assesses the difference a relation between relatively cool coronal plasma ($1.5 ~\mathrm{MK}$)  and much hotter plasma at about $10 ~\mathrm{MK}$.

In recent years observations of non-flaring ARs with X-ray instruments
has intensified with the purpose of better constraining this slope
\citep[e.g.][]{miceli12, sylwester10, shestov10}.  Additional
constraints at high temperatures have been obtained combining EUV and
Hinode/XRT soft-X ray emission \citep{golub07} for an on disk AR
\citep[e.g.][]{reale09, testa11, petralia14}. The results found were
about two orders of magnitude variation of the EM from $3-10
~\mathrm{MK}$.  RHESSI, even though it is not sensitive to this faint
plasma, has revealed its presence in the $6 - 10 ~\mathrm{MK}$ 
range \citep{mctiernan09, reale09b}. Most recently, \cite{hannah16},
using the NUSTAR hard X-ray telescope, found an even larger decrease 
of the EM by imposing upper limits on this quantity due the absence 
of observed emission at high temperature. Similarly, the FOXSI 
hard X-ray emission sounding rocket \citep{ishikawa14} provided 
an upper limit to the DEM above  $8 ~\mathrm{MK}$ for an on disk AR.

In this paper we address the issue of spatially and temporally
characterizing the high temperature emission of a non-flaring active 
region with the aim of providing further constraints to the heating 
mechanism responsible for its formation and maintenance. 
Even though such plasma has been observed in several active regions, 
at present very little is known about the spatial and temporal 
distribution within the same active region. To our knowledge, this is 
the first time that the spatial distribution of this very hot
component is given at different heights above the limb. 

To minimize uncertainties in quantifying this plasma and its
temperature, we used spectroscopic data from a single element (Fe). 
For the first time this type of analysis is carried out combining 
the SOHO/SUMER and Hinode/EIS spectra allowing the observation of 
iron lines in contiguous ionization stage from Fe\,{\footnotesize X} 
to Fe\,{\footnotesize XIX} to be used.  This choice of instruments combination 
is at present unique: in provides a line sequence of a single element in a broad temperature range and lets our thermal analysis be independent of the plasma composition.
We provide evidence of
persistent emission from Fe\,{\footnotesize XIX} high in the corona, 
above the limb.

After presenting our observations in Section \ref{sec:data} and 
the data reduction in Section \ref{sec:prep}, we introduce the 
diagnostic technique in Section \ref{sec:diagn}. Section
\ref{sec:temporal} reports the results on the temporal analysis, 
Section \ref{sec:int_cal} presents the inter calibration of the two
instruments, while the results from the thermal analysis is given in 
Section \ref{sec:therm}. The conclusions are summarized in Section \ref{sec:concl}.

\section{Observations} \label{sec:data}

\begin{figure*}
\includegraphics[scale=.35,trim=0 0 250 250  ]{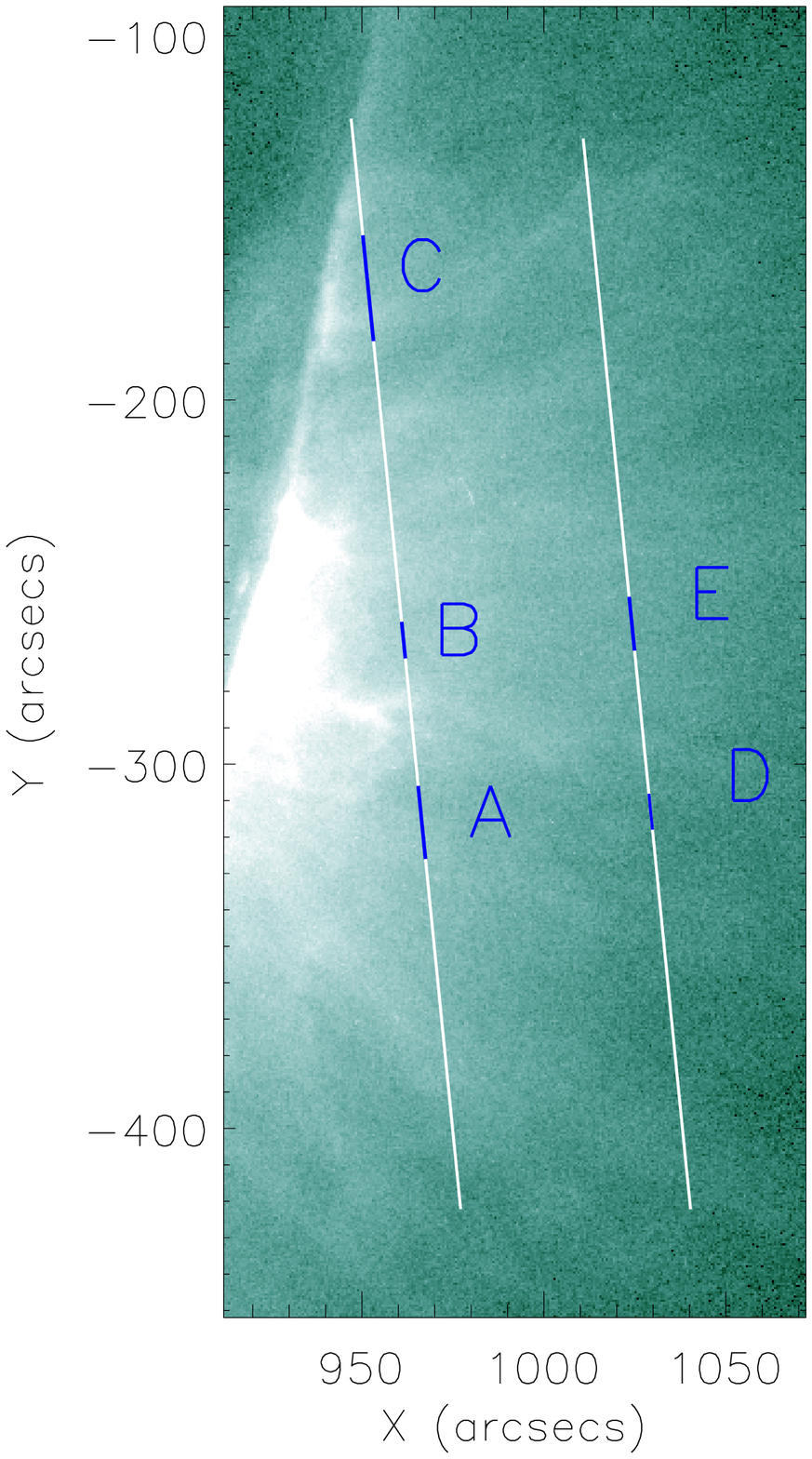} 
\includegraphics[scale=.35,trim=0 0 250 250  ]{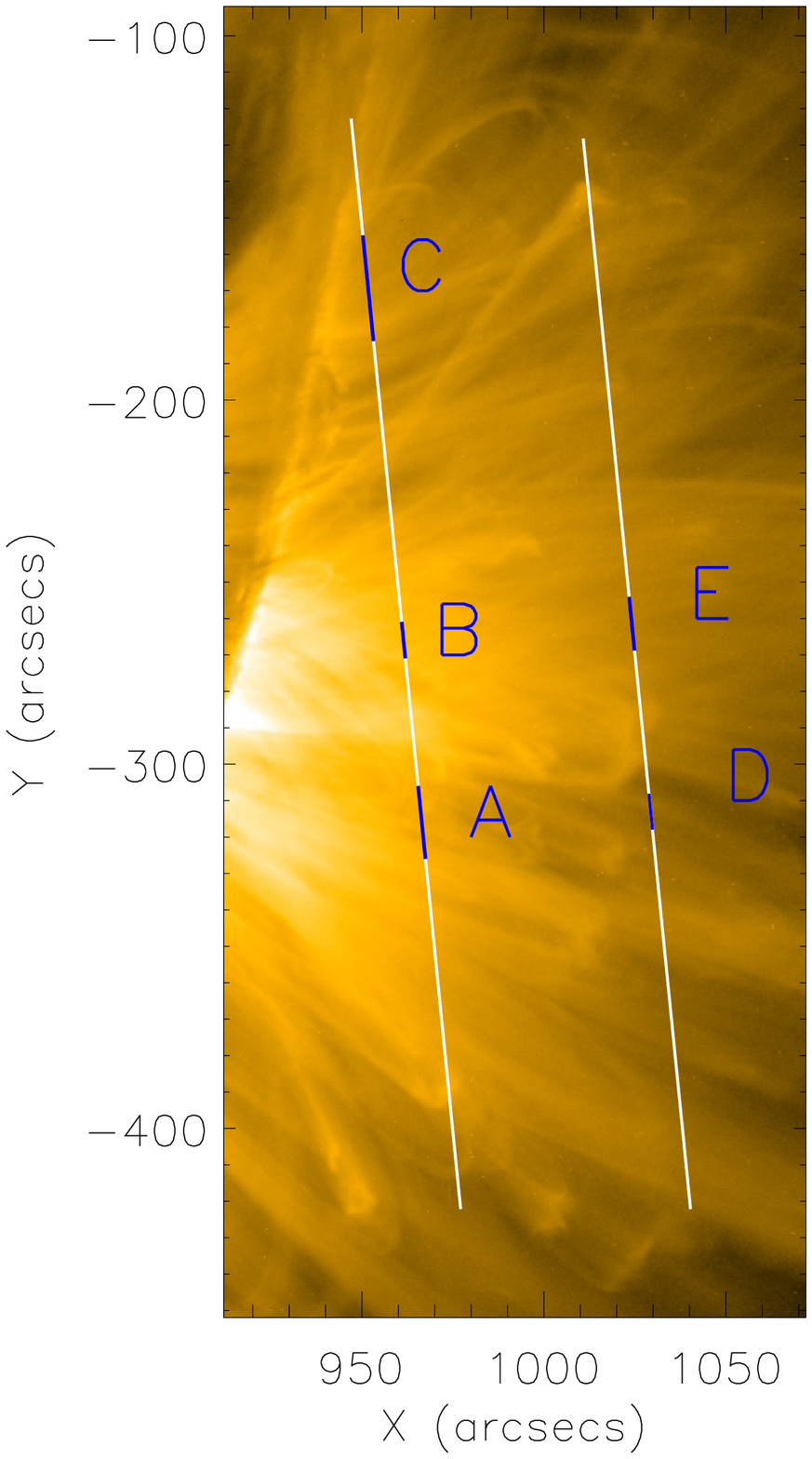}
\includegraphics[scale=.35, trim=0 0 250  0 ]{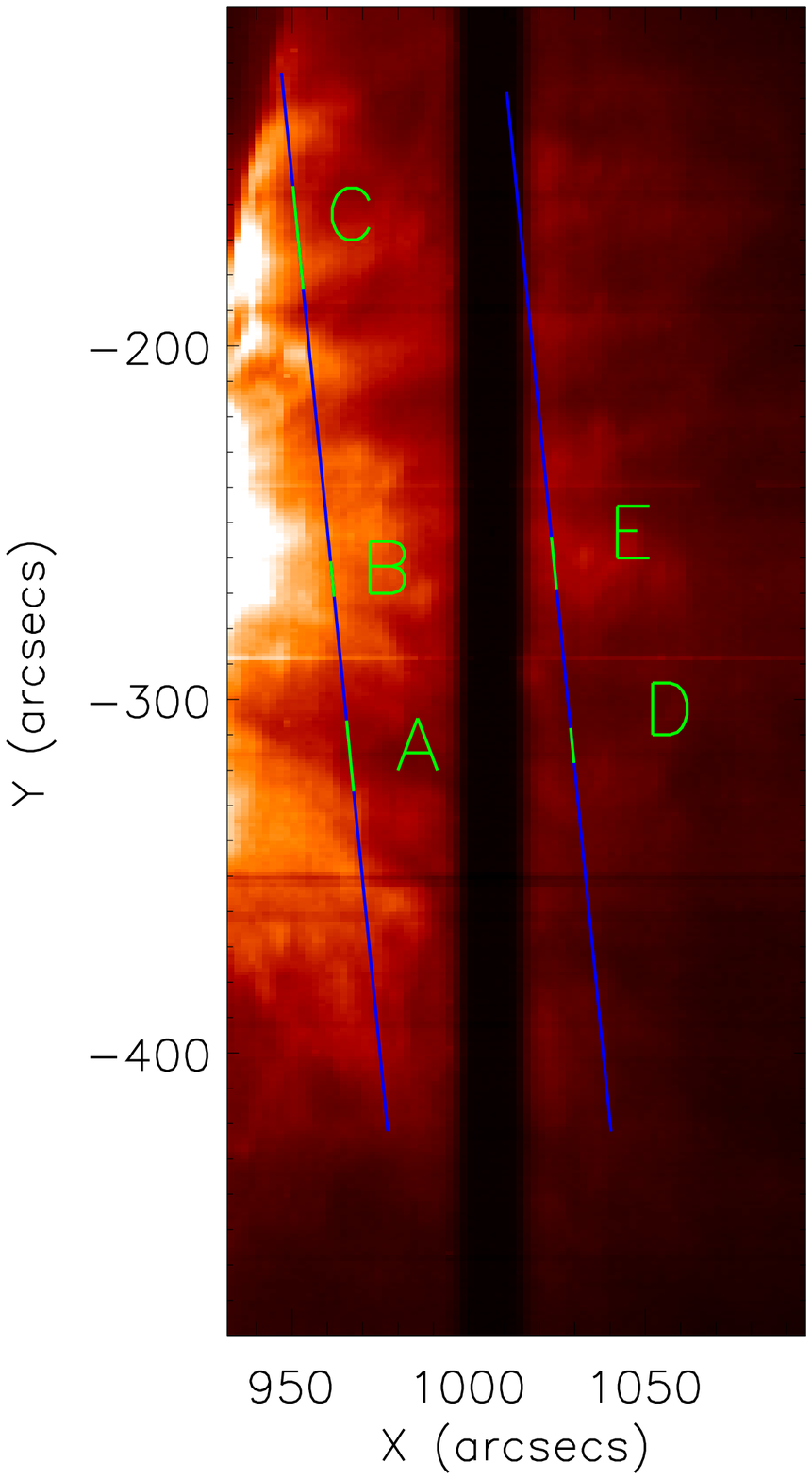}
\caption{ AIA 94 (left) and 171 (middle) images obtained by
  averaging two exposures taken between 03:06 and 03:10 UT on April 28. The SUMER slit in both positions 1 and 2 is marked by the white line. The masks selected for the analysis are also marked in color. Right: EIS raster image in Fe\,{\footnotesize XII} 192.39 \AA~ intensity which started at 23:01 UT on April 27. The SUMER slit and the masks are also marked in color. The EIS raster is affected by the  spacecraft eclipse.
}\label{fig:f1}
\end{figure*}

We ran HOP  (Hinode Operation Plan) 211 on active region 11459 for several hours at the west
limb between the 27 and 28 April 2012 using both the SOHO/SUMER
\citep{wilhelm95} and the Hinode/EIS \citep{culhane07} spectrometers (Figure \ref{fig:f1}). 

{\it SUMER}: After the loss of Detector A in 2006, due to failure of
the readout electronics, only the B Detector remained available.
In mid-2009, a degradation was observed in the center of the active area of the detector. 
As a protective measure, it was necessary to reduce the high voltage and the consequent reduction of the overall gain led to a drop of sensitivity in the central, KBr-coated, part of the photocathode, where it fell below detectable thresholds. Only the two uncoated (bare) areas of the detector (about 200 pixels each) were usable at the time of our observations.
Additionally, since the resistivity of the microchannel plate (MCP)
was observed to decrease with increasing temperature, to avoid a
runaway effect, pauses were added in the observing sequences. All
SUMER data sequences discussed here are made of sequences of 60
exposures of 75s each. After each pair of exposures a pause was added
(of about 190s for the three sequences closest to the solar limb and
of about 210s for the three sequences at a greater distance above the limb). 
Thus, each sequence can be thought as 30 regularly spaced pairs of 75s exposures.

SUMER observed in a sit-and-stare mode using the $\mathrm{1\arcsec \times 300 \arcsec}$ slit  for about eight hours (16:02:08 UT to 00:28:58UT on the 27th) centered at a solar distance of $992 \arcsec$. In the following this will be called slit position 1. 
During this time SUMER scanned the wavelength range needed to cover the observation of Fe\,{\footnotesize XVII}, Fe\,{\footnotesize XVIII} (Ca\,{\footnotesize XIV}) and Fe\,{\footnotesize XIX} as listed in Table \ref{tab:obs}. As a result, each Fe line was observed for about two hours (60 exposures with 75 sec exposure time). 
An example of the spectra for these lines is shown in Figure \ref{fig:sp_125}. 

A second sequence of observations, starting on the 28th at 00:35:55UT,
was made by  moving the pointing west of about
$\mathrm{60\arcsec}$. After a first sequence which used the same slit, the slit was changed to $\mathrm{4\arcsec \times 300 \arcsec}$ when scanning Fe\,{\footnotesize XVII} and 
{\footnotesize XIX}  (in the following we call this outer slit position as position 2, see also Table \ref{tab:obs}).

\noindent {\it EIS}: EIS data consist of three rasters nominally pointed at the two SUMER slit center positions. Each raster was scanned in about two hours starting at 18:19:33 UT. The observing program used the  $2\arcsec$ slit, which scanned eastward over 82 positions with 2\arcsec step and 90 seconds exposure time. The  final field of view is $164\arcsec \times 376\arcsec$.
Details of the observations are given in Table \ref{tab:obs}.

Our observations occured during the Hinode eclipse season, with the result that not all the common field of view of SUMER and EIS was exploitable. This is shown in Figure \ref{fig:f1}.

\noindent  We point out that the data we used are not all
co-temporal. For this study, we tried to select quiescent areas of the
AR to minimize the temporal evolution of the plasma. However, this aspect has to be taken into account in the interpretation of our results. We discuss this aspect in Section \ref{sec:temporal}.

\noindent
Prior to and during our observations, AR 11459 flared a few times, with moderate intensity. 
These flaring events left some signatures on our datasets. Some small
activity had already happened while observing with SUMER in slit
position 1, while observing the Fe\,{\footnotesize XVIII} 1118.0575
\AA. A hot loop system passed through the SUMER slit as shown in
Figures \ref{fig:slit_time} and \ref{fig:aia_sum}. 
This was also imaged by the second raster of EIS. 
The first flare happened on the 27th of April with the peak at
21:04UT. A C2.2 flare followed in the same region peaking at 23:39UT
while SUMER was observing Fe\,{\footnotesize XVII} 1153.16 \AA~
(Figure \ref{fig:slit_time}). On the 28th there was a C1.5 flare (peak
at 0:43UT) and a C1.4 (peak at 1:54UT), but our data does not look to
be affected. 
We discuss this in more detail in the following sections. 
In our analysis we also used the continuous monitoring provided by SDO/AIA for reference and comparison.

As the aim of this work is to investigate the quiescent conditions of ARs, the data affected by the flares were discarded from our analysis. Details of our data selection are given in the next Section.


\floattable
\begin{deluxetable}{ccclcc}
\tablecaption{SUMER and EIS observations for 27/28th April 2012. \label{tab:obs}}
\tablecolumns{6}
\tablenum{1}
\tablewidth{0pt}
\tablehead{
\colhead{Inst.} &\colhead{start day, hour} & \colhead{end day, hour} &
\colhead{Center}&  \colhead{Main line} & \colhead{position n.} \\
\colhead{}&\colhead{[UT]} &\colhead{[UT]} & \colhead{[x\arcsec, y\arcsec]}& \colhead{[\AA]} &\colhead{}}
\startdata
SUMER & 27, 16:00:53 & 27, 18:47:11 & [962, -271.9]    & Fe\,{\footnotesize XVIII} 974.86, & 1\\
      &              &              &                   &              Ca \,{\footnotesize XIV} 943.59  & \\
SUMER  & 27, 18:51:49& 27, 21:38:05 & [962, -271.9]    &Fe\,{\footnotesize XIX} 1118.0575 &1 \\
SUMER  & 27, 21:42:39& 28, 00:28:58 & [962, -271.9]    & Fe\,{\footnotesize XVII} 1153.1653 &1\\
SUMER & 28, 00:34:40 & 28, 03:29:55 & [1025.5, -274.7] & Fe\,{\footnotesize XVIII} 974.86, &2\\
     &              &              &                   &              Ca \,{\footnotesize XIV} 943.59 & \\
SUMER  & 28, 03:34:57 & 28, 06:30:12 & [1025.5, -274.7]& Fe\,{\footnotesize XVII} 1153.1653 & 2\\
SUMER  & 28, 06:35:07& 28, 09:30:23 & [1025.5, -274.7] &  Fe\,{\footnotesize XIX}  1118.0575 & 2\\
EIS     & 27, 18:19:33& 27, 20:24:49 & [951.7, -300.5] & Ca \,{\footnotesize XIV}   193.87 &1\\
EIS     & 27, 20:24:55& 27, 22:30:11& [951.7, -300.5]  & Ca \,{\footnotesize XIV}   193.87 &1\\
EIS     & 27, 23:01:40& 28, 01:06:56&[1012.7, -300.3] & Ca \,{\footnotesize XIV}   193.87 &2\\
\enddata
\tablecomments{The table lists the coordinates of the FOV center obtained with the  co-alignement, as well as the high temperature lines observed.}
\end{deluxetable}

\section{Data reduction}
\label{sec:prep}

The SUMER data were decompressed, wavelength-reversed, corrected for dead-time losses, local-gain depression, flat-field, and image distortion induced by the readout electronics. All the above steps were performed by using the standard software provided in  $SolarSoft$ \citep{freeland98}.

SUMER off-limb data may be affected by stray-light. We estimated this contribution in two ways (detailed  in Appendix \ref{app:stray}). 
SUMER spectra are not wavelength calibrated. We performed this calibration (presented in Appendix \ref{app:sum_wcal}).

The EIS data were corrected for instrumental effects and calibrated
applying the $eis\_prep.pro$ (version 29-Jan-2015) available on
$SolarSoft$. The radiometric calibration has recently been further
improved by \cite{delzanna13} and \cite{warren14}. For our analysis we
used the \cite{delzanna13} calibration, but a discussion of the influence on the results obtained using a different calibration  is presented in Section \ref{sec:int_cal}.

\noindent
Before performing the data analysis, we processed the data to take
into account  various issues that could affect  results. These are
presented below.

We started by carrying out a co-alignment of the fields of views of
EIS and SUMER, which was done by using the SDO/AIA images for
reference. 
This was done for two reasons: the SUMER slit follows the SOHO roll,
which is not aligned North-South, and the EIS raster data has a temporal dependence.
The task is not easy as we were dealing with off-limb data where the
structuring and image contrast are less marked than on-disk. For the
first SUMER slit position 1, we found that the best way to proceed 
to co-align SUMER-AIA was to use the flare data (SUMER
Fe\,{\footnotesize XVIII} and AIA 94 channel).  For the slit position
2 we used the pair SUMER Ca\,{\footnotesize X} 557.766~\AA ~ with AIA
171, and SUMER Ca\,{\footnotesize X} 574.011 \AA~ with AIA 171. 
The two Ca lines are observed about two hours apart, but the result of 
these two alignments were similar within 2\arcsec.
Details of this work are given in Appendix \ref{app:align} and the
results are listed in column four of Table \ref{tab:obs}.

It is known that EIS is not aligned with respect to AIA. We first used
the $eis\_aia\_offset.pro$ IDL procedure available in the $SolarSoft$
database to correct our EIS data. However, we were not satisfied with
our results and we decided to proceed with an independent manual
co-alignment based on cross-correlation. We used the AIA 195 channel
and EIS Fe\,{\footnotesize XII} 192.394~\AA. The details of the method
are given in Appendix \ref{app:align} and the results are given in Table \ref{tab:obs}.

\subsection{Spectral  line selection}
\label{sec:lines_sel}

Table \ref{tab:lines} gives the list of the observed spectral lines and their fluxes used in this work.
To reduce the uncertainties in the results of the thermal analysis, we minimized the effect of the
elements abundance, by selecting only lines from Fe ions. We have
enough Fe ions to cover the temperature range $ 6 < \log T <
6.95$. In addition, as we shall show, we performed several tests to 
monitor the effects of elemental composition on our results.  
To this line list we added Si\,{\footnotesize VII} to constrain the DEM and EM at low temperatures and Ca\,{\footnotesize XIV}, which was used for the intercalibration EIS-SUMER (see Sec. \ref{sec:int_cal}). 
This is, in fact, the only ion which is present in both instruments whose lines are free from blends. 
Our results should not be affected by element abundance variations as the chosen ions are all 
from low First Ionization Potential elements. In addition, the
Ca\,{\footnotesize XIV} is formed at a temperature similar to Fe
\,{\footnotesize XV} and Fe \,{\footnotesize XVI} 
which were observed by EIS.

The EIS high temperature lines from ions above Ca\,{\footnotesize XIV}
and Fe\,{\footnotesize XVI} were absent or unusable because they
were too blended. We also tried to extract Ca\,{\footnotesize XVII}
192.853 ~\AA~ but with no success. To do this, we used the \cite{ko09}
method, which deblends this line from the Fe \,{\footnotesize XI} and
O V lines. We used the theoretical CHIANTI ratio between the  
Fe \,{\footnotesize XI} 188.1/192.8 to fix the Fe \,{\footnotesize XI}
192.8 ~\AA~ profile parameters, but could not find any good solutions
to the multi-line fitting. We then recalculated this ratio directly
for the data by selecting a quiet off-disk area far from the limb, 
where the 192.8 ~\AA~ line was very symmetric, and did not show any
evidence of the presence of O \,{\footnotesize V} and
Ca\,{\footnotesize XVII}. We then assumed that O \,{\footnotesize V}
and Ca\,{\footnotesize XVII} were absent. With this new ratio value we
found the best solution of the multi-Gaussian fitting was one in which
Ca\,{\footnotesize XVII} line was absent. We concluded that in our
data, this line was absent or too faint to be measured.
In conclusion, the high temperature plasma is only covered by the SUMER data.
Our analysis was carried out over several masks as described in the 
following Sections. We used the $eis\_mask\_spectrum.pro$ to select
the masks in the EIS rasters and $spec\_gauss\_eis.pro$ available in
$SolarSoft$ to perform a multi-Gaussian fitting of the resulting spectra.

\begin{figure*}[ht]

\includegraphics[scale= 0.35, angle=90]{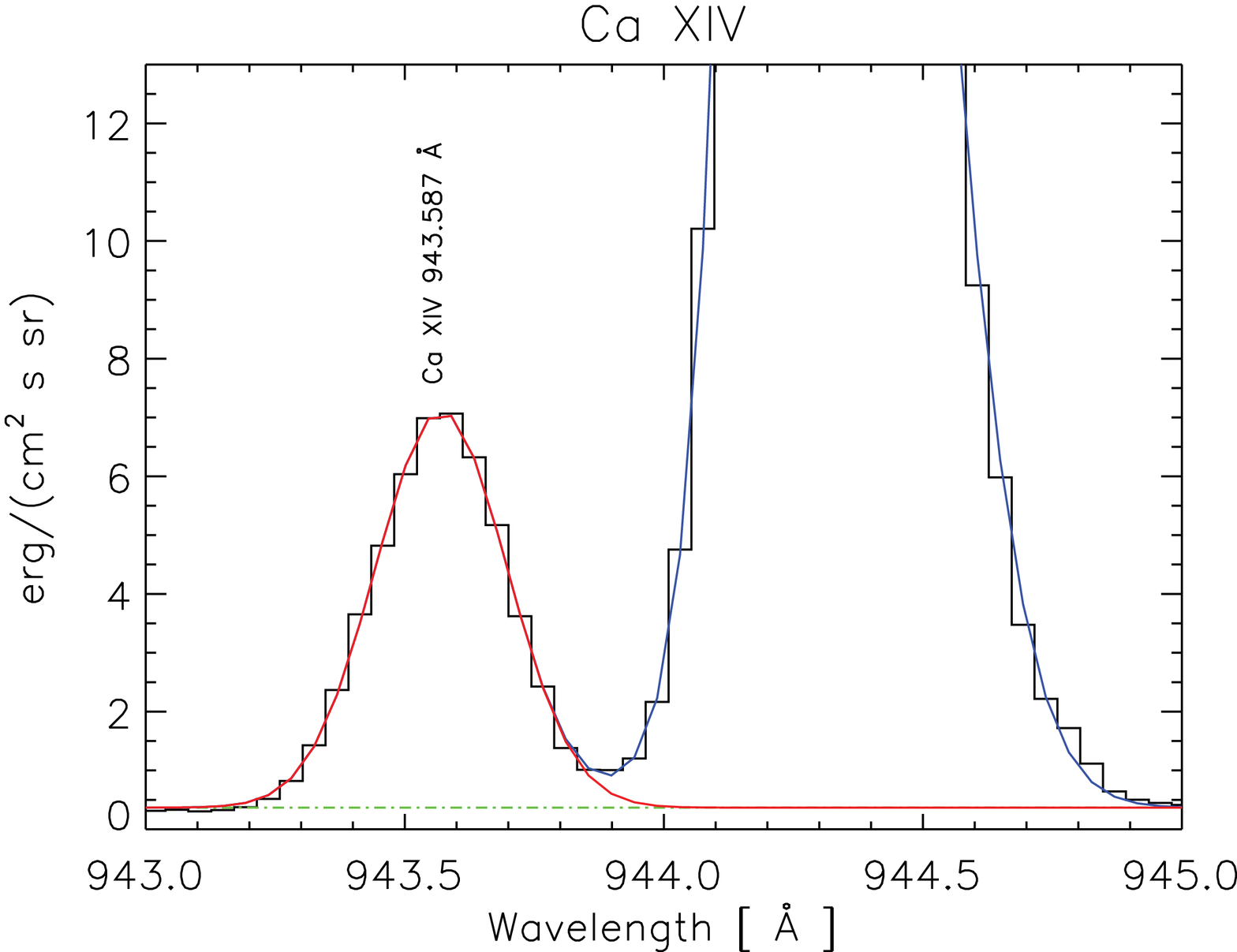}
\includegraphics[scale= 0.35, angle=90]{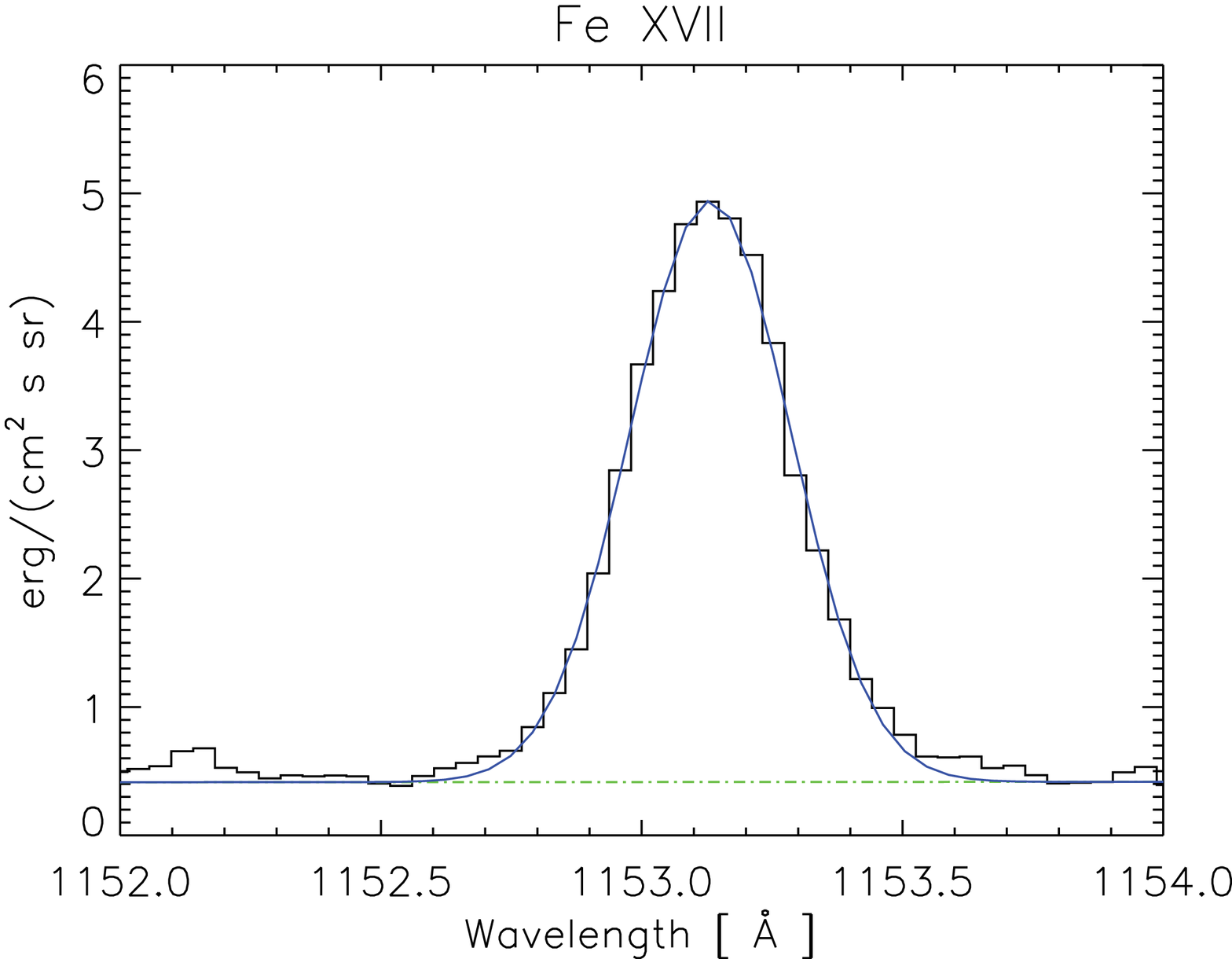}\\
\includegraphics[scale= 0.35, angle=90]{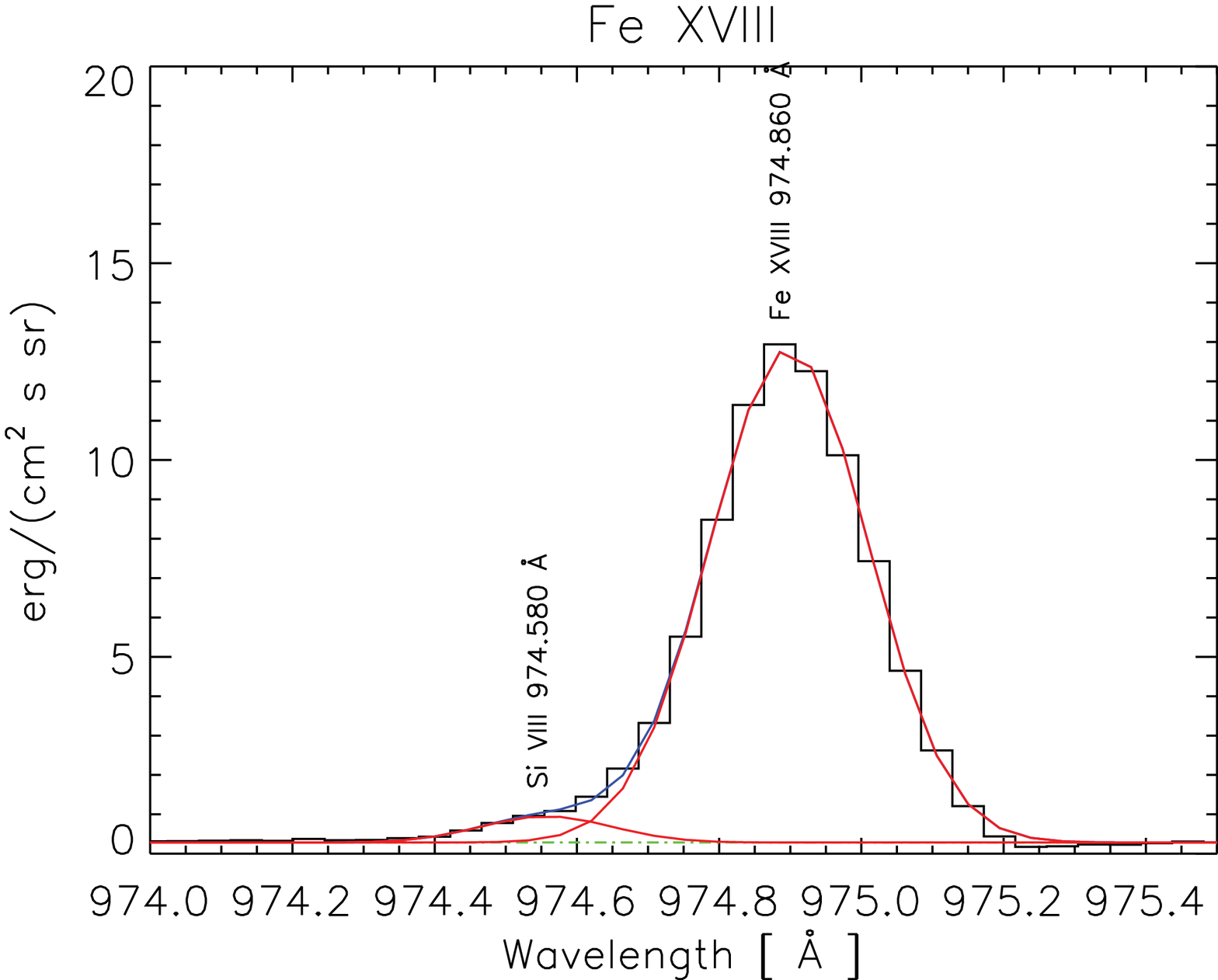}
\includegraphics[scale= 0.35, angle=90]{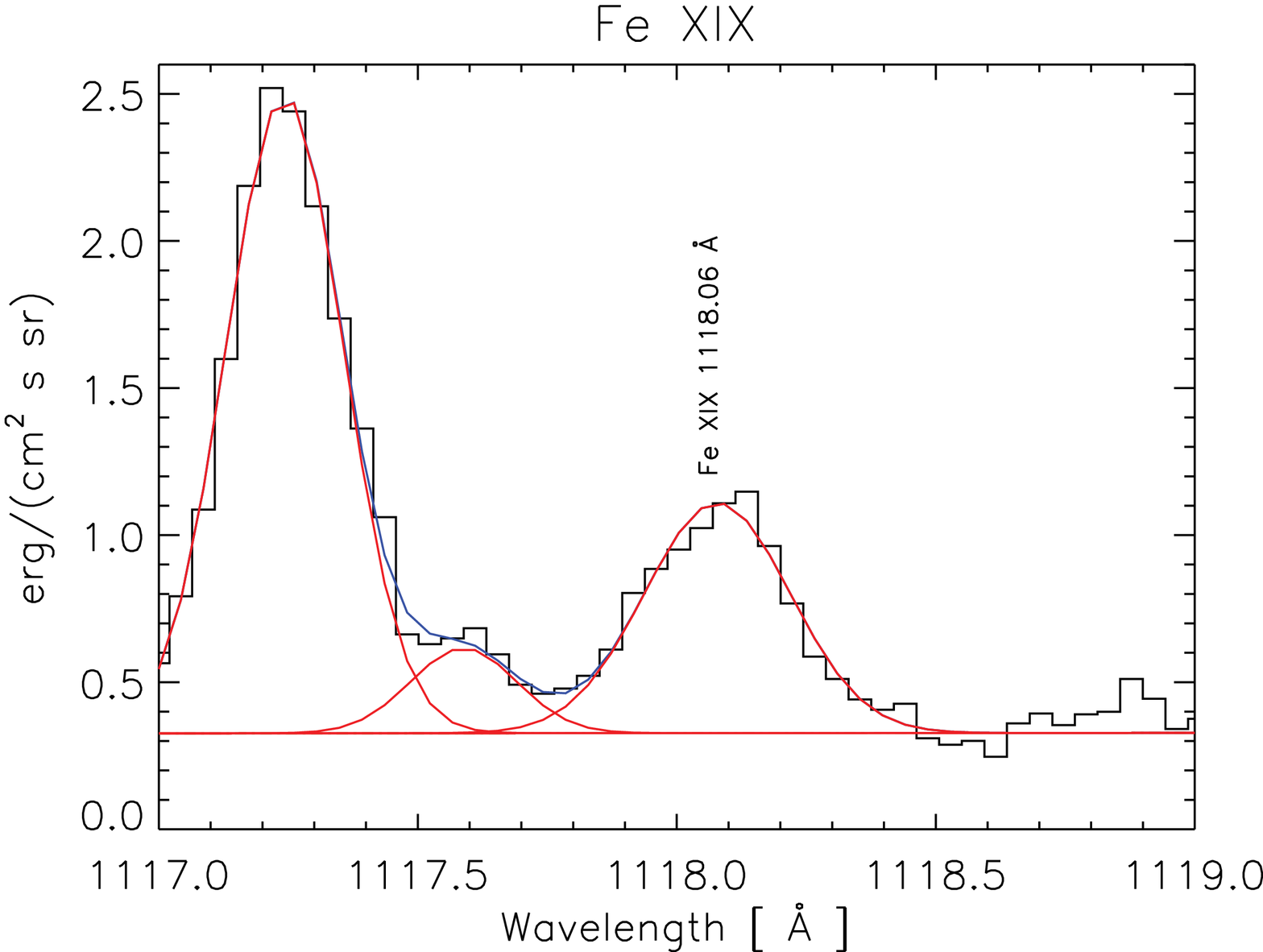}
\caption{Spectra of mask $A$ for the SUMER hot lines and their resulting fit. The red curve marks the single Gaussian fit, while the blue curve is the sum.}
\label{fig:sp_125}
\end{figure*}

Amongst the hot lines in the SUMER spectra, the Fe \,{\footnotesize
  XVIII} 974.860 \AA~ blend required particular attention (see Figure \ref{fig:sp_125}). 
This blend has previously been discussed by \cite{teriaca12} and is
due to the presence of a ghost image (caused by the electronics)  
of the H\, {I} 972.54 \AA ~line falling about 2.8 \AA~  red-ward of
this line,  and Si \,{\footnotesize VIII} 974.58 \AA. The amplitude of
the ghost has been estimated by these authors to be about 1/200 the
main line and it cannot easily be fitted within the blend. 
Having the H\, {I} 972.54 \AA ~line in our spectra, we could estimate
the ghost by adopting this value to correct our Fe line flux. We found
its contribution to be  only $1\%$ to $2\%$. The Fe \,{\footnotesize
  XVIII} line was fitted using two Gaussians profiles in order to 
deblend it from the Si \,{\footnotesize VIII}. However, the result for 
the different masks was not always satisfying. We then estimated the 
contribution of the Si \,{\footnotesize VIII} by calculating the
intensity ratio with the Si \,{\footnotesize VIII} 1182.48 \AA ~lines
we had in the spectrum, which is well isolated and has similar amplitude. 
We estimated this ratio from our data by selecting a quiet region
(mask $F$) where Fe \,{\footnotesize XIX} is absent and we expected to
have none or a very small contribution of Fe \,{\footnotesize XVIII} to the blend. 
The result of the double Gaussian fit to the spectra of mask $F$ gives 
the Si \,{\footnotesize VIII} 974.58/ 1182.48 \AA ~= 1.04. This value is consistent with  
\cite{feldman97} off-limb observation who found a value 1 for this
ratio, while \cite{curdt04} found a smaller value (0.7). We adopted
our result to correct the Fe \,{\footnotesize XVIII} in our masks.
Considering that in the masks chosen for the thermal analysis the 
Fe \,{\footnotesize XVIII} line is about 20 times brighter than the 
Si \,{\footnotesize VIII} 1182.48 \AA, our uncertainty on the blend
remains within the assumed $20\%$ uncertainty of the fluxes.

We indeed assumed an error of $20\%$ in the lines flux. This value
includes uncertainties in the atomic physics, ionization/recombination
rates calculations (which, for certain ions could be larger). We are
aware that other factors may arise to increase the uncertainty. 
For instance, the EIS calibration which will be discussed in
Sec. \ref{sec:int_cal}. The lack of co-temporal data may also imply
some inconsistency in the analysis. This will be discussed in Sec. \ref{sec:concl}.

Figure \ref{fig:sp_125} shows an example of the hot lines in the SUMER
spectra for mask $A$ together with the result of multi-Gaussian fitting.

\subsection{Upper limit on the flux for Fe \,{\footnotesize XIX} and Fe \,{\footnotesize XXIII}}
\label{sec:err}

Above $3~ \mathrm{MK}$ the non flaring plasma produces faint emission. To better constrain our thermal analysis we estimated an upper limit of the flux for those flaring lines falling in the spectral range of our data, but which were not visible. This is the case for SUMER Fe \,{\footnotesize XIX} 1118.06 \AA~~  in mask $B$ and for the EIS Fe \,{\footnotesize XXIII} 263.77 \AA~  in the other masks.

Assuming a Poisson background statistics approximated by a Gaussian,
we set a $3 \sigma$ confidence level as the minimum threshold for
detection of the line signal above the background photon counts. We
selected a wavelength interval similar to the expected line width. We
used this to calculate the minimum flux needed for the spectral line to be measurable.  The results are given in Table \ref{tab:lines}.

\floattable
\begin{deluxetable}{lrrrrrrrrc}
\tablecaption{Lines list and total fluxes (I).
\label{tab:lines}}
\tablecolumns{10}
\tablenum{2}
\tablewidth{0pt}
\tablehead{
\colhead{Ion} &\colhead{ $\lambda$}& \colhead{log $T_{\rm max}$} & \colhead{I(A)}& 
\colhead{I(B)} &  \colhead{I(C)}& \colhead{I(C)} & \colhead{I(D)}&\colhead{I(E)} & \colhead{Inst.}\\
\colhead{} &\colhead{[\AA]} &\colhead{[K]} &\colhead{} &\colhead{} &\colhead{[18:19UT]} &\colhead{[20:24UT]} &\colhead{} &\colhead{} &\colhead{}}
\startdata
Si \,{\footnotesize VII}    &  275.38 &  5.79 &   299.7 &  525.1 & 53.6   & 75.7   &  56   & 135.2  & EIS \\
Fe \,{\footnotesize X}(sbl) & 257.26  &  6.04 &  1347.4 & 2189.9 &  526.3 &  674.8 &  586.3  & 988  & EIS \\
Fe \,{\footnotesize XI}     & 188.30  &  6.13 &  1258.0 & 1764.1 & 868.5  & 917.2  &  640.4  & 935.9 &EIS \\
Fe \,{\footnotesize XII}    &  192.39 &  6.20 &   953.5 & 1283.0 & 878.4  & 926.8  & 446.2   & 656.6 &EIS \\
Fe \,{\footnotesize XIII}   &  202.04 &  6.25 &  2824.1 & 3279.1 & 2999.9 & 3041.3 & 1504.6 & 2025.3& EIS \\
Fe \,{\footnotesize XIV}    &  264.79 &  6.29 &  2264.3 & 2741.5 & 2532.4 & 2474.0 &  654.6  & 1038.07 & EIS \\
Fe \,{\footnotesize XV}     &  284.16 &  6.34 & 19078.8 &21675.8 & 18155.5& 17585  & 7848.5  & 10187.5& EIS\\
Fe \,{\footnotesize XVI}    & 262.98  &  6.43 &  1474.1 &1565.3  & 1154.9 & 946.7  &  384.0  & 517.7  & EIS \\
Ca \,{\footnotesize XIV}    &  193.87 &  6.57 &   112.1 &  92.3  & 63.8   & 45.5   &   28.6  & 32.8 &  EIS \\
Ca \,{\footnotesize XIV}    &  943.59 &  6.56 &     4.0 &  3.79  & 2.72   & 2.72 & 1.44  & 1.76 & SUMER \\
Fe \,{\footnotesize XVII}   & 1153.16 &  6.73 &     3.4 &   2.7  & 1.82   & 1.82 & 0.97  &  1.6  & SUMER \\
Fe \,{\footnotesize XVIII}  &  974.86 &  6.86 &     6.5 &   4.8  & 3.18   & 3.18 & 2.17  & 2.77 & SUMER \\
Fe \,{\footnotesize XIX}    &  1118.06&  6.95 &     0.5 & $<$ 0.11 &  0.76  & 0.76 & 0.22  &  0.22 &SUMER \\
Fe \,{\footnotesize XXIII}  & 263.765 &  7.16 &  $<$3 &        & $<$3  & $<$3 & $<$3 & $<$3 &EIS \\
\enddata
\tablecomments{SUMER and EIS lines list and total fluxes (I) for each mask (from $A$ to $E$) used for the analysis of Sec. \ref{sec:therm}. The fluxes are in [$erg/cm^2/s/sr$]. The theoretical position of the line and the temperature of maximum formation are given in columns two and three.
($<$ ): upper limit imposed as the line is not visible; (sbl): self-blended line.}
\end{deluxetable}

\section{Plasma diagnostics methods}
\label{sec:diagn}

We summarise the plasma diagnostics that are used in this work, while further details can be found in \cite{parenti15}. 

For the EUV-UV optically thin lines the total intensity is given by

\begin{equation}\label{eq:I2}
I(\lambda)= \frac{1}{4\pi}  \int_l{Ab~ G(T_e, n_e)~ n_e n_H dl}
\end{equation}

where $l$ is the line of sight through the emitting plasma,  $Ab$ is  the abundance of the element with respect to hydrogen, $G(T_e, n_e)$ is the $contribution ~function$
which contains all the atomic physics  parameters,   $n_e$ and $T_e$ are the electron number density and temperature and $n_H$ is the hydrogen number density.

For the thermal analysis we need to know the electron density distribution in the AR. 
One method of estimating the electron density is given by calculating
the ratio of two line intensities from the same ion, where at least
one involves a metastable level $m$. This make the ratio density
dependent, assuming the temperature of the maximum of the ionization fraction of the ion.
The density is then inferred by matching the ratio derived from observations with the  theoretical value calculated at different densities. The results of this analysis are presented in App. \ref{app:dens}.

To estimate the distribution of the plasma emission measure with the
temperature, there are various methods which differ in the
approximations applied. 
We introduce the {\it column emission measure}  \citep[EM,][]{Ivanov-Kholodnyi63, pottasch63} along the line of sight $l$, as 

\begin{equation}\label{eq:em}
 EM = \int_l n_e n_H dl
\end{equation}

If we assume that the emitting plasma along $l$ is isothermal at $T_c$ and  the electron density known ($n_c$), using Eq. \ref{eq:I2} we can write

\begin{equation}\label{eq:i_em}
EM (T_c) = \frac{4\pi I(\lambda)} {G(T_c, n_c)}
\end{equation}

If the density is unknown, a first approximation test on the temperature distribution of the plasma is given by plotting EM($T_e$) as function of $T_e$, for a set of lines formed in a wide range of temperatures. This is called the emission measure loci approach.  
A necessary condition for the plasma to be isothermal is that these curves cross at the plasma temperature. However, this plot gives only an indication of the plasma distribution as the uncertainties on the measures and data inversion can introduce additional solutions to the inversion \citep{guennou12}.

When a set of lines from optically thin plasma formed at different
temperatures is available, the differential emission measure (DEM)
inversion is more appropriate to probe the multi-thermal plasma. 
The DEM is proportional to $n_e n_H$ (variable with $T_e$) in the temperature intervals $dT_e$ and it is defined here as

\begin{equation}\label{eq_dem}
  \mbox{\it DEM}(T_e) = n_e n_H \frac{dl}{dT_e} 
\end{equation}

We also introduce the effective temperature, which is the DEM-weighted average: 

\begin{equation}
T_{eff} = \frac{\int DEM(T) \times T dT}{\int DEM(T) dT}. 
\end{equation}
 

The thermal analysis using these diagnostics is presented in Sec. \ref{sec:therm}.

For the diagnostic analysis in this work we used the CHIANTI v.8 atomic database and software \citep{dere97, delzanna15} to calculate the theoretical emissivities of the spectral lines. A particular case was the treatment of  Ca\,{\footnotesize XIV} 943.59 ~\AA, which will be discussed in Sec \ref{sec:int_cal}. 

We adopted the CHIANTI default ionization equilibrium, and \cite{feldman92b} elemental abundances. This latter choice was made after several tests we did changing element composition in the EM loci approach and DEM inversions. Even if the effect of composition on the result is small, we found \cite{feldman92b} data to produce the better observed versus theoretical fluxes ratio.  
We adopted a density of $10^9 ~\mathrm{cm^{-3}}$ for the SUMER slit in position 1 and $4 \times10^8 ~\mathrm{cm^{-3}}$ for position 2 derived from the analysis presented in Appendix \ref{app:dens}.

\section{Temporal variability and sub-data selection}
\label{sec:temporal}
We first made an investigation of the spatial and temporal properties
of the AR along the SUMER slit  in the hot lines. This has been useful
to the selection of the region of interest for detailed investigation. We mostly used SUMER data, where the signal is strong.

\subsection{Spatial structuring and temporal variability in hot lines}

\begin{figure}[h]
\includegraphics[width=\linewidth]{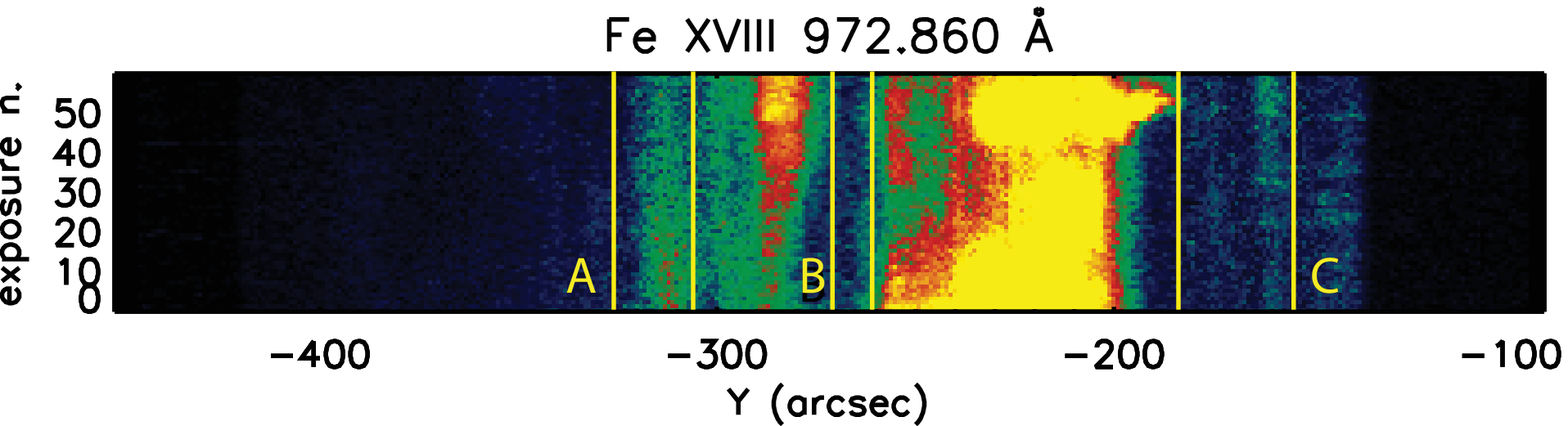}\\
\includegraphics[width=\linewidth]{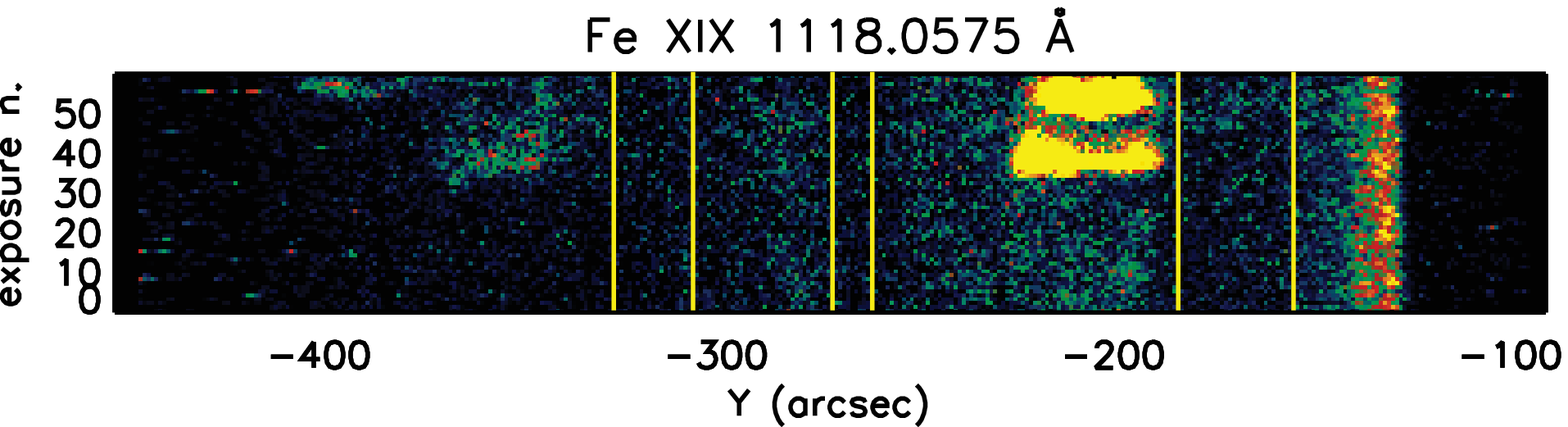}\\
\includegraphics[width=\linewidth]{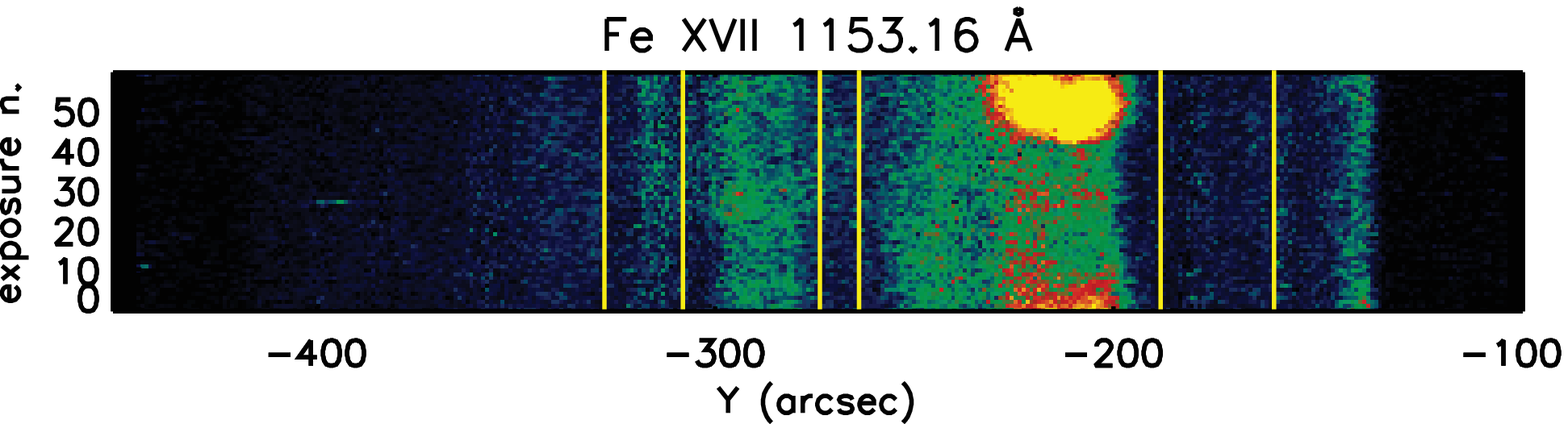}
\caption{SUMER Fe\,{\footnotesize XVII}, Fe\,{\footnotesize XVIII} and {\footnotesize XIX} intensities along the slit in position 1, plotted for the sixty exposures (with 75 seconds exposure time). The images have been saturated to enhance the fainter emission at the side of the flare. The colour range represents, from top to bottom, 0-99, 0-10, 0-40 counts/pixel. 
The three vertical pairs of yellow lines mark the three masks $A$, $B$ and $C$. The limb is on the right side of the plots.  }
\label{fig:slit_time}
\end{figure}

Figure \ref{fig:slit_time} shows the intensity of the SUMER Fe\,{\footnotesize XVII} -- {\footnotesize XIX} lines along the slit and for the sixty exposures on April the 27th, slit position $1$. From the top to the bottom, they have been ordered following the time sequence of data acquisition. 
The plots have been saturated to highlight the fainter structuring of the AR, and the brightest regions in yellow correspond to the flaring areas  (around $Y = -230\arcsec$).

The Fe\,{\footnotesize XVIII} sequence shows a very bright area
corresponding to the passage of a hot loop accross the slit (this is
the one used to co-align SUMER and EIS, see also Figure
\ref{fig:aia_sum} which is plotted using a different contrast). At
both its sides of the hot loop we can identify faint structures whose flux seems to
remain almost constant in time. This suggests that they are not
affected much by the flare. 
We  highlight them in the Figure \ref{fig:slit_time} by pairs of vertical lines (see also Table \ref{tab:masks}, masks $A$ and $C$). 
In line with our aim of  investigating  non-flaring regions, we selected
these two area as candidates to pursue our analysis. We also added a third region which was taken as a sample of the unstructured emission (mask $B$), which appears faint in all the plots of Figure \ref{fig:slit_time}.  The details of these masks are listed in Table \ref{tab:masks}. Some structuring is still visible in the AR core. 

A similar analysis has been carried out for the SUMER slit position 2
and is reported in Annex \ref{app:sum_p2}. Here we identified
two other candidate regions for the thermal analysis. 

The structuring of the AR in hot lines has been extracted by
temporally averaging the exposures. Reducing the background noise thus
makes it easier to see the fainter regions (section
\ref{sec:t_avg}). We also carried out a temporal analysis of these
regions to understand better how much their flux varies with the flaring activity (section \ref{sec:t_tot}). 


\subsubsection{Spatial structuring}
\label{sec:t_avg}

\begin{figure}[h]
\includegraphics[width=\linewidth]{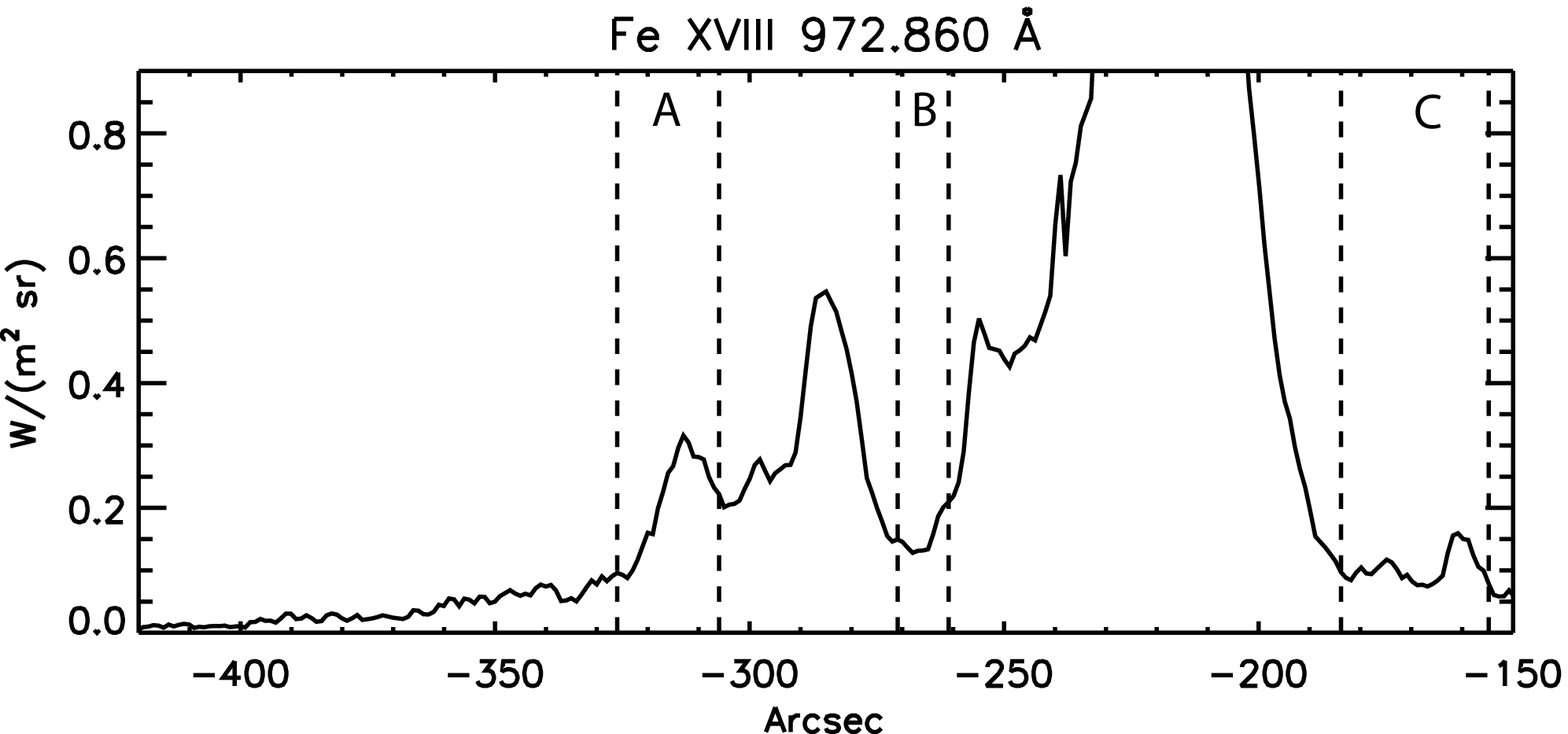}
\includegraphics[width=\linewidth]{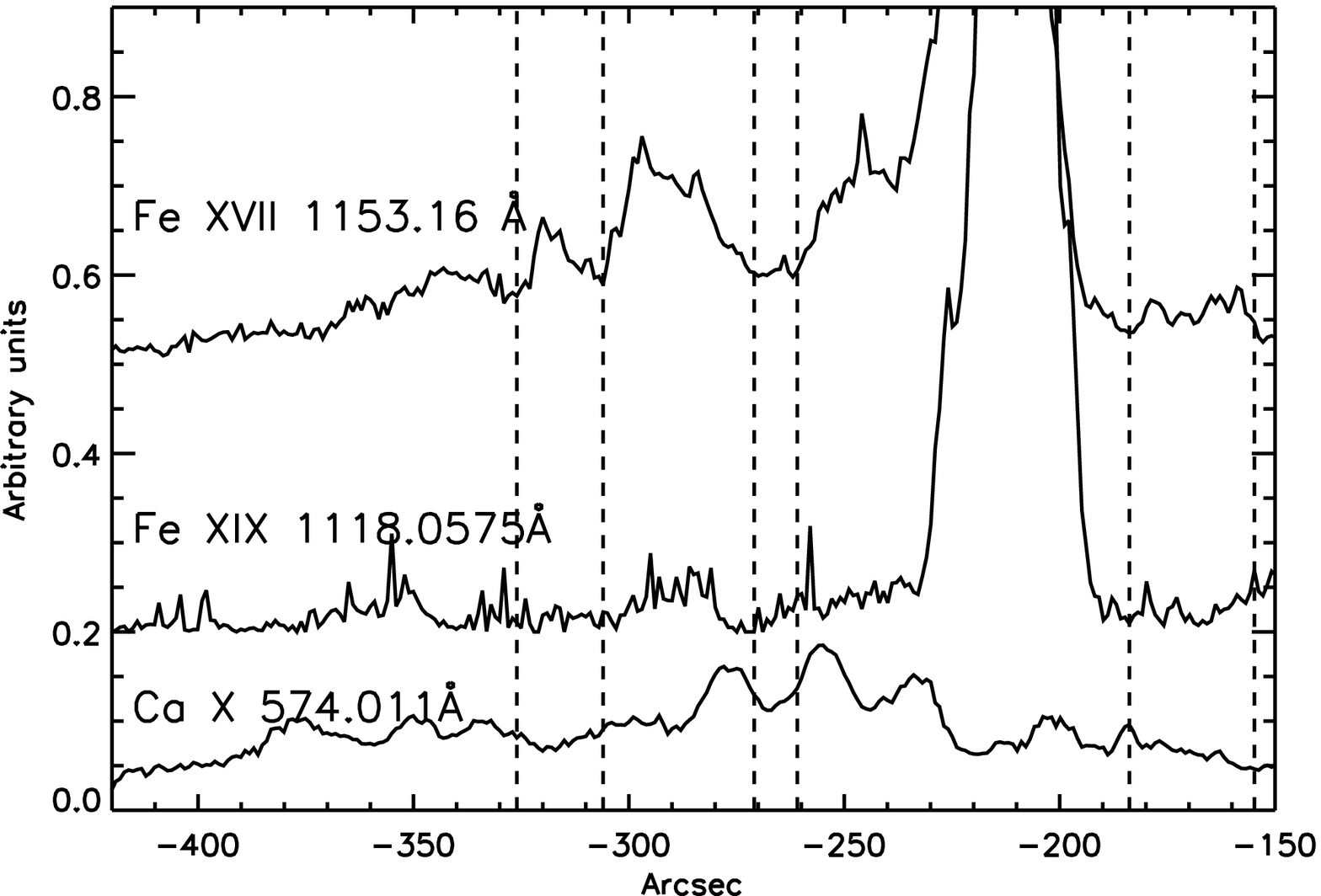}
\caption{Top: SUMER Fe\,{\footnotesize XVIII} intensity along the slit
  integrated over the sixty exposures. The pairs of vertical dashed
  lines mark the three areas selected for further analysis. Bottom: 
same as the top plot for Fe\,{\footnotesize XVII}, Fe\,{\footnotesize XIX} and Ca\,{\footnotesize X}.
\label{fig:fe18_slit}}
\end{figure}

Figure \ref{fig:fe18_slit} top shows the integrated intensity of the
Fe\,{\footnotesize XVIII} line profile for the spectra averaged over
the sixty exposures, plotted along the SUMER slit in position 1 using 
one pixel spatial resolution. The north is on the right of the plot.
This figure shows well the structuring of the AR. As in Figure
\ref{fig:slit_time}, the data in the range approximately Y=
[-260\arcsec, -190\arcsec] are due to the hot loop, while the vertical 
pairs of dashed lines mark the three areas (masks) selected for the analysis. 

\noindent Figure \ref{fig:fe18_slit} bottom shows the same integrated
intensities for the SUMER Fe\,{\footnotesize XVII}, Fe\,{\footnotesize
  XIX} and Ca\,{\footnotesize X}. The latter is used to represent the
$1 \mathrm{MK}$ corona. The three Fe lines are observed in three
different spectral windows, which means that they are not co-temporal 
(only Ca\,{\footnotesize X} is co-temporal to Fe\,{\footnotesize
  XVII}). Nontheless, the intense  Fe\,{\footnotesize XVII} and  
Fe\,{\footnotesize XVIII} lines have a similar pattern along the slit, 
which highlights brighter structures and fainter areas. 
On the other hand,  the Fe\,{\footnotesize XIX} intensity outside 
the flaring region is very weak and there is little we can say using 
the intensity of a single pixel. In order to be able to use this line, 
for each of the selected regions we have spatially averaged the
spectra. The three regions are marked by the pairs of vertical dashed lines.
Going from the bottom of the slit towards the top, we find mask $A$: 
this is one of the faintest and most persistent structures in 
Fe\,{\footnotesize XVII} and Fe\,{\footnotesize XVIII}. Mask $B$ is
located close to the slit centre and outlines a weak area between 
two bright ones. Mask $C$ is the closest to the limb and contains weak structuring.

\floattable
\begin{deluxetable}{ccccc}
\tablecaption{The masks along the SUMER slit selected for the analysis. \label{tab:masks}}
\tablecolumns{4}
\tablenum{3}
\tablewidth{20pt}
\tablehead{
\colhead{Mask} &
\colhead{Date} &
\colhead{$Y1\arcsec <Y< Y2\arcsec$ } & \colhead{$R_\sun\arcsec$} 
}
\startdata
A & 2012-04-27 & -326,  - 306 & 1007 \\
B & 2012-04-27 & - 271, - 261 & 997  \\
C & 2012-04-27 & -181.8, -154.8 & 958\\
D & 2012-04-28 &  -320.5, -310.5& 1075 \\
E & 2012-04-28 & -268.8, -254 & 1052\\
\hline
F & 2012-04-28 & -421.2, -401.5 & 1117.7\\
G & 2012-04-28 & -421.2, -352.5 & 1076\\
\enddata
\tablecomments{ The last two masks are used to test blends (see Sec. \ref{sec:lines_sel}) and perform the SUMER wavelength calibration  (described in Appendix \ref{app:sum_wcal}).}
\end{deluxetable}

\subsubsection{Temporal variability of the selected regions}
\label{sec:t_tot}

\begin{figure}[h]
\includegraphics[width=0.95 \linewidth]{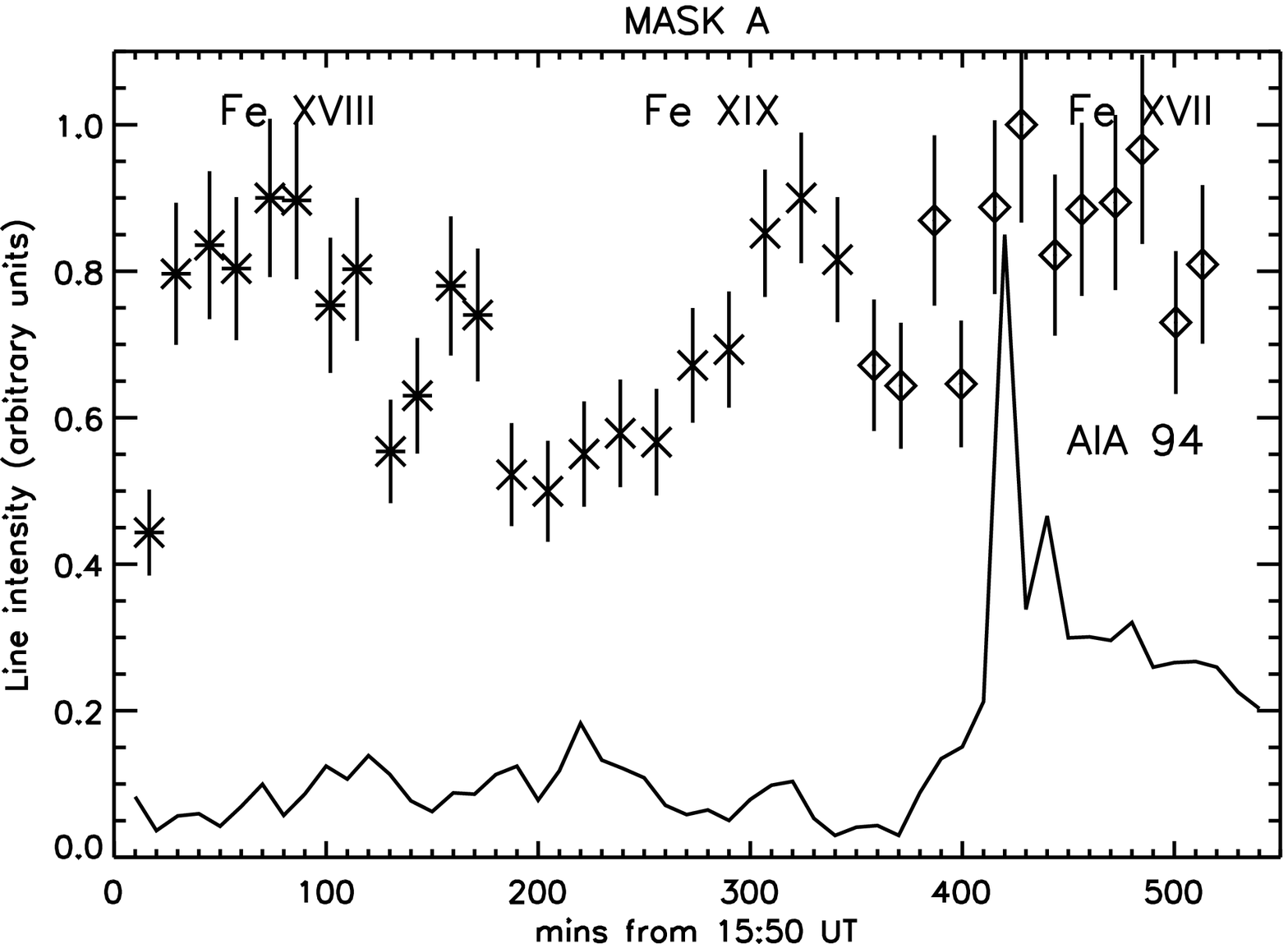}\\
\includegraphics[width=0.95\linewidth]{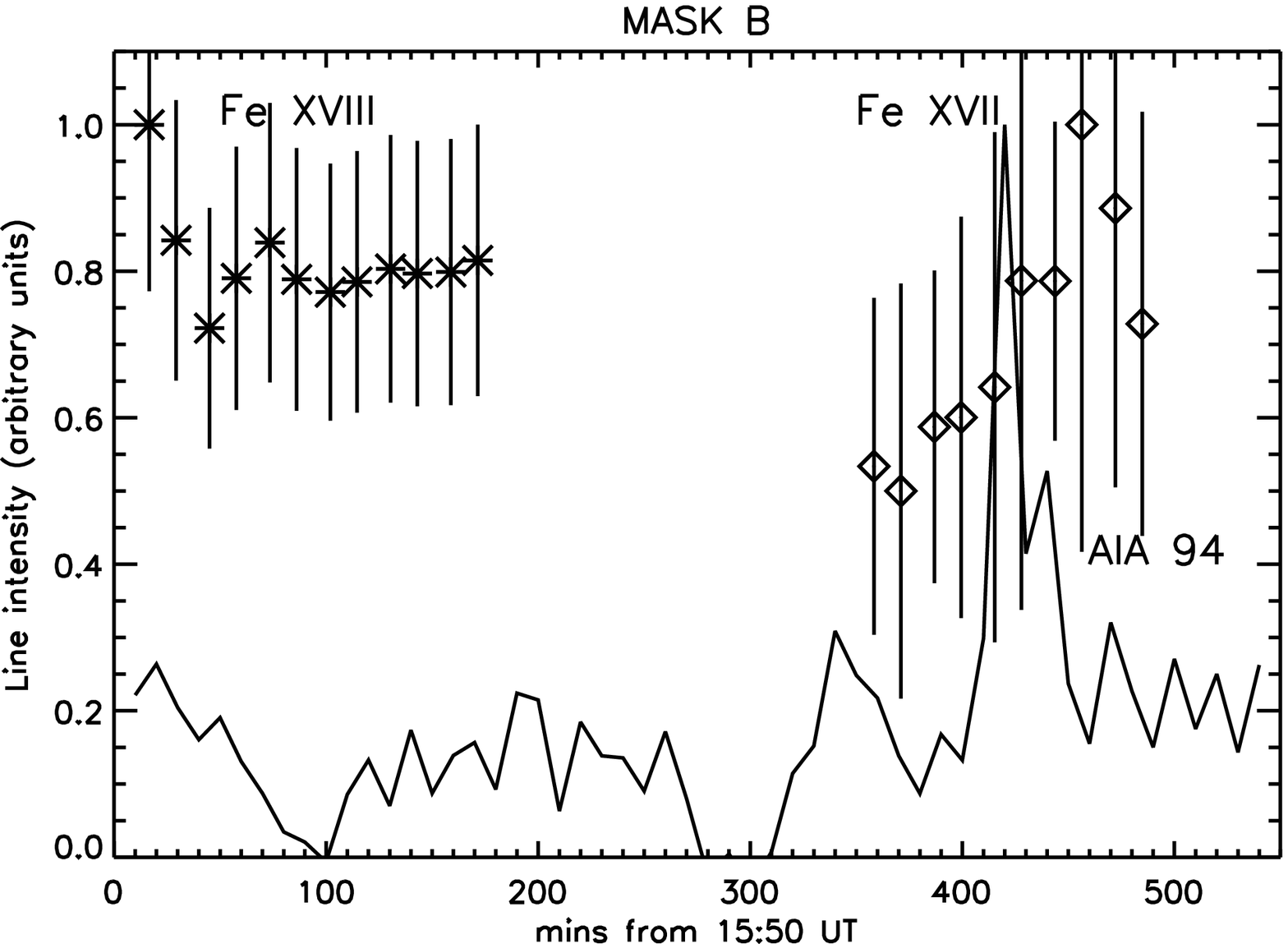}
\caption{Light curves for the SUMER hot lines for mask $A$ and $B$ for slit position 1 compared to the AIA 94 corrected with the 171 channel. The fluxes from the three SUMER lines have been plotted using a different symbols. The error bars are from the line profile fitting accuracy.
\label{fig:ligth_ab}}
\end{figure}

To further investigate the temporal variability within the selected regions, we spatially averaged the spectra and kept some temporal resolution by averaging the spectra over only five exposures (six for  Fe\,{\footnotesize XIX}). 
For example, Figure \ref{fig:ligth_ab} shows the resulting light
curves for mask $A$ and mask $B$. We compared these to light curves of
the AIA 94 channel for the same masks using data integrated over 10
mins. The intensity has been corrected for the  cooler component
contributing to AIA 94 has estimated using the 171 channel, as described in \cite{reale11}. 
For mask $A$, the hot component of AIA 94 is almost stable up to the
time of the first flare at about 21UT ($\sim$ 420 mins in the
plot). This component is mostly comparable to the  SUMER
Fe\,{\footnotesize XVIII} which, however, shows more variability. This
line is blended with the Si\,{\footnotesize VIII} 974.58 ~\AA~ which
has been removed. However the signal in a single point is weak and
it is possible  that some residuals of the Si line has affected the light curve. 
Some temporal variability is also visible in the Fe\,{\footnotesize
  XIX} which increases  during the observations, while
Fe\,{\footnotesize XVII} seems to be only marginally affected by the flare. 
We decided to keep this mask for our analysis and average out the temporal
variability by using an average spectrum.

Also mask $B$ shows some variability, even though  Fe\,{\footnotesize
  XIX} is always absent. This mask is closer to the flaring activity
which is clearly visible in the AIA 94 curve. For this mask a
signature is also visible in Fe\,{\footnotesize XVII} showing a peak
delayed with respect to the AIA one. We used this mask by discarding
the exposures of Fe\,{\footnotesize XVII} which were affected by the flare.

In conclusion, this analysis shows that there is some variability in
time of the hot lines, but only in one case can we see a clear link to
the flaring activity. We discarded the affected exposures. 
To average out the temporal variability and increase the signal to noise of the data, we pursued the analysis on these masks by averaging the spectra over all the exposures, as shown in Figure \ref{fig:fe18_slit} and \ref{fig:slit_time2}. From these figures we see that the selected structures (apart from mask $E$), stay stable during the whole period of the observations (that is the time delay between observing Fe\,{\footnotesize XVII} and Fe\,{\footnotesize XVIII}).

\section{SUMER-EIS cross calibration}
\label{sec:int_cal}

The analysis carried out using integrated fluxes derived from
different instruments could introduce inconsistencies due to the
absolute radiometric calibrations. We carried out tests using the EM
loci method on the Ca and Fe lines and found several issues that we had to solve. We used the data from some of the selected masks to address the problem. 

\subsection{Consistency in the EIS data} 

In addition to the pre-flight analysis \citep{lang06}, the absolute
radiometric calibration of EIS has been investigated post-launch by
\cite{delzanna13}, GDZ, and \cite{warren14}, NRL. These last two introduce a
different correction factor to the pre-flight calibration which, for
the period of our observations, can reach a factor of 2.4 (see Figure \ref{app:calib} and \cite{warren14}). 
These differences have to be taken into account in the analysis of the data. 
All three calibrations are available in the  $SolarSoft$.

\begin{figure}[h]
\includegraphics[width=\linewidth]{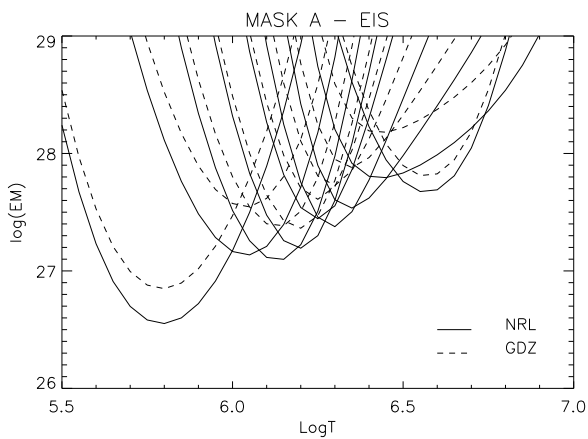}
\caption{EIS EM loci for mask $A$ using both GDZ (dashed curves) and NRL (solid curves) calibrations.}
\label{fig:gdz_nrl}
\end{figure}

Figure \ref{fig:gdz_nrl} shows the EM loci for the EIS data of mask
$A$. There is consistency between the two results (GDZ and NRL), even
though the use of the GDZ calibration results in larger EMs than in the NRL case, with a slightly lower peak temperature. 
In the following analysis we have used the GDZ calibration.

\subsection{Consistency in the SUMER data} 

\begin{figure}[th]
\includegraphics[width=\linewidth]{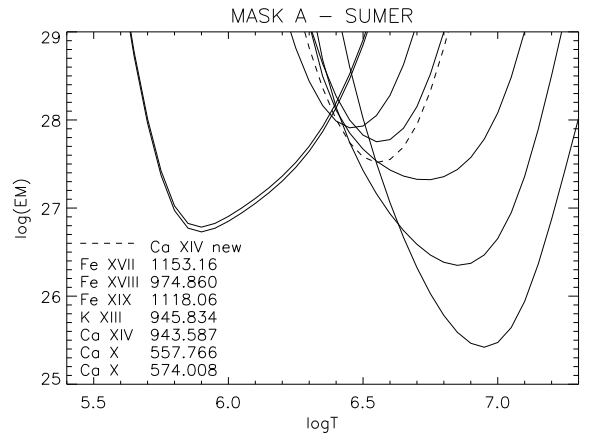}
\caption{EM loci for mask $A$ using SUMER data. The plot shows the change in the EM of Ca \,{\footnotesize XIV} using the new calculation (dashed curve).}
\label{fig:sum_ca}
\end{figure}

Figure \ref{fig:sum_ca} shows the EM loci for the SUMER data for mask $A$, obtained by using the lines having a formation temperature close to the peak of the emission measure. From this set, only the high temperature lines will be retained for the thermal analysis presented in Section \ref{sec:therm}. 

The EM peak is around $\log T = 6.4$, which is typical of ARs. Also we
note that the Ca \,{\footnotesize XIV} loci plotted with the solid
curved (CHIANTI v.8) is inconsistent with the rest of the curves,
being it too high. Having investigated the problem, we found this to
be an atomic data issue rather then a wrong choice of element
abundances: the CHIANTI v.8 emissivity is not consistent with the observed one.\\
The Ca XIV forbidden line observed by SUMER at 943.58~\AA\ is the
strongest line within the ground configuration of this ion, between
the ground state 2s$^2$ 2p$^3$ $^4$S$^{3/2}$ and the first excited
level, the 2s$^2$ 2p$^3$ $^2$D$^{3/2}$. The excitation data in CHIANTI
v.8 are from \cite{landi05}, and were calculated with the distorted
wave (DW) approximation. It is well known that this approximation
works very well for strong dipole-allowed transition, but typically
underestimates the electron excitation rates of the forbidden lines, especially within the ground configuration.
This was confirmed by a recent $R$-matrix calculation by \cite{dong12},
  where significant increases in several transitions rates were reported.
  The excitation rate for the  943.58~\AA\ forbidden transition is about a factor of two higher with the Dong et al. calculations. We have taken the  Dong et al. excitation rates and built a new CHIANTI
model ion to be used for the present analysis. We supplemented these data with A-values from the recent calculations by \cite{wang16}. The ratio with the strongest line, the resonance EUV line at 193.87~\AA\ observed with Hinode EIS,  is slightly temperature sensitive, but does not vary much with electron density. At log $T$[K]=6.4, the ratio between these two lines is a factor of 1.73 higher with the new model ion, which is significant.

The loci obtained with this new  model is plotted as a dashed line in
Figure \ref{fig:sum_ca}, where it has become consistent with the ensemble of the curves.

\subsection{Combining EIS and SUMER data}

When we plotted the EM locis from the two instruments together for the different masks, we found a systematic difference: at similar temperatures the SUMER emission measures were lower than the EIS ones. For instance, this is the case for two lines from Ca \,{\footnotesize XIV}.
We assumed this discrepancy was due to the SUMER degradation, discussed by \cite{teriaca12}. We then made tests to calculate a correction factor to be applied to the SUMER fluxes.  
The correction factor can be obtained using data from an isothermal plasma, 
by comparing the measured to the theoretical lines ratio predicted by the CHIANTI database at the given temperature. 

While a detailed discussion on the EM loci analysis will be given in the next section, here we just mention that 
the curves from the selected masks are all similar around the peak  located between $\log T = 6.4-6.5$. For our lines ratio analysis 
 we selected the temperature obtained in mask $B$, as it is the one
 where the Fe \,{\footnotesize XIX} is absent. This means that most of
 the Ca line emission is formed close to the EM peak
 temperature. Assuming this plasma temperature, we obtained a SUMER
 correction factor 1.8 using the GDZ EIS calibration (that is consistent with the expected
 degradation of the instrument performances estimated by
 \cite{teriaca12}). We also carried out the same analysis using the NRL EIS calibration and found a factor 1.4.
For the following thermal analysis we used the GDZ factor.

\section{Thermal analysis}

This section presents the results from the thermal analysis using different DEM and EM methods. 

\label{sec:therm}
\subsection{EM Loci}

\begin{figure*}[ht]
\includegraphics[width=0.45\linewidth]{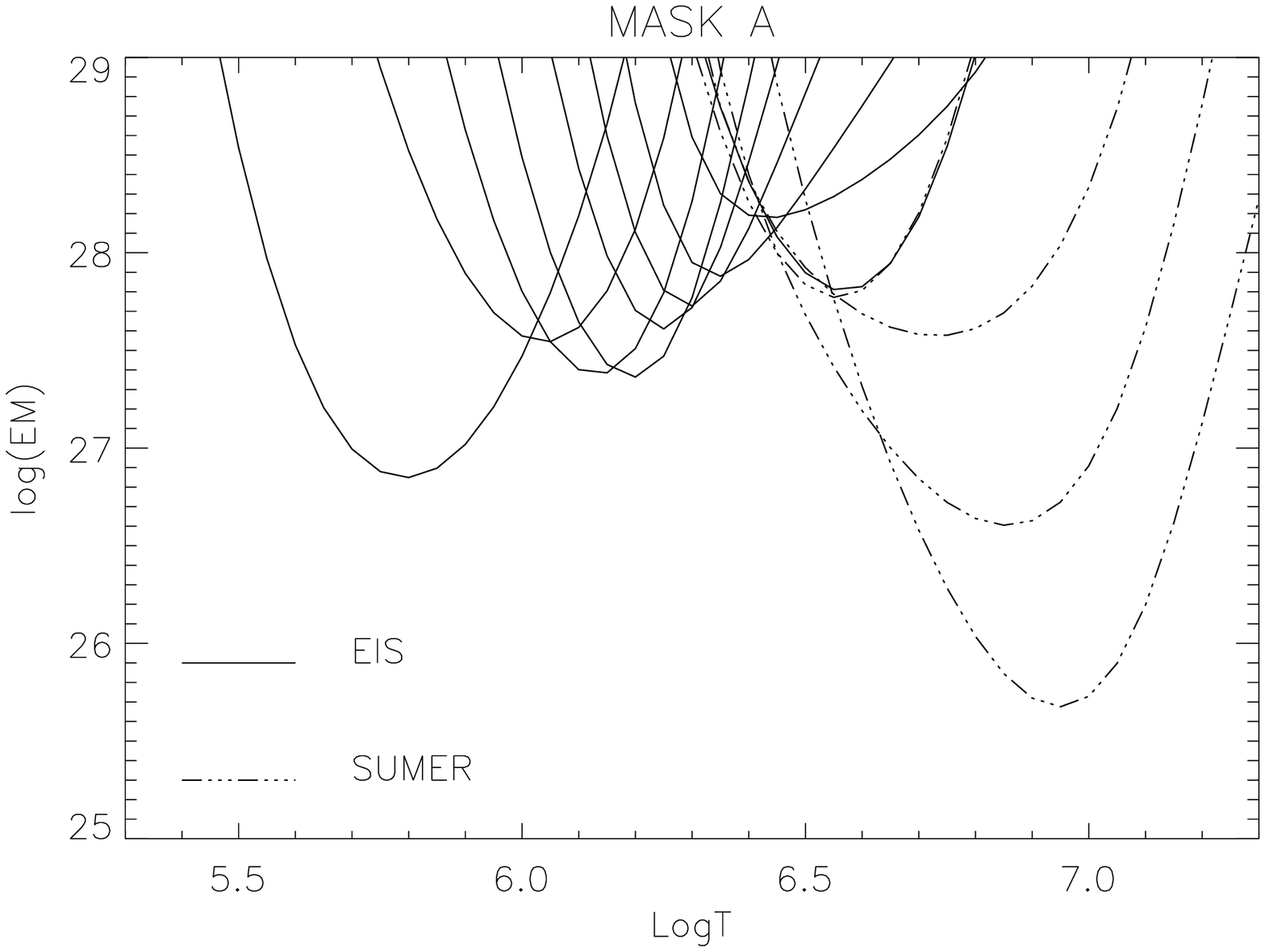}
\includegraphics[width=0.45\linewidth]{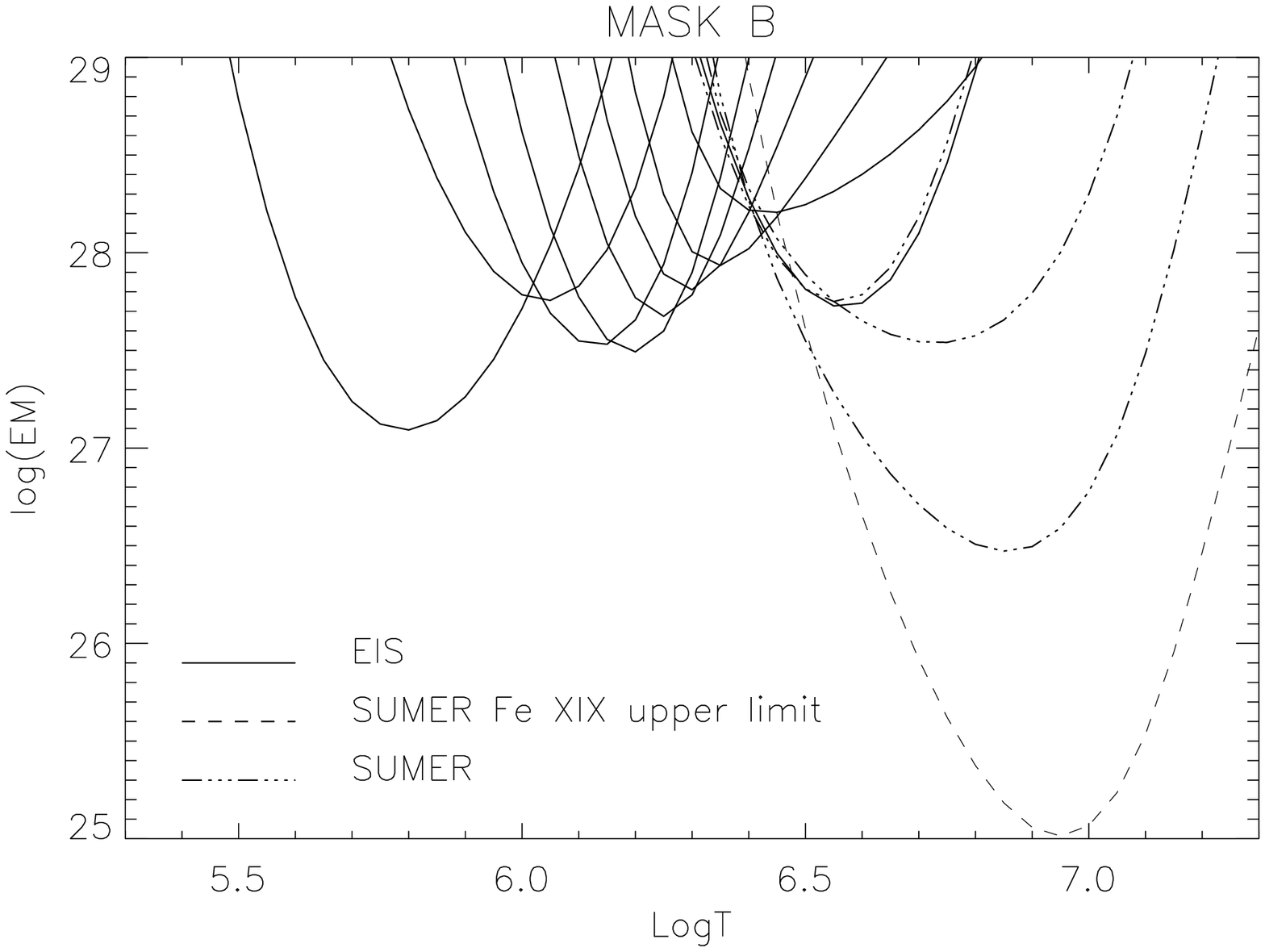}\\
\includegraphics[width=0.45\linewidth]{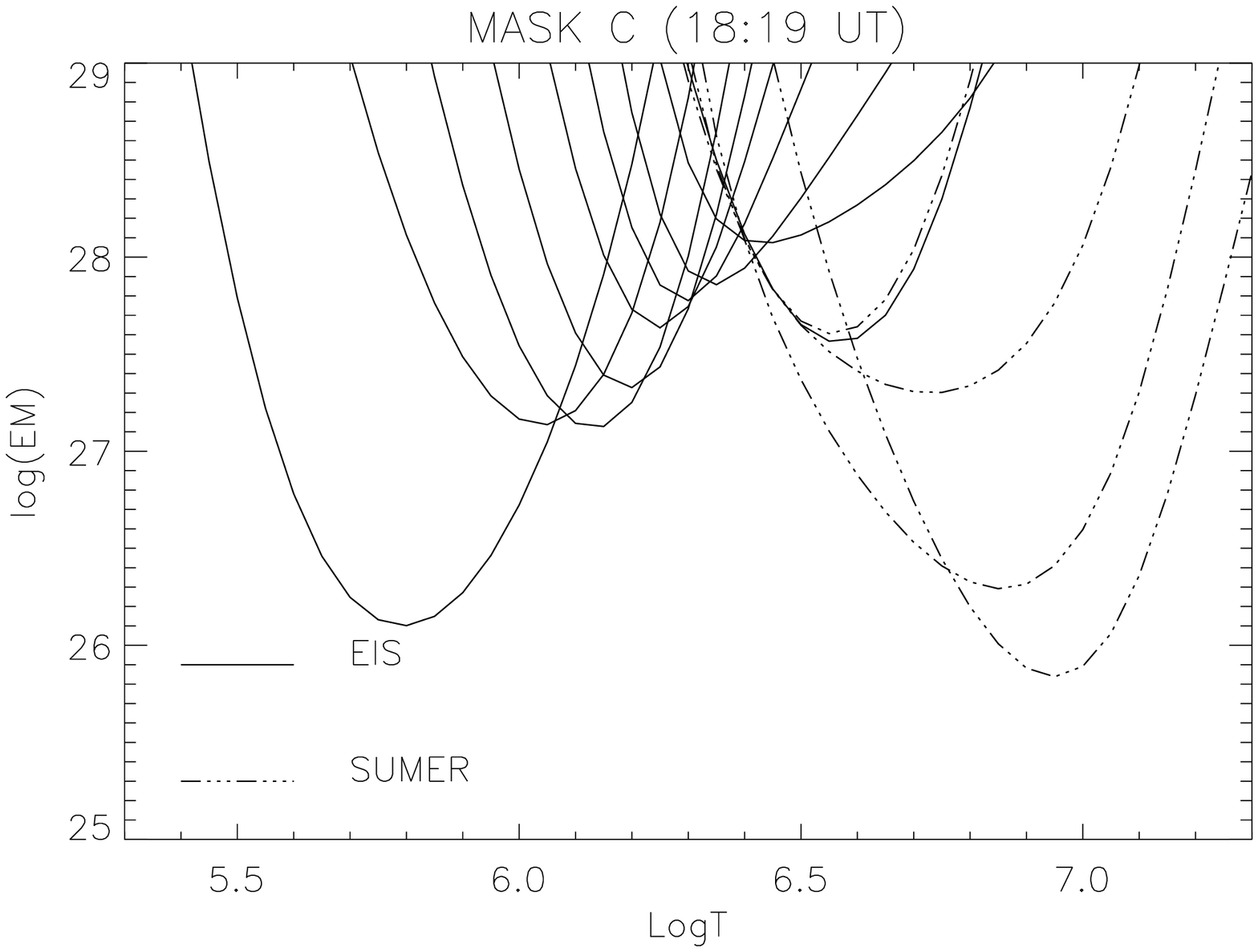}
\includegraphics[width=0.45\linewidth]{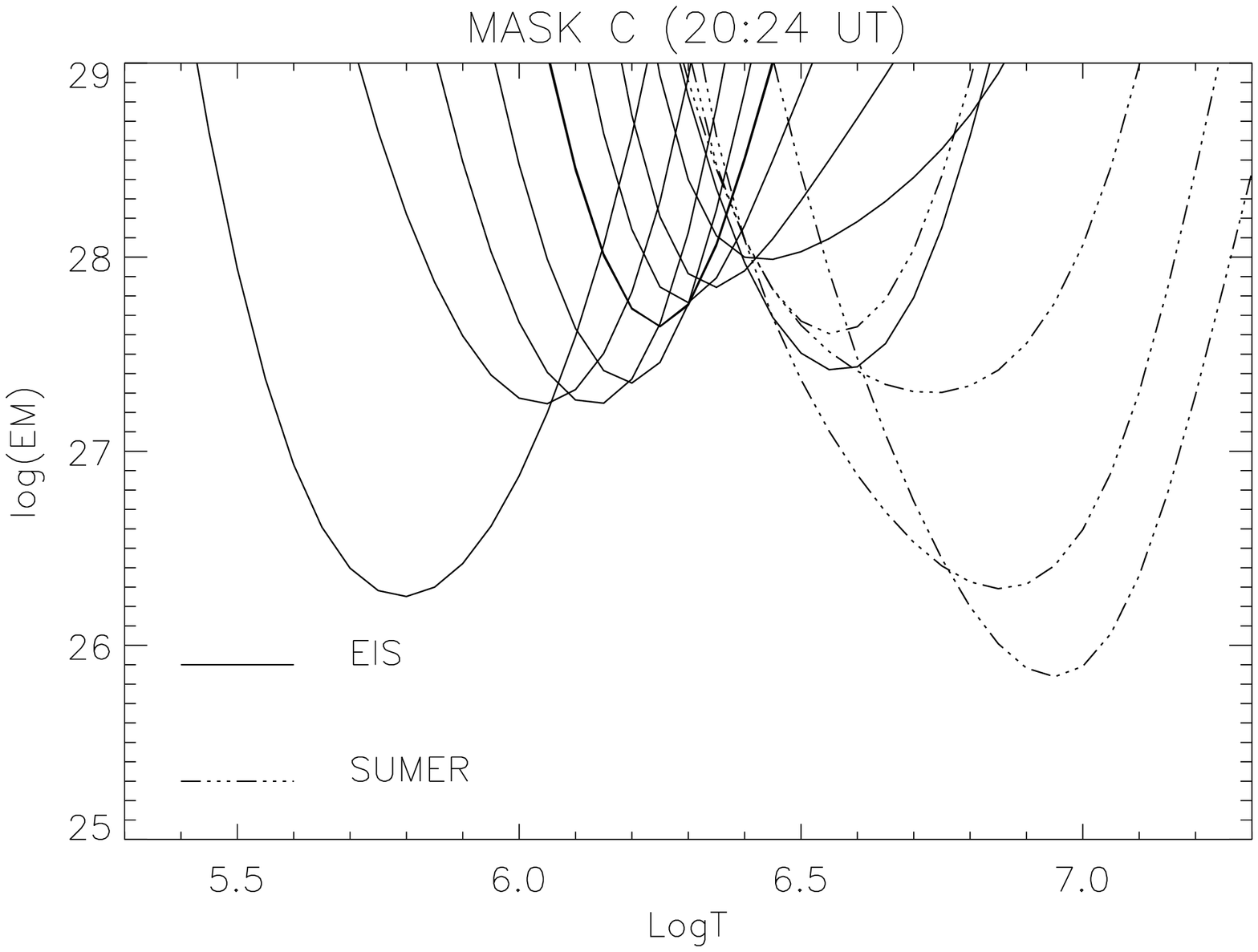}\\
\includegraphics[width=0.45\linewidth]{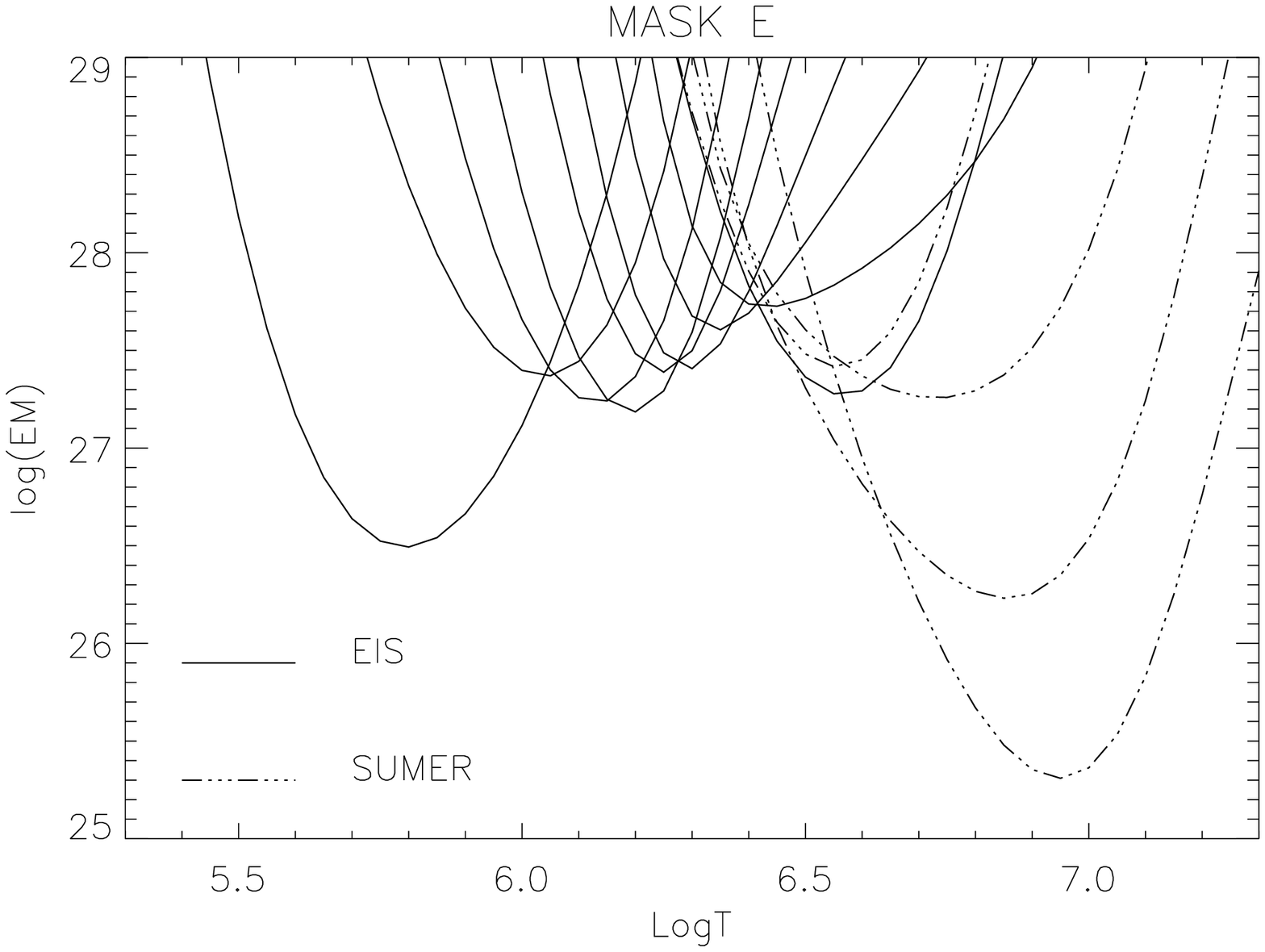}
\includegraphics[width=0.45\linewidth]{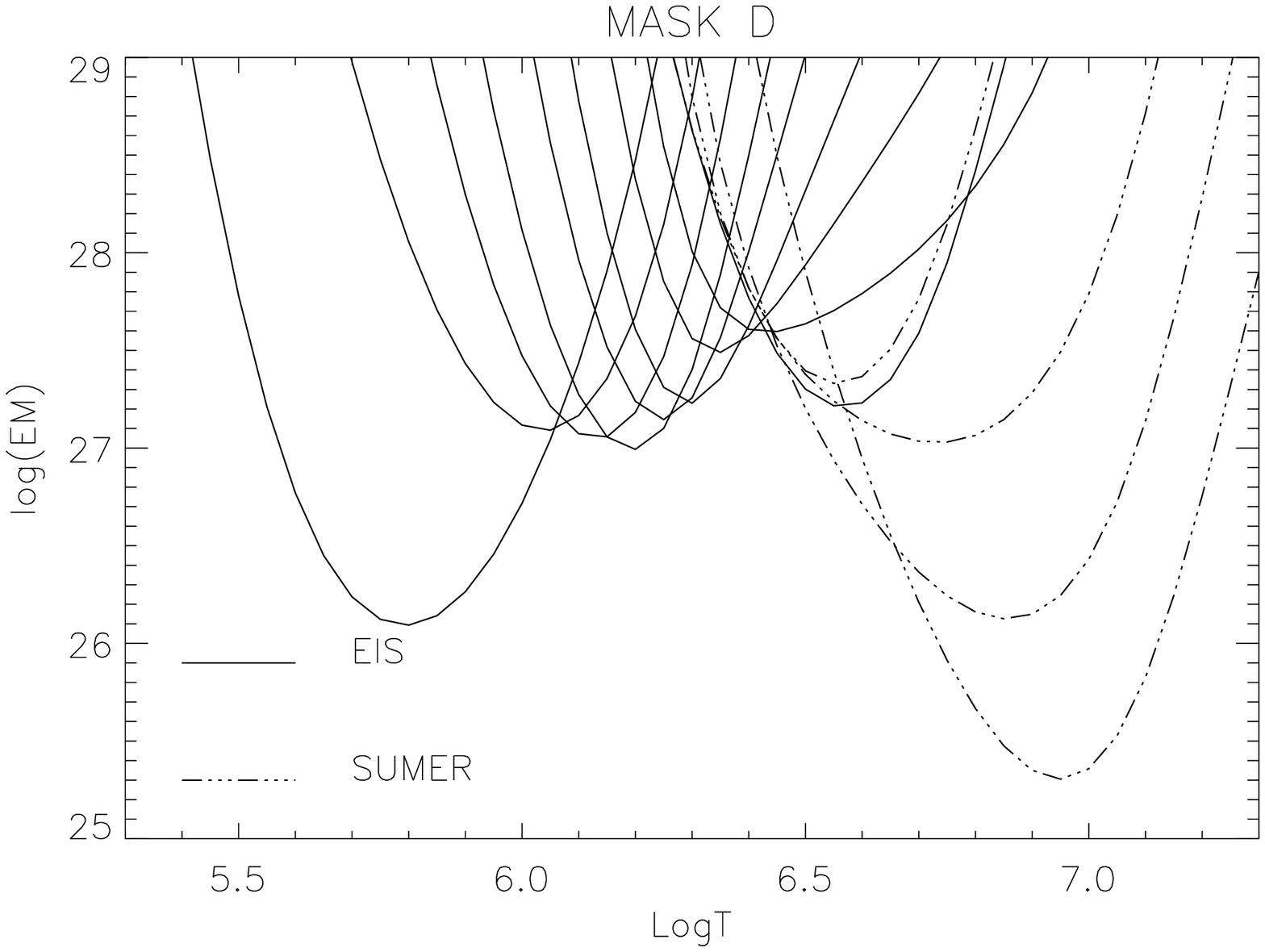}
\caption EM loci for the three masks of SUMER slit position 1 (first and second lines) and the two for SUMER slit position 2 (bottom line). The SUMER fluxes have been multiplied by 1.8. The two plots for mask $C$ are obtained using the data from the two EIS rasters with the same coordinates.
\label{fig:loci27}
\end{figure*}

 Figure \ref{fig:loci27}  shows the combined EIS and corrected SUMER loci EMs  for the selected masks for,  SUMER slit position 1 (top and middle lines) and  2 (bottom line).

As already pointed out, there are similarities between these plots
which all show the bulk of plasma at about $3~ \mathrm{MK}$; which is
a well known result (e.i. \cite{parenti10}). Additionally, the
decrease of the EM is noticeable with the increase in solar
height. The difference between them is mainly found in the absence of
Fe \,{\footnotesize XIX} for mask $B$ (in the figure we plotted an upper limit used for the DEM analysis). The cooler emission for this region might be representative of background AR plasma.
 
This first analysis suggests a new result: from above the limb to about $91 ~\mathrm{Mm}$ and over about $1.5\times 10^2 ~\mathrm{Mm}$ across the AR, 
the thermal properties above $3~\mathrm{MK}$ are similar almost everywhere. 


For the region covered by the mask $C$ we can also provide temporal
information, because this is the only mask that could be applied to
both EIS rasters in position 1. The EM loci plots for this mask are
shown in the middle row of Figure \ref{fig:loci27}. Only the EIS
curves can show differences (for both masks we used the same SUMER
data). We see no significant change, so we conclude that the EM loci
plots do not show evidence of spatial and temporal variations.



\subsection{Differential emission measure}
\label{sec:dem_gdz}

The DEM vs T curves (see Eq. \ref{eq_dem}) have been obtained with a method based on a simple chi-square minimization.
We essentially used a modified version of the  \textit{xrt\_dem\_iterative2.pro} DEM inversion routine
 \citep{weber04}  in order to have more flexibility
in the choice of input parameters. The standard routine, widely used in solar physics
and available within $SolarSoft$, is based on 
the robust chi-square fitting routine \mbox{(\textit{mpfit.pro})}.
The DEM is modelled assuming a spline, with a fixed selection of the nodes.
Since it turns out that the DEM solutions are quite sensitive to the choice of nodes,
 we modified the program to allow for the definition of the number and location 
of the spline nodes.  We also introduced the option to  
 input minimum and maximum limits to the DEM spline values,
which are passed to \mbox{(\textit{mpfit.pro})}.
This was found to be particularly useful for constraining the upper limits of the highest
temperature values. We used upper limits for the DEM values which provide 
radiances in the Fe\,{\footnotesize XIX}  and  Fe\,{\footnotesize XXIII} lines as given in Sec. \ref{sec:err}. 
 
\begin{figure}[h]
\includegraphics[scale=.36, angle=90]{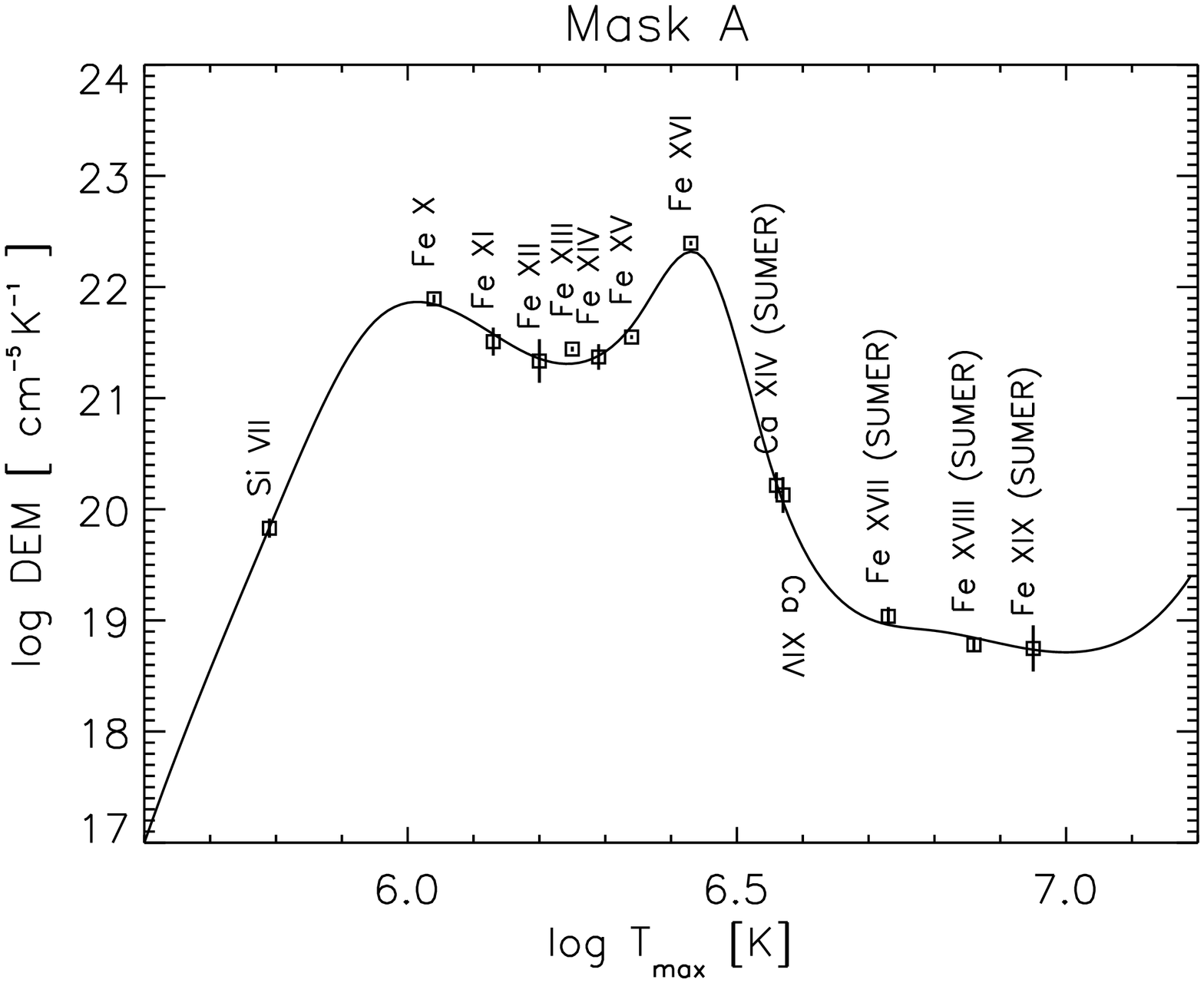}\\
\includegraphics[scale=.4]{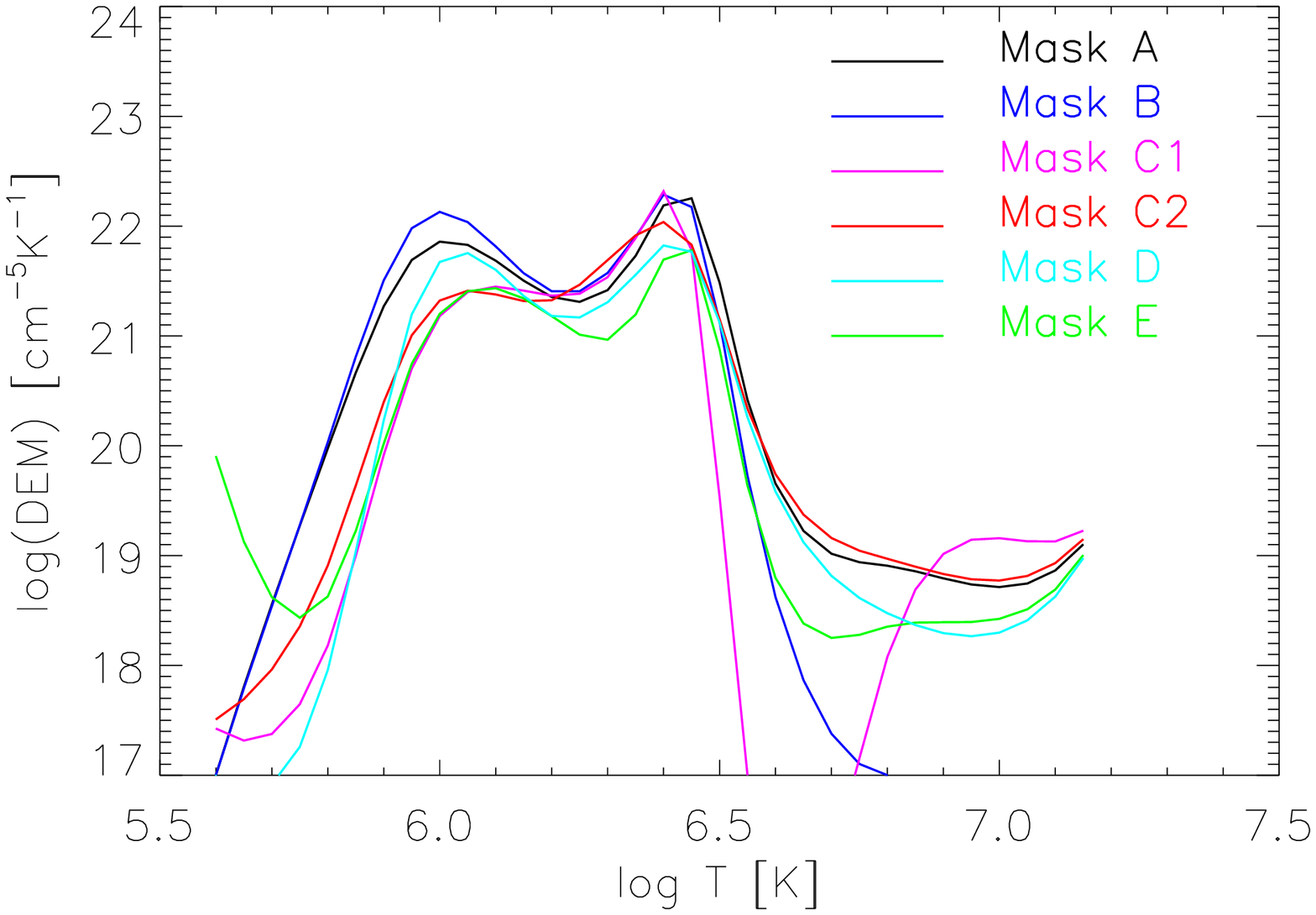}
\caption{Top: Results from the DEM inversion for position 1 on mask $A$. The points are plotted at the temperature of the maximum of the G(T), at the theoretical vs. the observed intensity ratio multiplied by
the DEM value. Bottom: The DEM for all the masks analyzed in this work.}
\label{fig:dems}
\end{figure}

Figure \ref{fig:dems} shows the results from this inversion. The top
plot shows the resulting DEM for mask $A$, overplotted with the  points at the temperature of the maximum of the G(T), which represent  the theoretical vs. the observed intensity ratio multiplied by the DEM value. The bottom plot contains all the masks together.

The temperature range for the inversion has been set to 
log $T$[K]=5.6 -- 7.2. For all the DEM inversions we selected 
spline nodes at log $T$[K]= 5.6, 5.8, 5.95, 6.2, 6.35, 6.45, 6.55, 6.8, 7.2, 
which provide relatively good agreement (within 20--30\%) between 
predicted and observed intensities, as shown in Table~\ref{tab:dem_ratio}. The resulting DEM values 
are relatively smooth. Adding a few extra spline nodes in the 1--3 MK range 
can improve the agreement between observed and predicted intensities,
but the DEM would be less smooth (as found for instance by \cite{landi08}).

Consistency was found between these results and the EM loci
approach, the DEM curves are similar for all the masks. As expected,
there is a decrease in amplitude  of the DEM with an increase in the
solar height. Some differences are found, but only at high temperatures.

The $1-2.5~ \mathrm{MK}$  corona is represented by a double-peaked DEM. The hotter peak is at around $2.5~ \mathrm{MK}$ and it is higher that the cooler peak for mask $A$, $C$ and $E$.
The $3~ \mathrm{MK}$ peak of the DEM of the off-limb corona is already known in literature \citep{reale14}. What is most interesting is the plateau above $5~ \mathrm{MK}$ due to the observation of the high temperature lines.
The DEM values at the main peak are  so high that the effective
temperature of emission for lines such as Fe \,{\footnotesize XVII}  and Fe \,{\footnotesize XVIII}  is about $3~ \mathrm{MK}$, i.e. these lines are mainly formed
far away from their peak formation temperatures (5 and $7~ \mathrm{MK}$ respectively using the 
CHIANTI v.8 charge state distributions in ionization equilibrium). 
This issue is quite typical for active regions, as described e.g. in \cite{delzanna13b} and \cite{delzanna14}.
The peak at $2.5~ \mathrm{MK}$ is very well constrained by the 
 Fe \,{\footnotesize XVI}  and Ca \,{\footnotesize XIV}  lines.  The fact that the DEM 
values around 3 MK are sufficient to explain the intensity
of the  Fe \,{\footnotesize XVII}  and Fe \,{\footnotesize XVIII} lines means that the DEM above $3~ \mathrm{MK}$ 
has to drop significantly by several orders of magnitude. 
However, whenever the Fe \,{\footnotesize XIX}  line is observed, its 
intensity requires a plateau in the DEM. 

A further constrain at higher temperatures comes from the fact that 
the EIS Fe \,{\footnotesize XXIII}  263.75~\AA\ is not observed (we note that weak unidentified lines are present in active region EIS spectra, close to this line).  
We measured the variation of the EIS background near the line to estimate a 
3 $\sigma$ value for the intensity of this line, as described in Sec. \ref{sec:err}. We put an upper limit
at log $T$[K]= 7.2 accordingly (the reason why the DEM increases slightly above $10~ \mathrm{MK}$). From Table \ref{tab:dem_ratio} we see that we have been quite conservative in choosing 3 $\sigma$, as the flux ratios between theoretical and measured values are well below 1.

The $10~ \mathrm{MK}$  DEM is about $10^3$ times smaller than the $2.5~ \mathrm{MK}$ peak, and even less for the cooler mask $B$ (here the Fe \,{\footnotesize XIX} is not visible). 
This presence of very hot plasma has already been identified in some areas
of ARs. Here we are able to extend our knowledge: this very hot plasma
seems to be in all locations where we have a structured AR, with
emission in the  Fe \,{\footnotesize XVIII} line, and it appears also
to be persistent with time (at least within our observation time). In
fact, the bottom of Figure \ref{fig:dems} tells us that there is a
temporal variation of the DEM above the $3 ~ \mathrm{MK}$ between
masks $C1$ and $C2$. This is the area very close to the limb and
probably subjected to more AR variability. However, the $10 ~ \mathrm{MK}$ DEM is not affected.

\subsection{Emission measure}

\begin{figure*}[th]
\includegraphics[scale=.4]{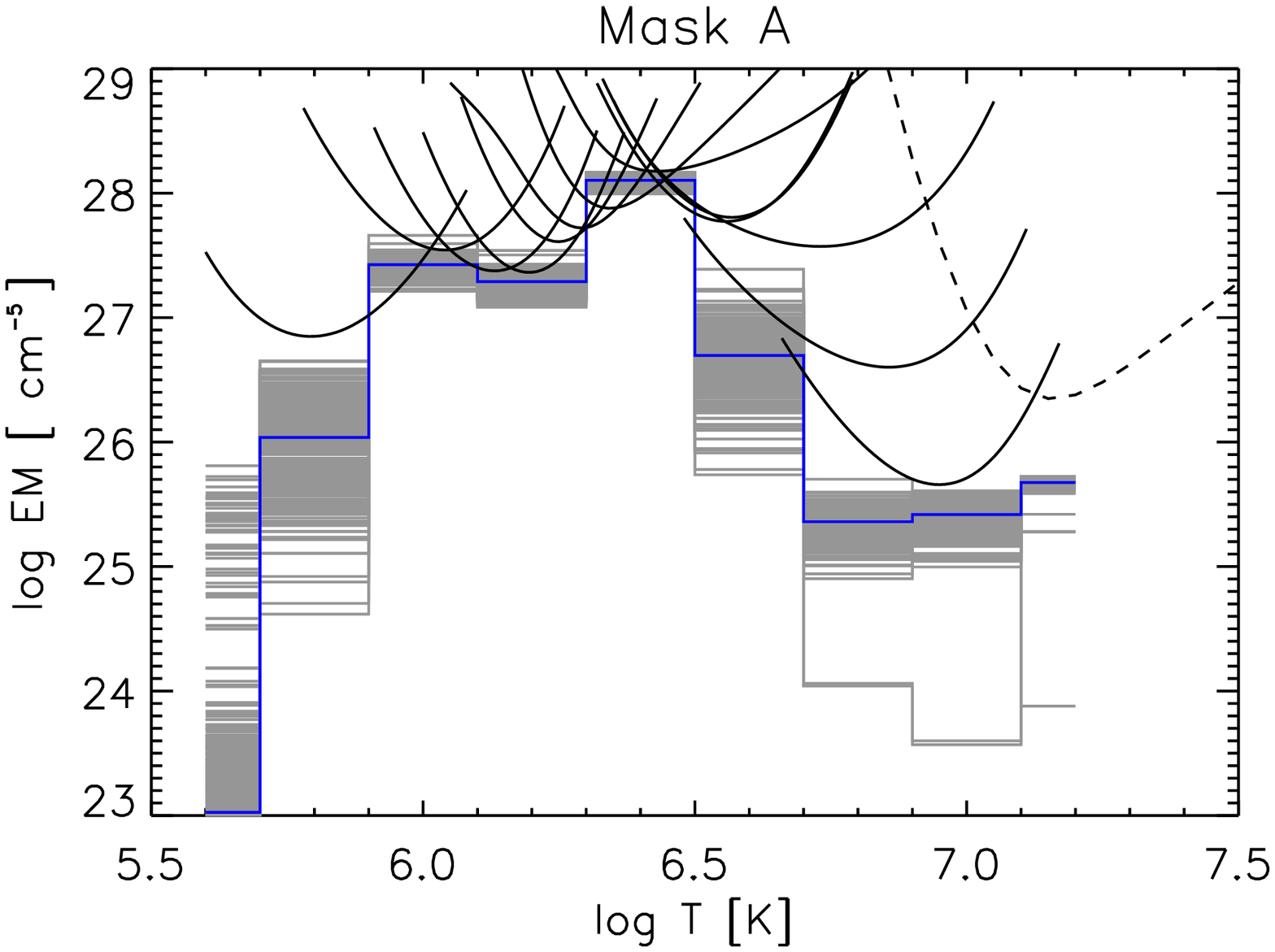}
\includegraphics[scale=.4]{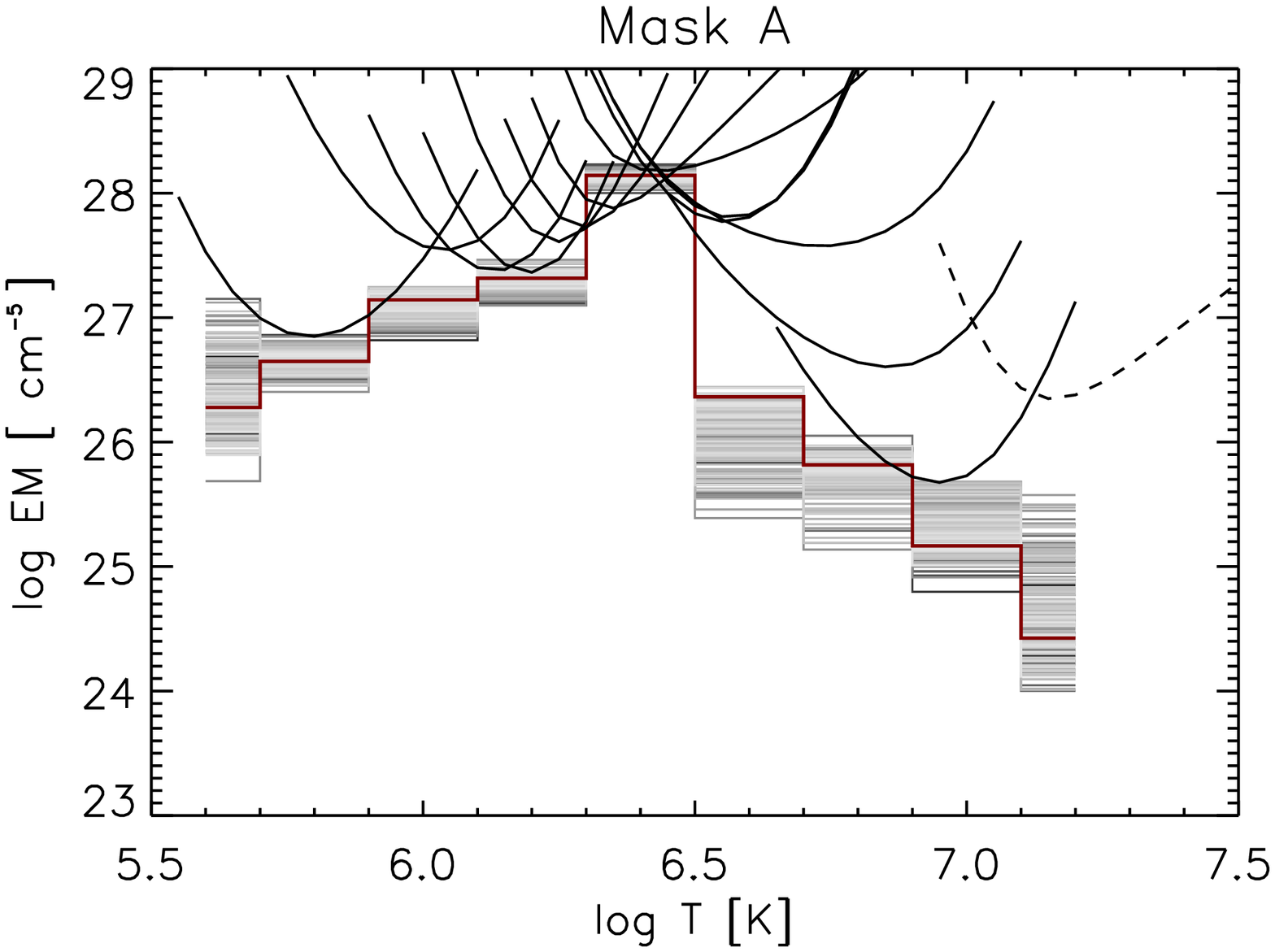}\\
\includegraphics[scale=.4]{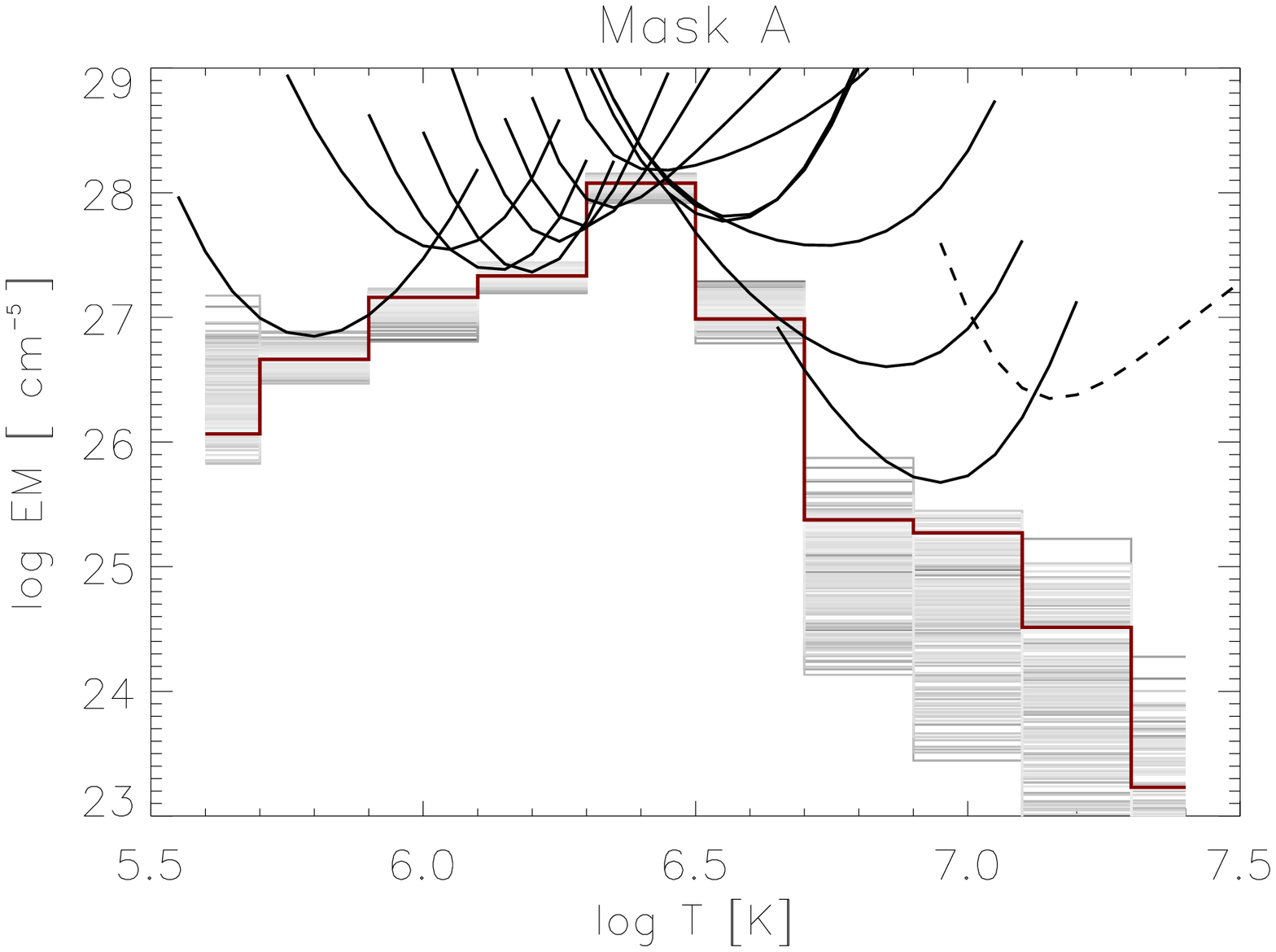}
\includegraphics[scale=.4]{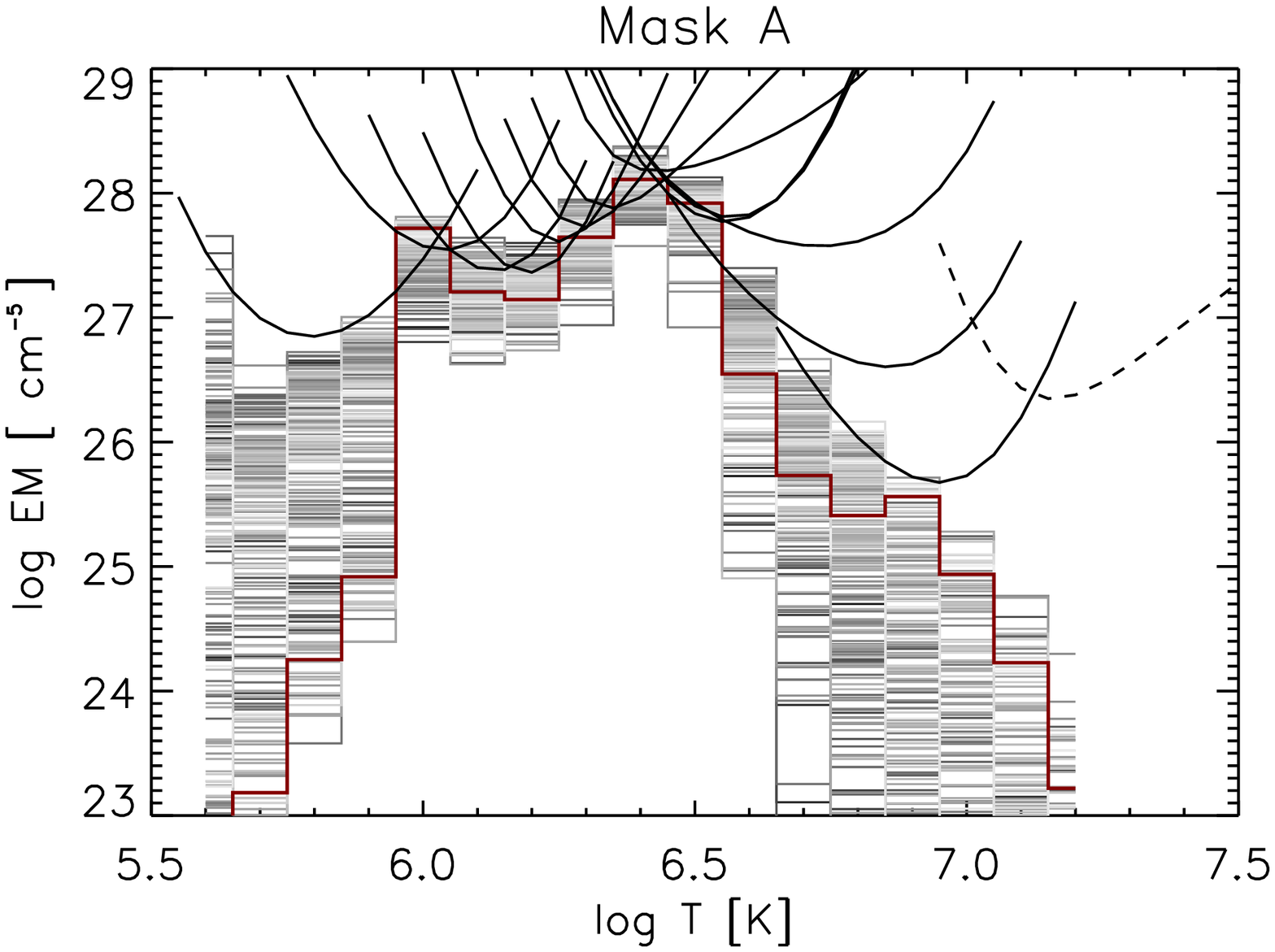}
\caption{ Mask $A$ EM loci. Top-left: in blue the EM derived from Sec. \ref{sec:dem_gdz}. Overplotted are the results from 200 runs varying the lines flux (see text). The EM loci are overplotted. The loci of the upper limit Fe \,{\footnotesize XXIII} line is plotted with the dashed line. 
Top-right: results from the MCMC inversion (red) with their clouds of solutions using the same parameters than the left plot (a temperature binsize set to 0.2 and the maximum temperature to $\log T = 7.2$).  
Bottom-left: result by extending the temperature range to $\log T=7.4$; Bottom-right: result by assuming  a binsize of $\log T=0.1$ and maximum temperature of $\log T = 7.2$. \label{fig:mcmc_A}}
\end{figure*}

Figure \ref{fig:mcmc_A} top-left shows the emission measure (solid-blue
line) for mask $A$ obtained by integrating the DEM of Figure \ref{fig:dems} over a temperature bin of size 0.2 (in logarithm scale). In the Figure we also overplotted the EM resulting from 200 inversions by the DEM code, by randomly varying each flux within $20\%$. These are shown as a light gray cloud of solutions within each temperature bin.   
The bin size $d \log T = 0.2$ is a good compromise to maximize the temperature resolution and to minimize the spread of the solutions within each temperature bin. The degradation of the solution with a smaller temperature bin is illustrated by the bottom-right plot of this figure.

We compared this calculation with the EM derived applying a different method. 
We used the Monte Carlo Markov Chain (MCMC) differential emission measure algorithm distributed with the  $PINTofALE$ \citep{kashetal98, kashetal20, drake10} spectral analysis package  testing different input parameters. 
  
\begin{figure}[h]
\includegraphics[scale=.4]{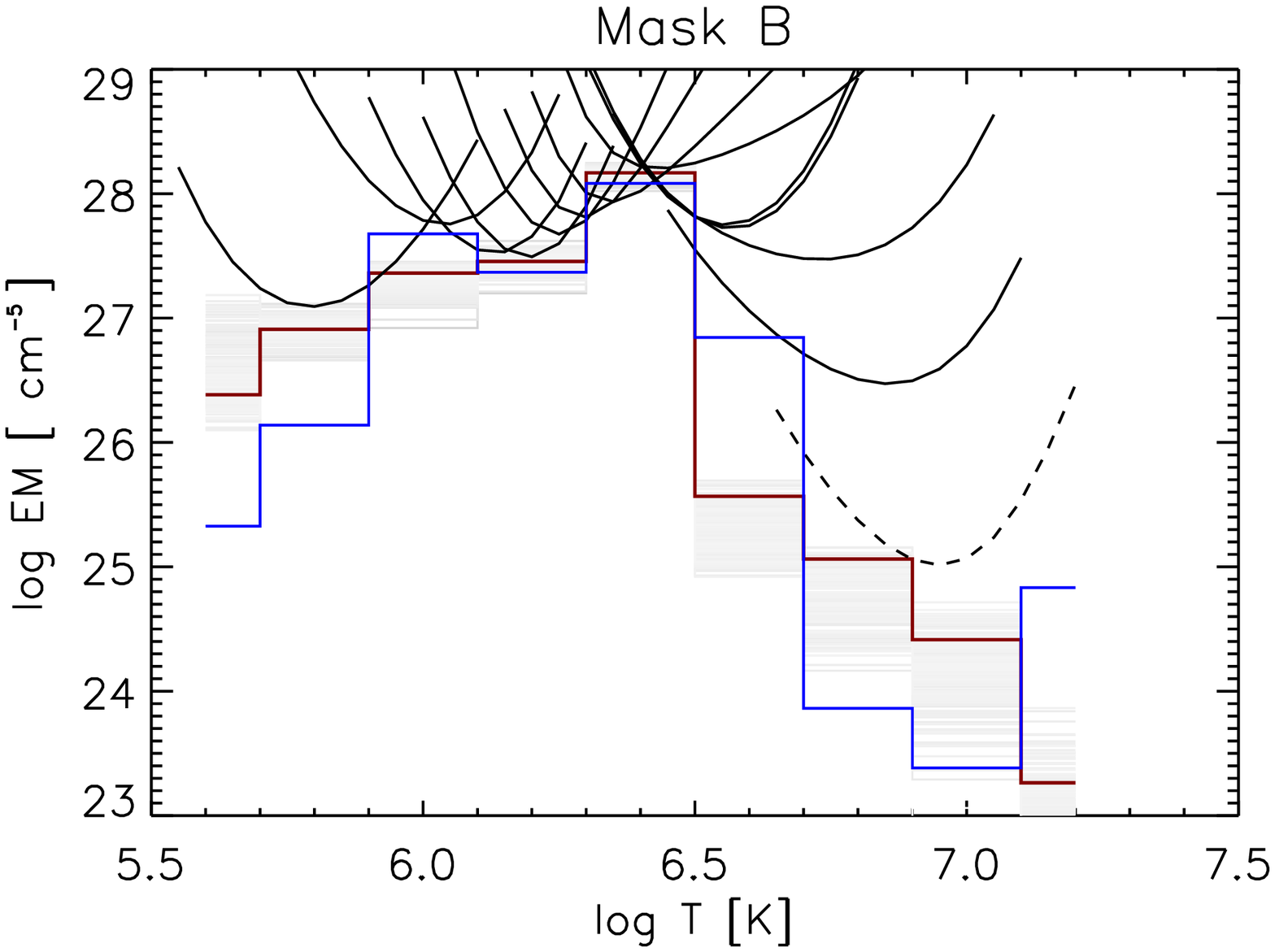}
\caption{Result from the MCMC inversion (red line) for mask $B$ with the cloud of solutions, assuming the temperature binsize of 0.2 and the maximum temperature to $\log T = 7.2$. The EM integrated from the DEM of Figure \ref{fig:dems} for the same mask is plotted with the blue line. The loci EM are overplotted. The loci of the upper limit line is plotted with the dashed line.
\label{fig:mcmc_B}}
\end{figure}

  With respect to several other methods, the MCMC method has the advantage of not imposing a predetermined functional form for the solution, and, most importantly, it provides confidence limits on the most probable DEM, thus allowing a determination of the significance of apparent structures that may be found in a typical reconstruction. The algorithm assumes that the uncertainties in the intensities are uncorrelated so that systematic errors in the calibration, which could depend on the wavelength, or in the atomic data, which could vary by ion, are not accounted for.
The code has been set to perform 200 explorations (batches) of the parameter space, and 200 Monte Carlo realizations for each exploration. 
Figure \ref{fig:mcmc_A} top-right shows the resulting EM from this method applying the same parameters of the left plot. 
In a similar way, Figure \ref{fig:mcmc_B} shows the MCMC EM for mask $B$ (red line) with the cloud of solutions. The EM obtained from the DEM applying the chi-square minimization method (presented in Sec. \ref{sec:dem_gdz}) is overplotted in blue. 

As seen in Sec. \ref{sec:dem_gdz}, the plasma distribution is dominated by a main peak around $3 \mathrm{MK}$, followed by a drastic drop of the EM at high temperature. Overall, in the interval $6.5 < \log T < 7$ the EM looks to drop more (by about one order of magnitude) in mask $B$ than in mask $A$.


When we compare the solutions for one mask for the two methods, we see that they are consistent in most temperature bins. More discrepancies are found at low temperature, where the inversion is less constrained, and for the $\log T = 6.8$ bin. In general the high temperature tail drops more rapidly with the first method.


As mentioned, to better constrain the solutions we decided to adopt a binsize of $\log T=0.2$, as a smaller one produces solutions with greater dispersion in each temperature bin. This is shown for instance for mask $A$ in  Figure \ref{fig:mcmc_A} bottom right.
We checked the effect of extending the temperature range to higher
values for those masks where the Fe \,{\footnotesize XXIII} was used
as upper limit. This is a flare line whose temperature of formation extends beyond $10 ~~\mathrm{MK}$. This inversion is shown for mask $A$ in Figure \ref{fig:mcmc_A} bottom left. We see again that at high temperature the solutions in each bin spans a larger interval.


\floattable
\begin{deluxetable}{lrcrrrrrr}
\tablecaption{Ratios (R) of predicted vs. observed radiances of the selected lines for the various regions.
 \label{tab:dem_ratio}}
\tablecolumns{9}
\tablenum{4}
\tablewidth{0pt}
\tablehead{
\colhead{Ion } &\colhead{$\lambda$[\AA]} & \colhead{log $T_{\rm eff}$[K]} & \colhead{R(A)}& 
\colhead{R(B)} &  \colhead{R(C1)}& \colhead{R(C2)} & \colhead{R(E)}&\colhead{I(D)} \\
}
\startdata
Si \,{\footnotesize VII}   &  275.38 &  6.00 &  1.01 (1.16) & 1.02 (1.15) &  1.06 & 1.02 & 1.06 & 1.03  \\   
Fe \,{\footnotesize X} (sbl)& 257.26  &  6.05 &  0.90 (0.54) & 0.88 (0.52) &  0.94 & 0.75 & 0.77 & 0.85 \\
Fe \,{\footnotesize XI}     & 188.30  &  6.12 &  1.17 (0.89) & 1.19 (0.91) &  1.09 & 1.0 & 1.12 & 1.17 \\
Fe \,{\footnotesize XII}    &  192.39 &  6.23 &  1.05 (1.13) & 1.02 (1.12) &  1.06 & 1.0 & 1.15 & 1.09 \\
Fe \,{\footnotesize XIII}   &  202.04 &  6.33 &  0.74 (0.89) & 0.81 (0.92) &  0.78 & 0.76 & 0.58 & 0.64 \\
Fe \,{\footnotesize XIV}    &  264.79 &  6.37 &  1.03 (1.21) & 1.04 (1.10) &  0.99 & 0.91 & 1.26 & 1.21 \\
Fe \,{\footnotesize XV}     &  284.16 &  6.40 &  1.25 (1.54) & 1.23 (1.45) &  1.26 & 1.06 & 0.99 & 1.07  \\
Fe \,{\footnotesize XVI}    &  262.98 &  6.42 &  0.84 (0.91)  & 0.80 (0.90) &  0.84 & 0.84 & 1.04 & 1.0 \\
Ca \,{\footnotesize XIV}    &  193.87 &  6.45 &  0.86 (0.62)  & 0.91 (0.78) &  0.84 & 1.11 & 1.04 & 1.14 \\
Ca \,{\footnotesize XIV}    &  943.59 &  6.45 &  1.04 (0.80) & 0.97 (0.87) &  0.87 & 0.81 & 0.91 & 0.93 \\
Fe \,{\footnotesize XVII}   & 1153.16 &  6.46 &  0.84 (0.66) & 0.91 (0.82) &  0.88 & 0.83 & 0.91 & 0.70 \\
Fe \,{\footnotesize XVIII}  &  974.86 &  6.50 &  1.16 (0.87) & 1.14 (0.86) &  1.17 & 1.32 & 1.07 & 1.09  \\
Fe \,{\footnotesize XIX}    & 1118.06 &  6.97 &  0.97 (1.03) &0.54  (0.98)$\ast$&0.91 & 0.66 & 0.97 & 0.97 \\
Fe \,{\footnotesize XXIII}$\ast$ & 263.765 &  6.95 &  0.15 (0.02)&  -  & 0.25 & 0.16& 0.1& 0.1\\
\enddata
\tablecomments{In the first two columns we list the dominant contribution to the  observed spectral lines. Note that the Fe \,{\footnotesize X}  257.26~\AA\ is a self-blend of two transitions.
As an indication of where the lines mainly form, we list in column 3 the log $T_{\rm eff}$[K] values for the mask $A$. Values in parentheses are the ratios obtained from the MCMC program using a temperature bin of $\log T=0.2$ and a maximum temperature of $\log T= 7.2$. 
The $\ast$ marks the imposed upper limit.}
\end{deluxetable}

\subsection{High temperature tail of the EM}
\label{sec:slope}

The different thermal analysis methods presented here converge in finding the well know peak of emission measure at around $3 ~\mathrm{MK}$. In addition, the long integration time and the low noise level of SUMER have revealed the persistent presence of a small amount of very hot plasma. The DEM is very similar everywhere suggesting common heating process at work. 
The amount of very hot plasma has been quantified with an EM which reaches at maximum about $0.1\%$ at $10~ \mathrm{MK}$ of the main EM peak value. 
Such a small ratio was previously found in on disk and limb quiescent ARs using soft \citep{reale09b, delzanna14} and hard X-ray \citep{miceli12, hannah16} data.
This ratio is  also consistent with \cite{parenti10} findings on the
averaged EM in a pre-flaring area using the HINODE/XRT hard filter
ratio. As presented in Sec. \ref{sec:intro}, a few other X-ray measurements found about only two orders of magnitude variation of the EM in the $3-10 ~\mathrm{MK}$ \citep[e.g.][]{reale09, testa11}.
To our knowledge this is the first time the EM above $5 ~\mathrm{MK}$
has been quantified over three orders of magnitude in an off-limb AR
observation by using measured lines profile from EUV spectroscopic data.

The spatial distribution in an AR of Fe \,{\footnotesize XIX} from on
disk observations was reported by \cite{brosius14} using the EUNIS-13 sounding rocket. They found a Fe \,{\footnotesize XIX} /Fe \,{\footnotesize XII} emission measure ratios (assuming a temperature formation of $\approx 8.9~\mathrm{MK}$ and $\approx 1.6~\mathrm{MK}$, respectively) of $\approx 0.59$ in the AR core, $\approx 0.076$ in the outer part, while they established a limit of $0.0081$ in a quiescent area. For our data this ratio is below 0.005, depending on the mask. Our upper limit is found for mask $C1$ (from a well detected Fe \,{\footnotesize XIX} line) and it is close to their upper limit for Fe \,{\footnotesize XIX}. Our lower values for this ratio are probably due to a different line-of-sight integration path, suggesting that the most intense hot plasma is  concentrated lower down in the corona. 


We also made a linear fit to the logarithm of the EM to establish the power law  index above the EM peak, which can be compared to other published results. We have to remember that other results suggest a different EM profile at these temperatures with possibly a secondary small peak around $10 ~\mathrm{MK}$ \citep[e.g.][]{reale09, shestov10, sylwester10}. 
For the fitting we set the temperature range between 2.5 and 10  $\mathrm{MK}$.
Table \ref{tab:slopes} summarizes our results and lists other finding from the literature. 
For our work we used data from mask $A$ and $B$, as representative of our dataset, as we can see from Figure \ref{fig:dems}.

Table \ref{tab:slopes} lists quite different values of the power law index. Contrary to mask $A$, our mask $B$ results are quite uncertain due to the different solutions found using the two inversion methods. As a general trend, we have the impression that the off-disk slopes are shallower than on disk and limb data. However, several elements linked to the inversion method could affect this, such as: the inversion method constraints, the temperature bin, the temperature range, the upper limit of the data. Some of these effects have been shown in Figure \ref{fig:mcmc_A}. For instance, if we were using a smaller temperature bin, our EM solutions would spread more in a way to give steeper slopes. 
We have also discussed the differences between the NRL and GDZ EIS
flux calibrations. The use of the NRL calibration would probably
produce a lower  EM peak value at higher temperatures. The EM slope
could be different. This is because the two calibrations do not have the same wavelength dependence (see Figure \ref{app:calib} and Figure \ref{fig:gdz_nrl}). The EM peak is defined by the Fe \,{\footnotesize XV}, Fe \,{\footnotesize XVI} and Ca \,{\footnotesize XIV} which fall in different parts of the spectra. 

From the physical point of view, in the case of the same heating mechanism being common to all ARs, 
it is also possible that changes may occur with the age of the AR, affecting the EM shape in time \citep{miceli12, ko16}. 
This is a very interesting topic that needs further investigation. 
Different heating mechanisms acting on ARs should leave, eventually, different signatures in the DEM. For instance, for those  results suggesting  a secondary peak around $8-10 ~\mathrm{MK}$, this could be explained by a secondary population of flare-like events with different initial energy and frequency from the one populating the bulk of the DEM \citep[e.g.][]{argiroffi08}.

However, the uncertainties on the measures are very high and a clear statement cannot be given yet.
Finally, we have to remember that all these measures are taken through
different line of sights integration, which results in weighted
emission measure information. In our case we have seen that the
$<1~\mathrm{MK}$ (the Ca\,{\footnotesize X} in Figure
\ref{fig:fe18_slit}) plasma has a different morphology than the hot
plasma. Certainly we are crossing different bundles of loops along the
line of sight. And it is possible that we are in presence of
independent populations of plasma, heated through a different process,
one of which maintains the plasma at high temperature. In the absence of further simulation tests, we leave this option open.

With our results we think to have provided further important observational constraints on AR heating. The hottest plasma is probably concentrated in the low lying part of the AR core, however its presence in small amount in the upper part of loops suggests a continuous energy injection also at these heights. 
Considering that the cooling  timescales of a 10MK plasma in equilibrium conditions is of the order of minutes, the temporal persistence of such temperatures  also imposes constraints on the way each spatial area (our masks) is heated. In the nanoflares scenario, the frequency of heating in the area covered by each of our masks should be higher than such a timescale.

\begin{deluxetable*}{lrccccc}
\tablecaption{List of the recent inferred power law index of high
  temperature EM for ARs.   
  \label{tab:slopes}}
\tablecolumns{7}
\tablenum{5}
\tablewidth{0pt}
\tablehead{
\colhead{Data} &\colhead{$\Delta$T [MK]}  &\colhead{$\alpha$} & \colhead{$d$logT [MK]} & \colhead{limit [MK]} &\colhead{Location} & \colhead{Type/method}}
\startdata
Mask A (mpfit)   & 2 - 10 & -4.7    & 0.2  &14.4 $\ast$ & off-limb & EUV spectra\\
Mask A (mcmc)    & 2 - 10 &  -4.4   & 0.2  &14.4 $\ast$& off-limb & EUV spectra\\
Mask B (mpfit)   & 2 - 10 & -8.5    & 0.2  & 9 $\ast$ &off-limb & EUV spectra\\
Mask B (mcmc)    & 2 - 10 & -5.3    & 0.2  & 9 $\ast$ &off-limb & EUV spectra\\
NuSTAR           & 5 - 12& $<$ -8       & 0.1  & 12 $\ast$ & disk, limb & Soft X-ray imaging \\
PA               & 3 - 10 & -5.4    & 0.1  &            & disk     &Soft X-ray  imaging \\
GDZ              & 3 - 10 & -14     & 0.1 & 10           &disk     & Soft X-ray spectra\\
HW               & 4 - 10 & -(6.1,10.3)& 0.05 & 8         & disk &  EUV spectra, imaging\\
\enddata
\tablecomments{The second column gives the temperature range used to fit the EM, the third one lists the fitted index ($\alpha$) of the power law, the fourth column gives the size (in logarithm scale) of the temperature bin. For spectroscopic data, 
 the fifth column lists  the formation temperature of the hottest line
 falling in the observed waveband. If the line is not observed and an
 upper limit is used for the EM, this is marked by  $\ast$.
 For the  NuSTAR data, this was the imposed temperature limit. 
PA: \cite{parenti10}, GDZ: \cite{delzanna14}, HW:\cite{warren12} over 15 ARs. \cite{warren14} reports slightly steeper slopes with the new NRL calibration. }
\end{deluxetable*}

\section{Summary and conclusions}
\label{sec:concl}

In this paper we have described the analysis of off-limb observations
of AR 11459 observed on the 27th and 28th of April 2012 with both the
SUMER and EIS spectrometers. This, to our knowledge, is the first
study addressing the thermal analysis of off-limb observations of an AR with spectroscopic constraints up to $10 \mathrm{MK}$, given by the observation of the Fe \,{\footnotesize XIX} spectral line.

After preparation of the data, we have provided the spatial distribution of the hot  lines emitted above $3 ~\mathrm{MK}$ along the SUMER slit. We also investigated the  light curves for selected areas. 
One initial result is that the intensity distribution along the slit
of Ca \,{\footnotesize X} ($\log T\_{max} = 5.9$) does not follow the
spatial distribution of hot lines (Fe \,{\footnotesize XVII} in this case) observed at the same time. This suggests we are looking to different thermal structures along the line of sight. 

As we are sensitive to the issue of flare contamination in our data, we also checked this aspect.
\cite{parenti10} investigated the effect of a flare on the neighboring loops. 
That flare had a similar intensity (class C) to those arising during
our observations. \cite{parenti10} found the flare to have a little
effect on the 
EM time evolution of the individual neighboring loops, even though
these had a footpoint location in common with the flaring loops. This
seems not to be the situation in our case, which  
reassures us about the influence on the EM of the small flares happening during our observations. 
To cross check, we analyzed the light curves of the hot lines and we excluded those datasets that may have been affected.

We then selected different masks, and we spatially and temporally
averaged the data in each of them to increase the signal to noise
ratio and minimize any temporal change. In particular, our Fe
\,{\footnotesize XIX} could be measured only after these procedures,
suggesting that there was only a small amount of very hot plasma. 

We performed EM and DEM analyses, optimizing the inversion
parameters to minimize the spread of the solutions. We found
consistent results with previous work: we confirm a small hot
component up to $10~ \mathrm{MK}$. In addition we were able to further
extend our knowledge of this very hot component.

In conclusion we can summarize our results as follows:

\begin{itemize}

\item Very hot plasma (above $3 \mathrm{MK}$) is present and
  persistent almost everywhere in the off-limb observations of the AR. 
In particular, we measure this up to at least $9.1 ~ \mathrm{Mm}$ above the limb and for $1.5 \times 10^2~ \mathrm{Mm}$ across the AR. 

\item  Apart from a cooler region (mask $B$), we found very similar DEM distributions for the different masks. 
In the hottest regions we found an EM of about $0.1\%$ at $10~
\mathrm{MK}$ with respect to the bulk of the plasma at $3~ \mathrm{MK}$ .  
In spite of a factor of about 2 difference in the peak of the EM
(constrained by the EIS long wavelengths lines) between the GDZ and NRL EIS radiometric calibrations, this main result still stands.  
We stress that these measurements were possible only using spatially and temporally averaged  deep exposures. 

\item It is interesting to see the consistency of the results for Fe
  \,{\footnotesize XVII} - {\footnotesize XVIII} - {\footnotesize XIX}
  listed in Table \ref{tab:dem_ratio}, considering that the SUMER
  observations lasted over 17 hours. In our analysis this implies that
  the above ratio, on average, does not change with time. Our data do
  not allow us to say anything further concerning shorter time scales.

\item The similarity  within the AR in the thermal properties of the
  different masks is accompanied to both the presence (masks $E$) or not (masks $A-D$) of spatial changes observed in the hot lines during few hours time.

\item The detection of a persistent Fe \,{\footnotesize XIX}  line in
  one of the analyzed regions determined a shallower trend in the hot
  side of the EM distribution. We fitted the high temperature tail of
  the EM with a single power law and found a power law index between
  -4 and -5 in that region (mask $A$), depending on the inversion method
  used. This is less steep than other values found previously, but we
  found a steeper trend in other regions (between -5 and -9). Although
  this puts constraints on the possible presence of an impulsive
  heating component, the resolution is not good enough to ascertain
  whether the shallow trend is really monotonic or we might have
  another minor peak around $10^7$~K, so this aspect deserves further
  investigation. We also note that the EIS NRL and GDZ radiometric
  calibrations can result in somewhat different values for the slopes.

\item We provide a new Ca \,{\footnotesize XIV} 943.59 \AA~ atomic
  model which replaces that included in the CHIANTI v. 8. This new
  data increases the intensity of the line and gives consistent EM
  results compared to other  lines formed at similar temperatures.

\item We provide SUMER-EIS cross calibration for the period of our observations. We found the SUMER intensities to be a factor 1.8 lower the EIS one when using the GDZ calibration, and 1.4 when using the NRL calibration. Previous work using data and calibrations from different periods found factors 1.5 and 1.2 \citep{giunta12, landi10}, suggesting that the relative calibration of the two instruments has not changed much with time. 

\end{itemize}

We conclude by suggesting that we have reached the limit of the EUV diagnostics possibilities for this topic. We encourage spatially resolved X-ray and multi-waveband systematic observations as the next step. New high sensitivity X-ray spectroscopic instruments should also be proposed for the next generation of missions. 
 Our analysis has been carried out assuming ionization equilibrium, as suggested by the relatively high electron density derived by our data. However, further investigation on this topic will be considered in our future work.\\

\acknowledgments

SP would like to thank H.P. Warren for the helpful discussion and for the tests he provided on the EIS radiometric calibration. SP also acknowledges the support of P. Young for the use of the EIS software and the Ca XIV deblending method. The authors thanks P. Lemaire for providing the SUMER point spread function.
We would like also to acknowledge the SUMER, Hinode planners and instruments teams for the extremely valuable support in the preparation and planning of HOP 211.
SP acknowledges the funding by CNES through the MEDOC data and operations centre.
This work used data provided by the MEDOC data and operations centre (CNES / CNRS / Univ. Paris-Sud), http://medoc.ias.u-psud.fr/.
GDZ and HEM acknowledge support from STFC (UK).
CHIANTI is a collaborative project involving George Mason University, the University of Michigan (USA) and the University of Cambridge (UK).
Hinode is a Japanese mission developed and launched by ISAS/JAXA, with NAOJ as domestic partner and NASA and STFC (UK) as international partners. It is operated by these agencies in co-operation with ESA and NSC (Norway).
SOHO is a mission of international cooperation between ESA and NASA.
Courtesy of NASA/SDO and the AIA, EVE, and HMI science teams.

\software{SolarSoft \citep{freeland98}, PINTofALE \citep{kashetal98}, IDL}

\appendix

\section{Coalignment}
\label{app:align}

\begin{figure}[h]
\includegraphics[width=.9\linewidth]{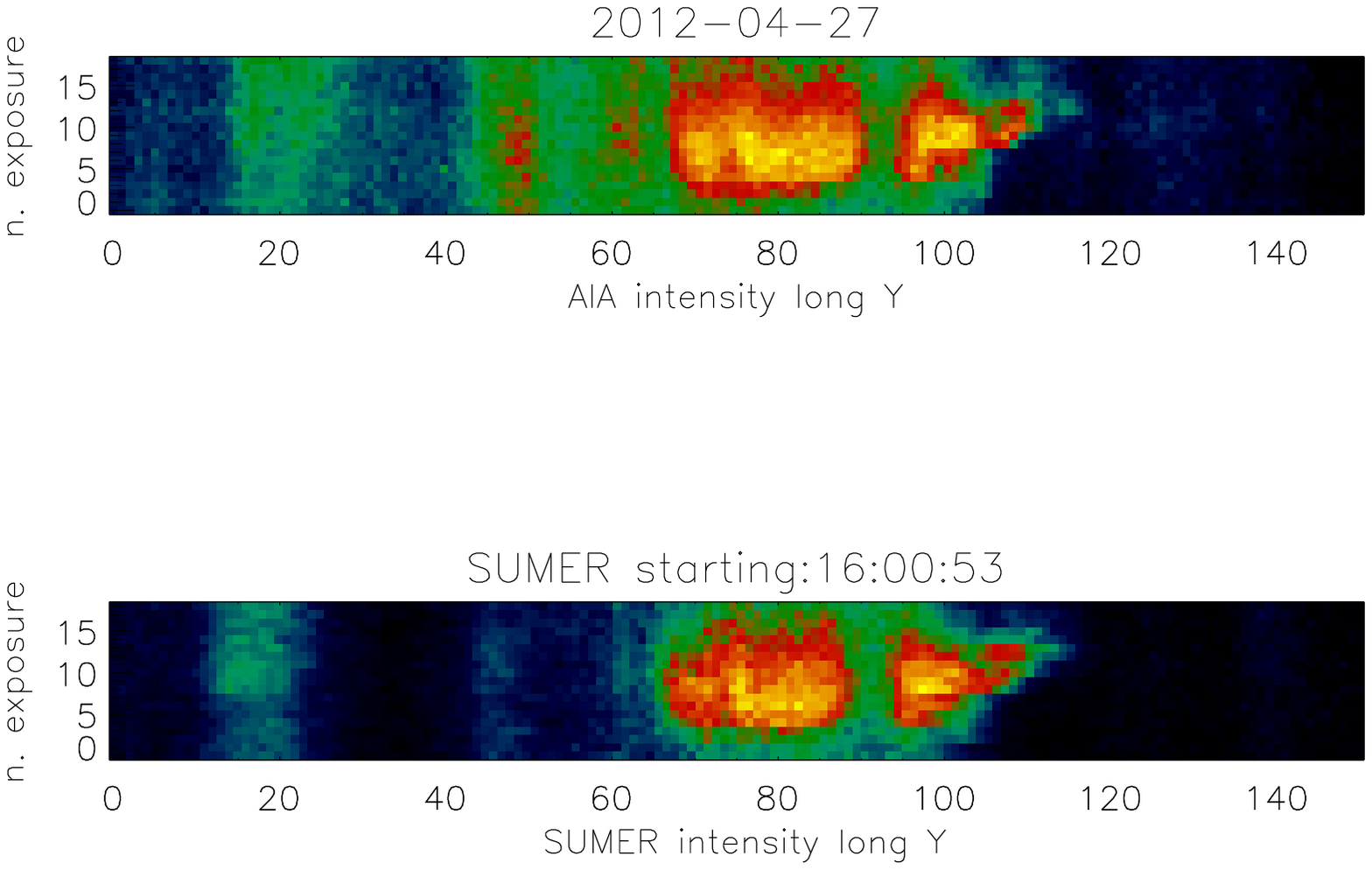}
\caption{ Top:  AIA 94 channel intensity along the SUMER slit (x-axes in the figure) as function of exposure number (y-axes in the figure). This is the best solution for the co-alignement.
Bottom: Similar to the top plot, but for SUMER Fe\,{\footnotesize XVIII} 974.86 \AA.  Note that the plot does not include the pause between each couple of exposures.  
The solar limb is on the right  side of the image.
\label{fig:aia_sum}}
\end{figure}

For the first slit position, we found that the best way to proceed was to use the temporal sequence of SUMER exposures which shows a strong emission in Fe\,{\footnotesize XVIII}, corresponding to the passage of a post-flare loop across the slit, as shown in Figure \ref{fig:aia_sum}.

The SUMER sub-time and sub-spatial sequence selected is shown in
Figure \ref{fig:aia_sum} bottom (the solar limb is on the right  side of
the image). The SUMER slit is not aligned to the AIA image columns, as
the SOHO spacescraft was rolled with respect to SDO. Moreover the
spatial resolution of the two instruments is not the same. We took
into account of these elements and proceeded to follow the steps
detailed here. We first selected an AIA data cube co-temporal to the
time series of SUMER. The data cube was rotated by the SOHO roll
amount in order to have the AIA image's columns aligned N-S with the
SUMER slit, we then selected a subfield and we spatially binned to the
SUMER spatial pixel. We then had AIA subfield images comparable to the SUMER exposures.
At a given time, each spatial column of the AIA-subfield  was cross-correlated with the corresponding SUMER exposure. 
 The best solution for the coalignment is given by the AIA column which maximizes the correlation in time and space, as shown in Figure \ref{fig:aia_sum} top.

\begin{figure}[h]
\includegraphics[width=0.5 \linewidth ]{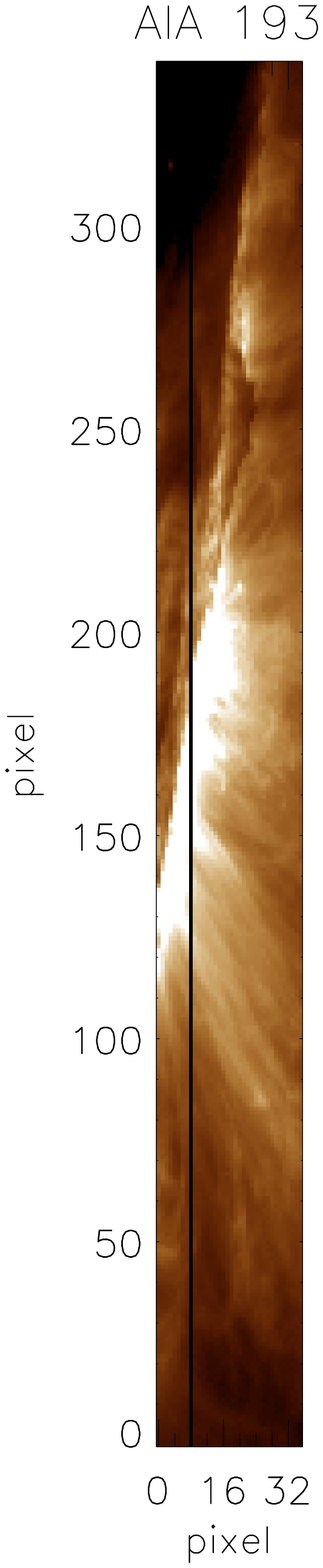}
\includegraphics[width=0.5 \linewidth ]{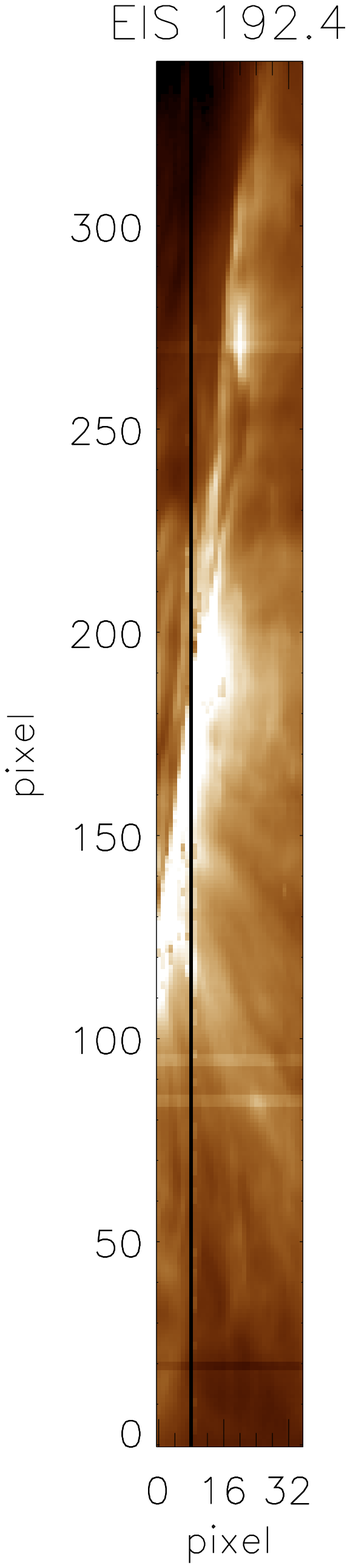}
\caption{ Result of the AIA 193 (left) and EIS 192.4\AA~(right) alignment for April 27.}
\label{fig:aia_eis}
\end{figure}

To co-align EIS with AIA we chose to use the AIA 195 and EIS Fe\,{\footnotesize XII} 192.394~~\AA, as there is no  Fe\,{\footnotesize XVIII} in EIS and the Fe\,{\footnotesize XVII} lines are too faint to be used.
We proceeded by selecting a cube of AIA images co-temporal to the EIS
raster sub-field close to the limb (see Figure \ref{fig:aia_eis}). We
built an AIA raster cube, each column made of data taken at the
equivalent EIS exposure time. The other two dimensions were filled by
columns of the same image  taken across a spatial lag of about 10
pixels, with steps of one. This method allowed us to extract the best correlation for the whole sub-raster in space and time. The AIA subfield which gave the best coalignment result is plotted in Figure \ref{fig:aia_eis}.

\section{SUMER stray light}
\label{app:stray}

\begin{figure}[h]
\includegraphics[width=.5\linewidth]{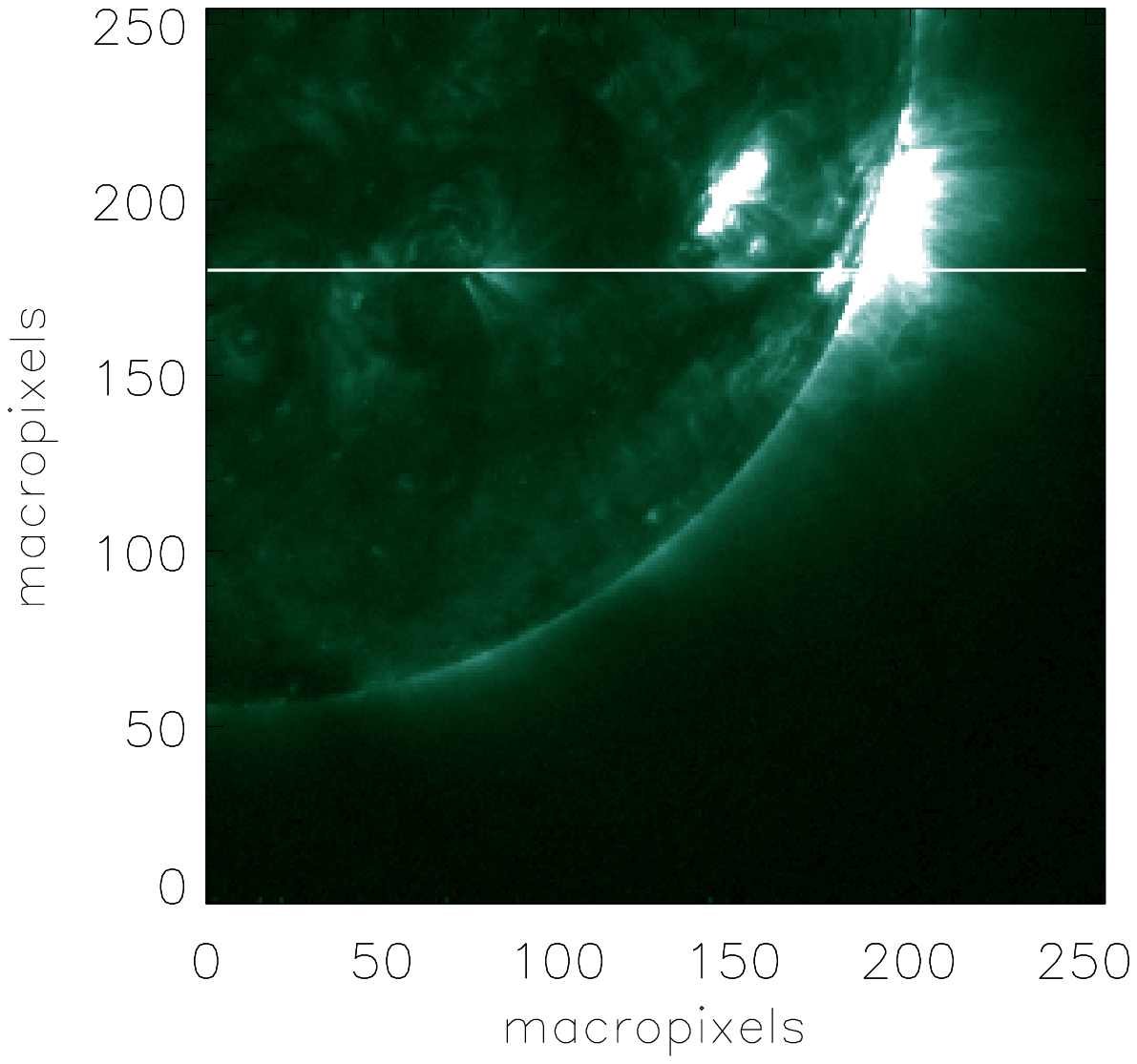}
\includegraphics[width=0.5\linewidth]{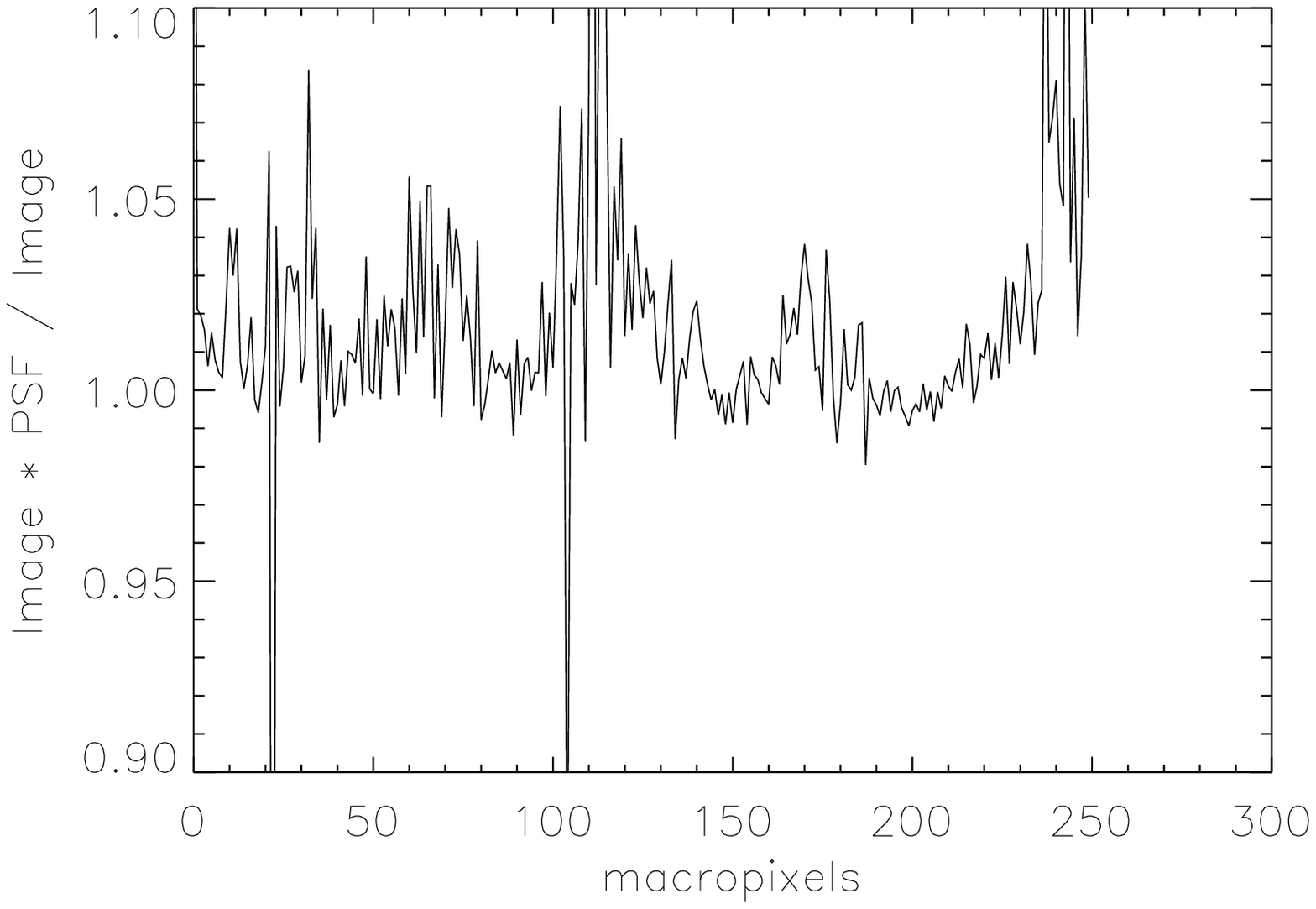}
\caption{Left: result of the AIA 94 image convolved with the SUMER PSF. This image was divided to the original AIA image to obtain the percentage of stray light in the measured flux. The right plot shows a horizontal cut of this images ratio at the vertical position which is sampled by the SUMER slit. This position is marked on the left plot with a white line.  }
\label{fig:aia_stray}
\end{figure}

Due to our interest in the hot emission, we estimated the stray light by using the AIA 94 channel data convolved with the SUMER point spread function. 
The reference image was the integration of two AIA 94 images taken on
the 27 April 2012 at 21:42 UT. This resulting image, binned to the
SUMER pixel size, was convolved with the instrumental profile (P. Lemaire, private communicaton) as shown in Figure \ref{fig:aia_stray} left. On the right plot of Figure  \ref{fig:aia_stray} we show the horizontal cut (marked by the white line) on this image, scaled to the original AIA image, which crosses the active region and the SUMER field of view. We see that the off-limb stray light is only few percent of the original flux.

\noindent We also tested stray light in cool lines. Our data also
contain the O\,{\footnotesize I} 1152.15~\AA~ which is a good
stray-light marker close to the limb. We integrated over the 60
exposures and extracted the line intensity along the SUMER slit
position 2 in those pixels where the SNR is above 10. Assuming that
the observed intensity is only due to scattered light in the
instrument, we calculated its ratio to the solar value
\citep{curdt04,parenti04, parenti05a}, and found that it is always below $3\%$.

\section{SUMER wavelength calibration}
\label{app:sum_wcal}

SUMER spectra are not wavelength calibrated and the grating dispersion is wavelength dependent.  Ideally, to perform the calibration  we need to have reference lines profiles (that are emitted by static features) along the whole waveband. 
The process is quite straightforward once we have observations
targeting the quiet Sun: the SUMER wavebands include a long list of
chromospheric lines, including neutrals or singly ionized ions, for
which we assume a mean zero velocity (see details of this calibration
in \cite{parenti04}). This is not the case when we are dealing with
off-limb observations in active regions, where the coronal emission
dominates, and plasma flows are more common. We tried to minimize the
effects of possible local flows applying the following two steps. We
used the data from the slit positioned further out in the corona
(position 2, mask  $G$), averaging the spectra over seventy pixels  in
the southern part of the slit (that is the area at the greatest
distance and in the AR periphery). To produce an average effect which
minimizes the detection of flows, we chose as reference lines all the
bright ones, independently of their formation temperature. Because of the wavelength dependence of the dispersion, we performed an independent calibration in our three spectral windows. We proceeded by measuring the position of our spectral lines in the pixel dimension on the detector, and used a linear relation and reference positions to convert it in the wavelength space.

\section{Electron density along the SUMER slit}
\label{app:dens}

\begin{figure}[th]
\includegraphics[scale=.4]{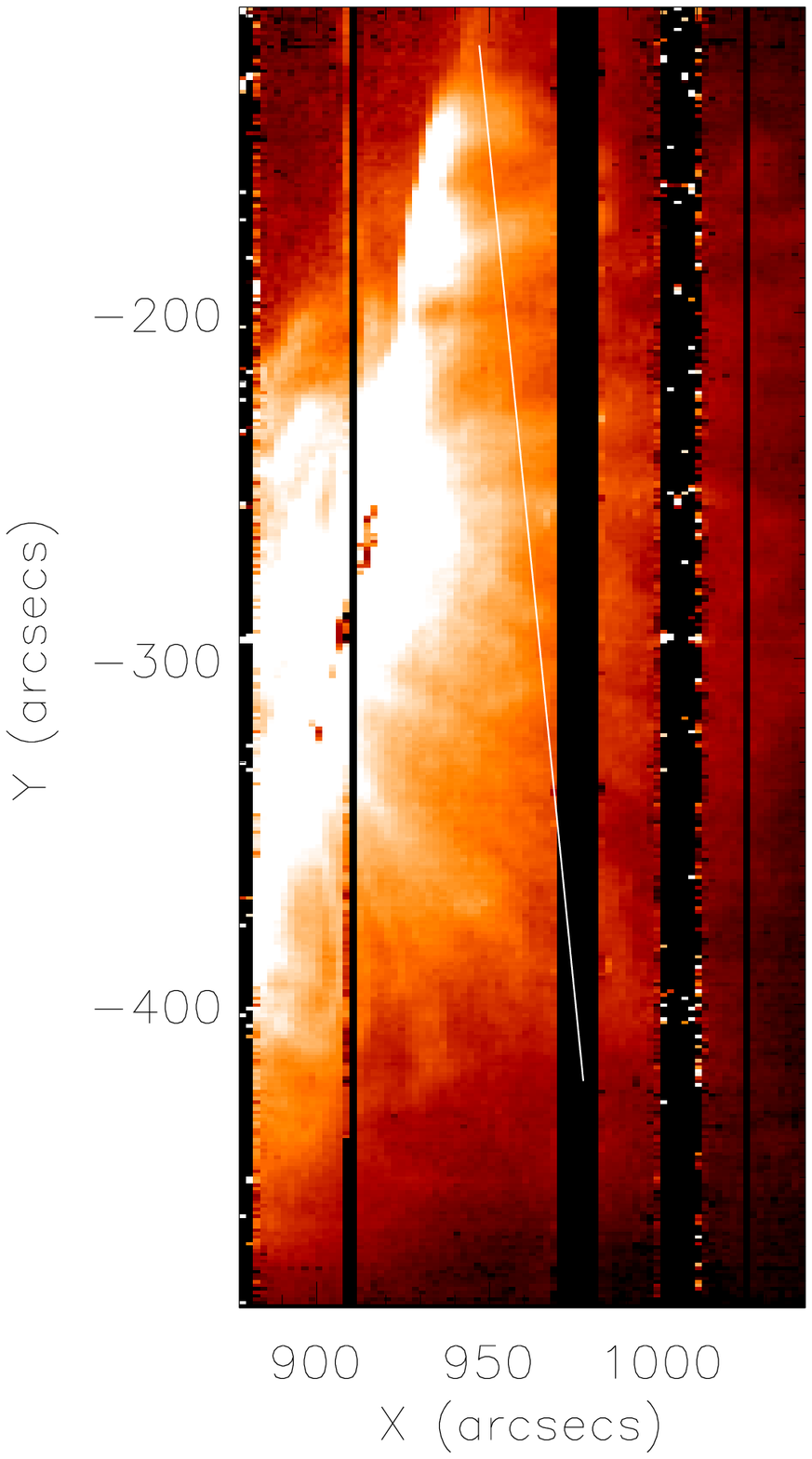}
\caption{ EIS map of density derived from the Fe\,{\footnotesize XIII} lines ratio for the raster at 20:24 UT. The intensity has been saturated to highlight the faint off-limb structuring. The SUMER slit is superimposed.}
\label{fig:n_map}
\end{figure}

\begin{figure}[th]
\hspace*{-.4cm}\includegraphics[width=.36\linewidth]{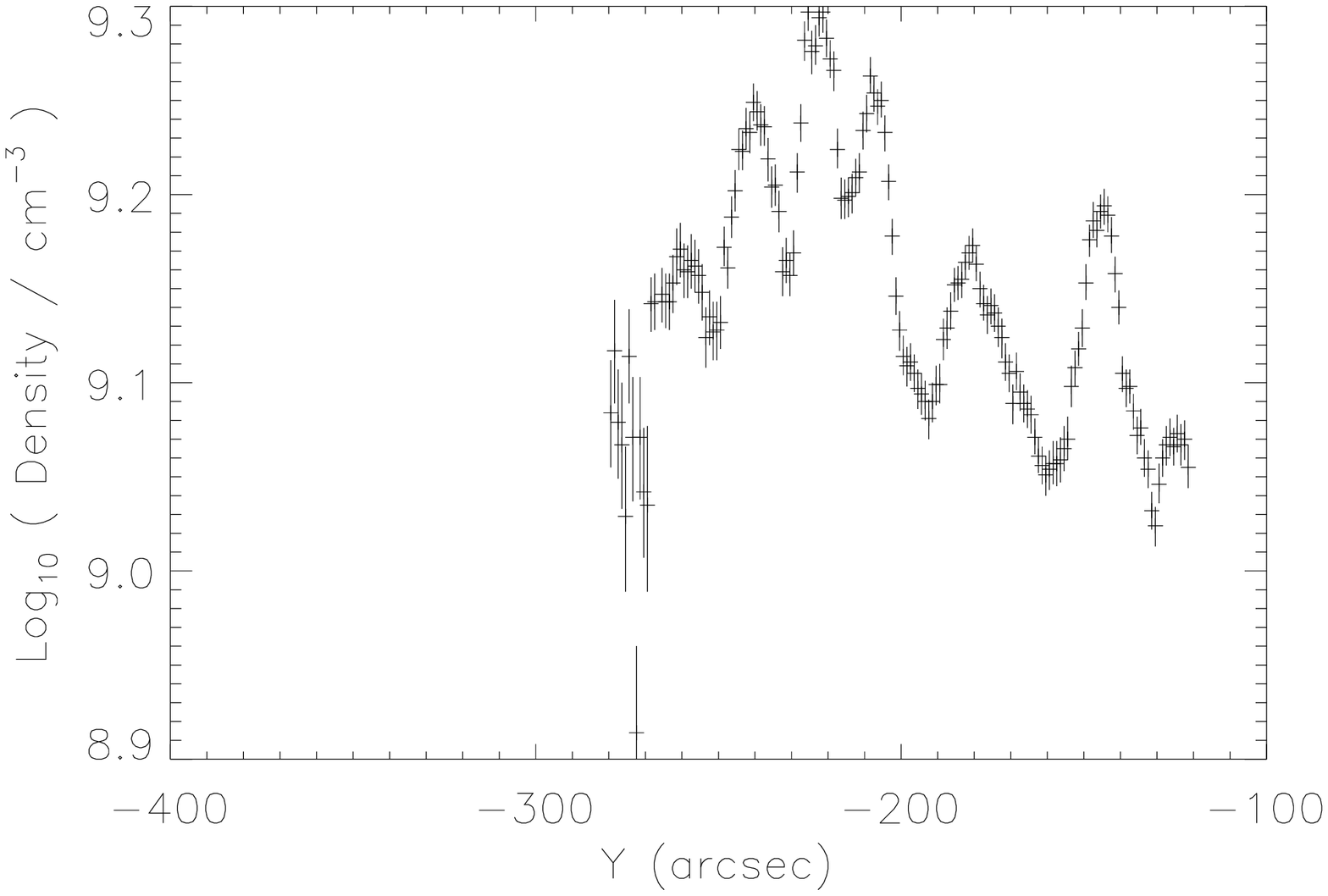}
\hspace*{-.5cm}\includegraphics[width=.36\linewidth]{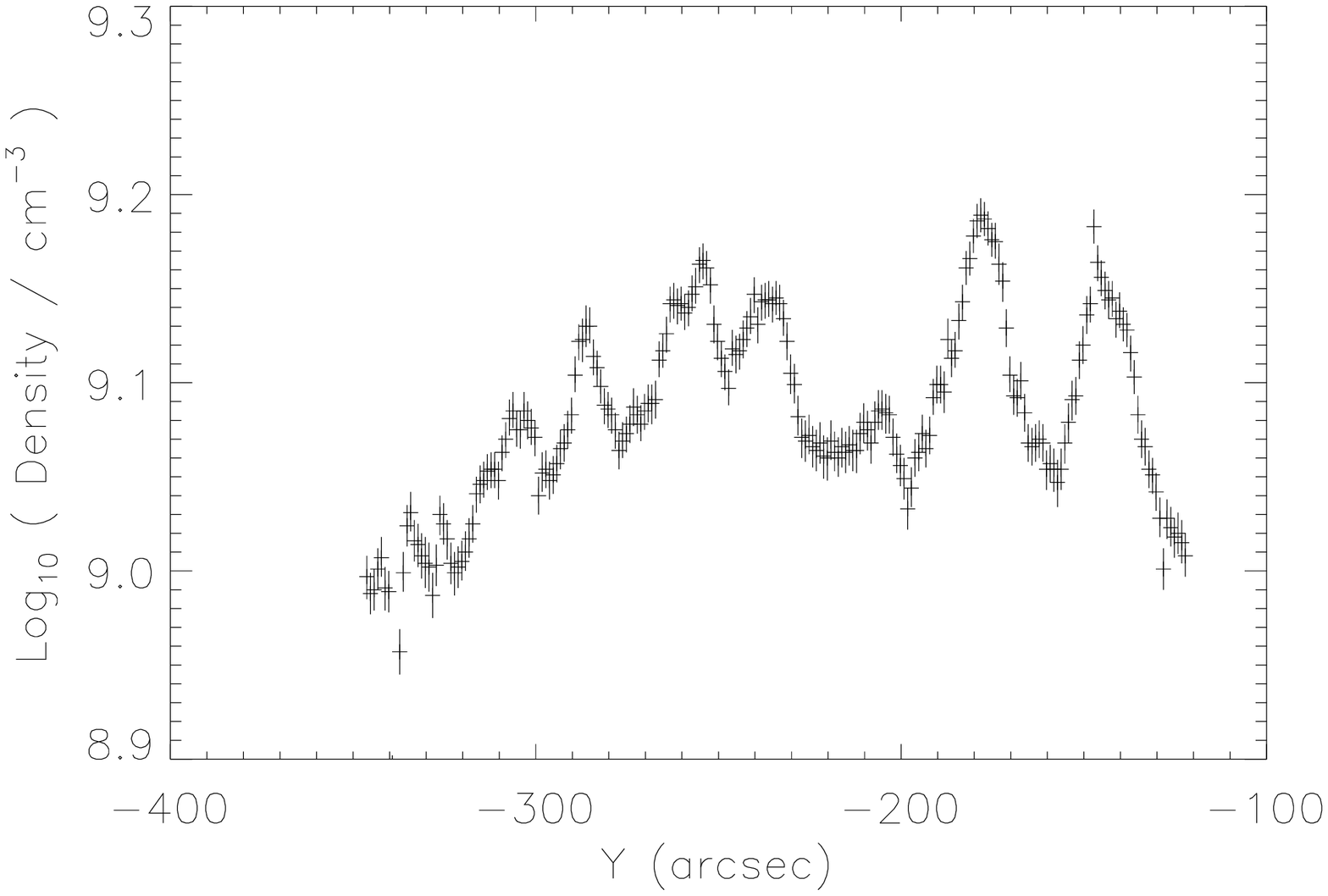}
\hspace*{-.4cm}\includegraphics[width=.36\linewidth]{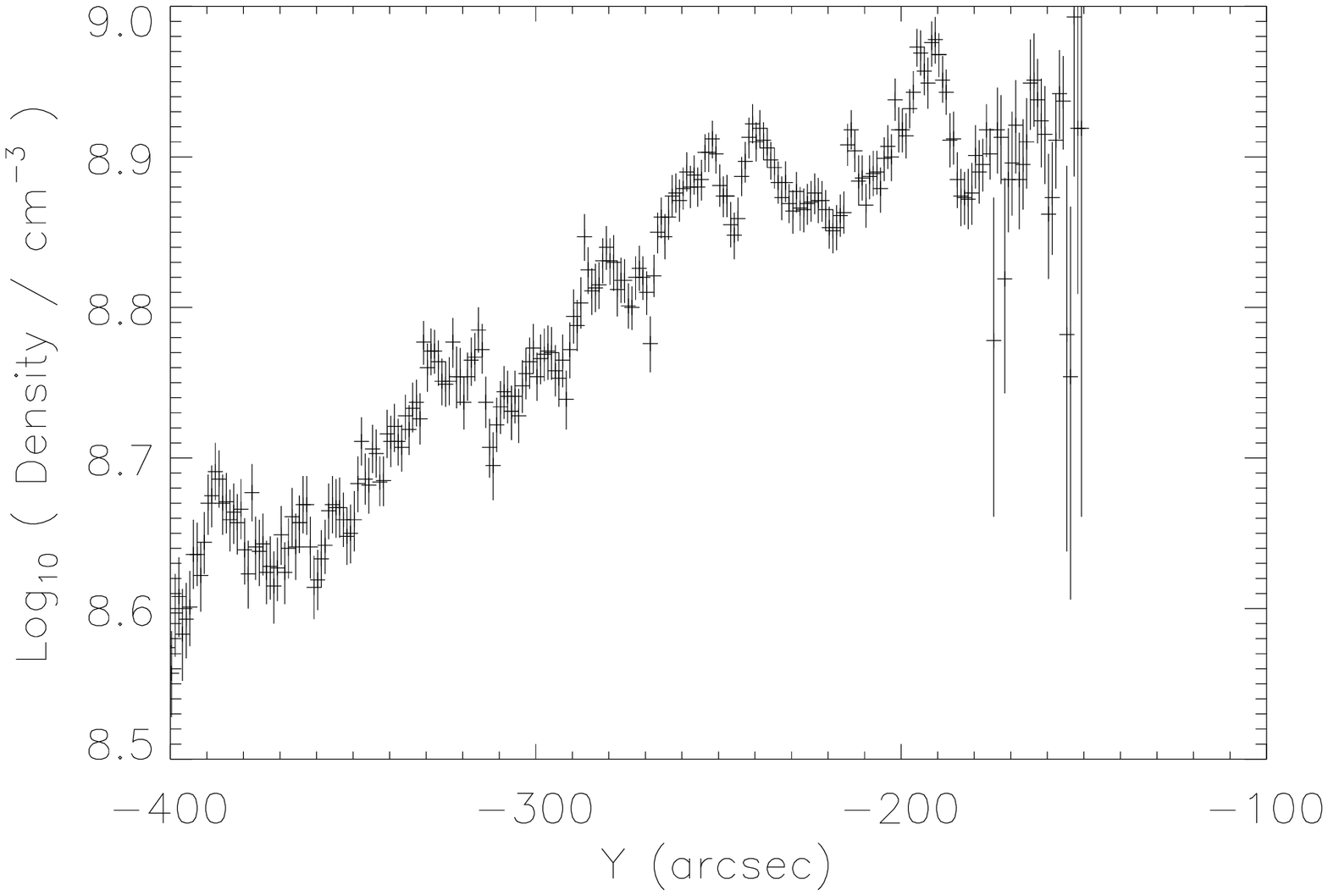}
\caption{Left and center: electron density derived from the EIS Fe\,{\footnotesize XIII}
  lines ratio along the SUMER slit for position 1 (EIS rasters at   18:18 UT and 20:24 UT  respectively). Right: the same but for SUMER slit position 2 (EIS raster starting at 23:01 UT).
\label{fig:n_prof}}
\end{figure}

We derived the electron density maps from the EIS rasters using the electron density diagnostics of lines ratio described in Section \ref{sec:diagn}. We used Fe\,{\footnotesize XIII} $(203.82+203.8)/202.04$ \AA ~($\log T = 6.25$), which is sensitive in the range $N>10^8 ~\mathrm{cm^{-3}}$. 
 Figure \ref{fig:n_map} shows the density  map from the raster starting at 20:24 UT on April the 27th (SUMER slit position 1). The density along the SUMER slit derived from this map is shown in the central panel of Figure \ref{fig:n_prof}.
For SUMER slit position 1 we also have an earlier EIS map and the density profile along the SUMER slit is shown on the left of Figure \ref{fig:n_prof}. The right plot in this figure gives the density derived along the SUMER slit for slit position 2.

The two profiles on SUMER position 1 reflect the changing of the
structuring of the AR at this temperature about two hours apart. We
can identify the same main features with a change in the relative
density amplitude, which is most evident in the AR core ($-250\arcsec
<Y<-200\arcsec$). This area is the one occupied by the flaring region
and by the hot loop visible in the first raster (Figure
\ref{fig:aia_sum}). Note how little these profiles resemble the intensity profiles along the slit of the hot lines, suggesting we are observing different thermal structures along the line of sight. 
Figure \ref{fig:n_prof} right shows a more important drop in density as Y decreases, due both the increase height above the active region and the increasing distance from the AR core.

The density values found here have been used for the thermal analysis.

\section{SUMER temporal variability for slit position 2}
\label{app:sum_p2}

\begin{figure}[th]
\includegraphics[width=0.5\linewidth]{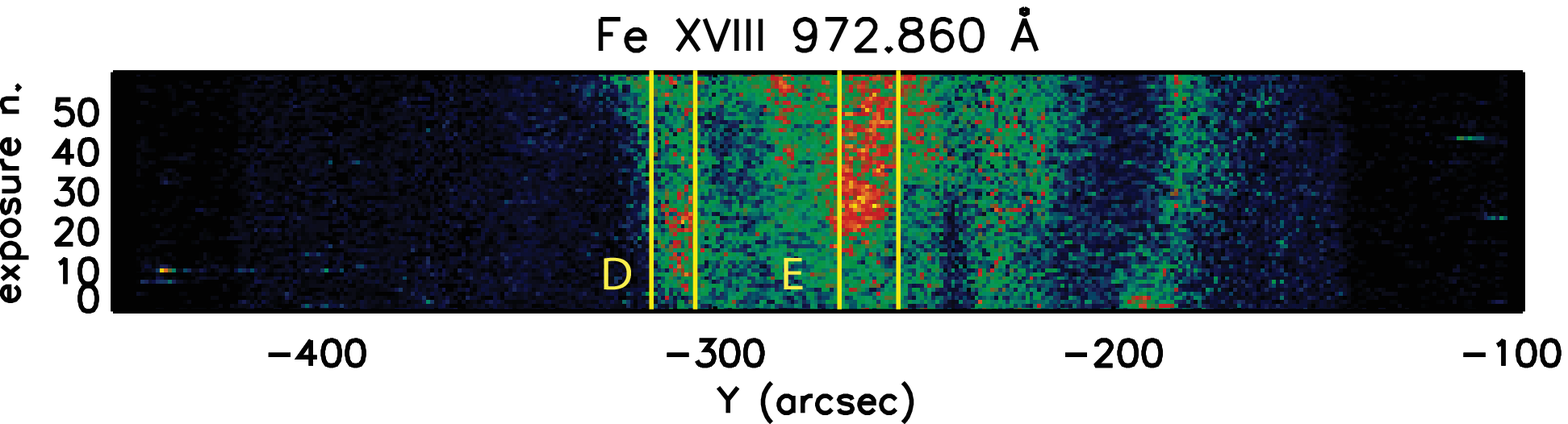}
\includegraphics[width=0.5\linewidth]{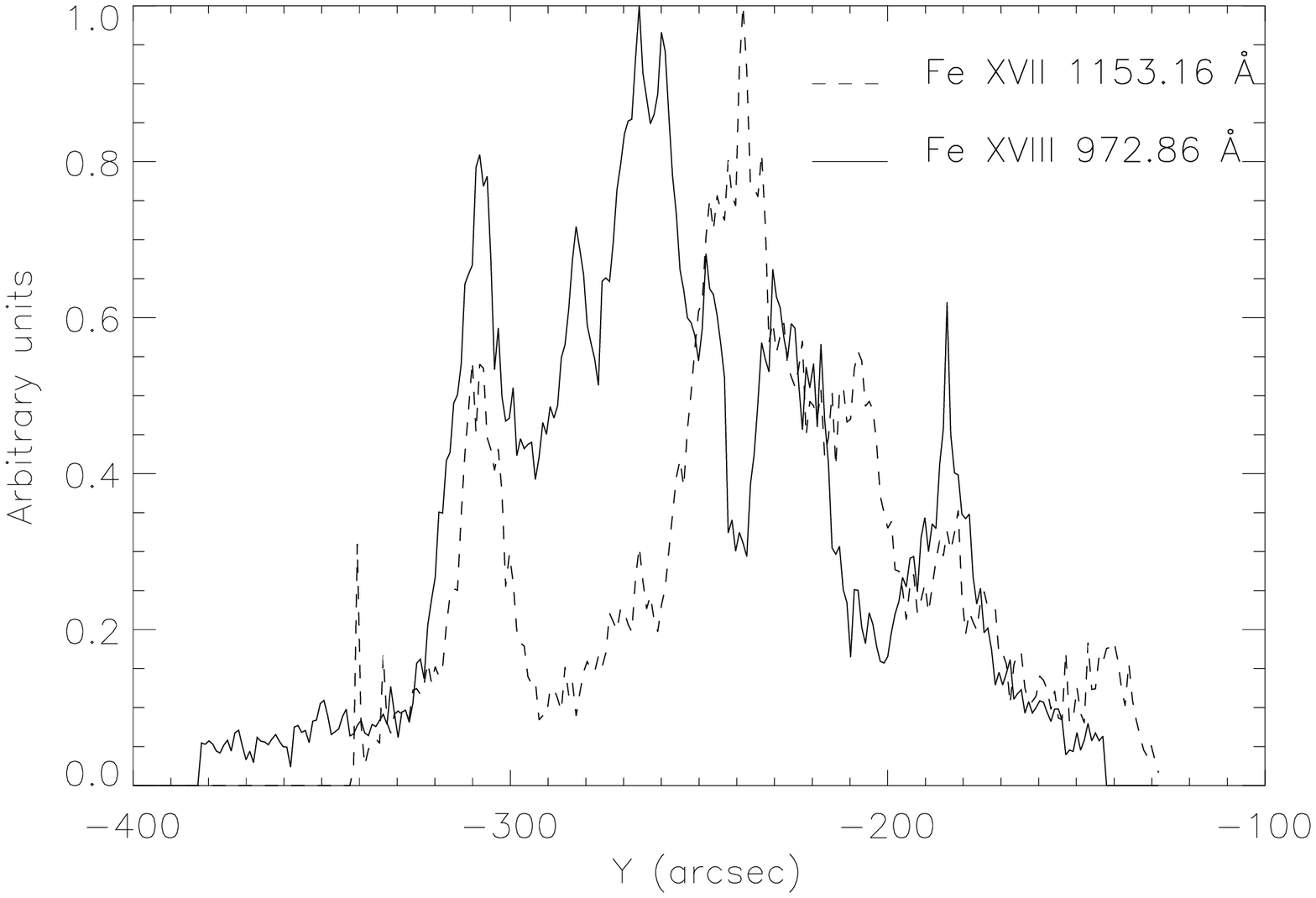}
\caption{Left: SUMER Fe\,{\footnotesize XVIII} intensity along the slit in position 2, plotted for the sixty exposures. Right: time integrated Fe\,{\footnotesize XVII} and Fe\,{\footnotesize XVIII} intensities along the slit in position 2. The north is on the right side of the plot. }
\label{fig:slit_time2}
\end{figure}

\begin{figure}[th]
\includegraphics[width=.45\linewidth]{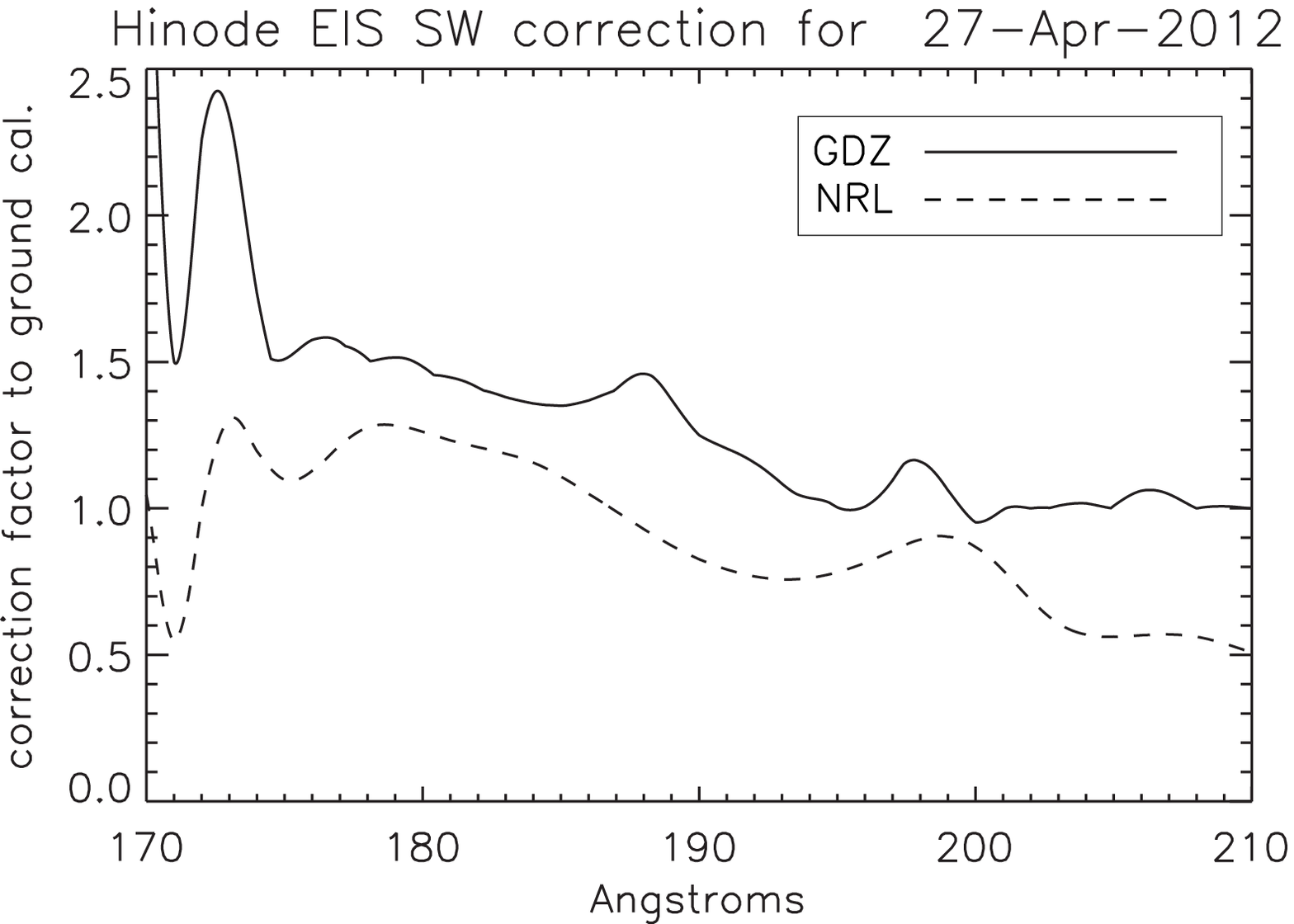}
\includegraphics[width=.45\linewidth]{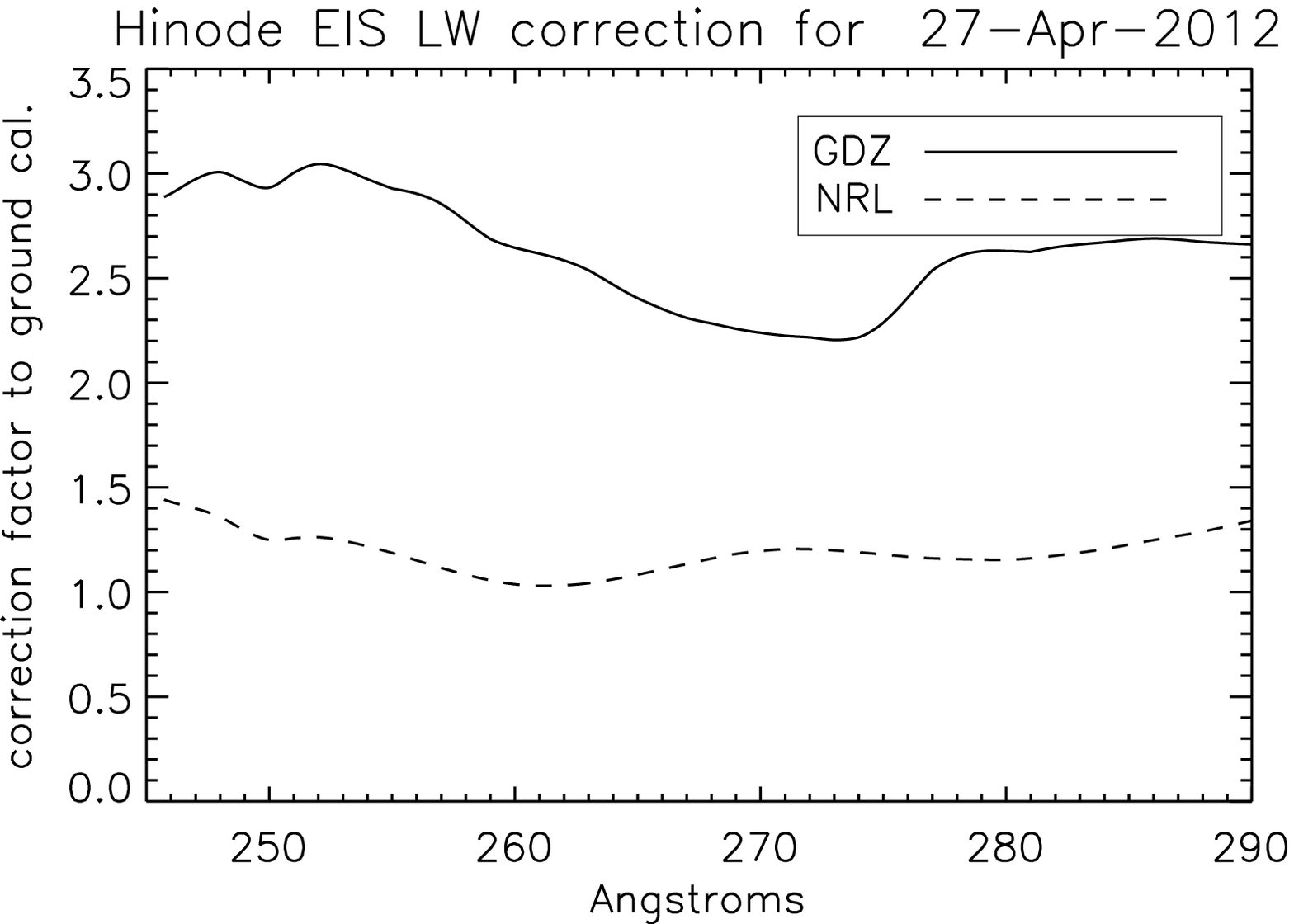}
\caption{Correction factor to the pre-flight radiometric calibration derived by \citep{warren14} (NRL in the figure) and \cite{delzanna13} (GDZ in the figure).}
\label{app:calib}
\end{figure}

Figure \ref{fig:slit_time2} left shows the temporal variation of the  Fe\,{\footnotesize XVIII} along the SUMER slit in position 2. 
We can notice a general temporal variation of the structures which tend to spatially spread along the slit as the timeline increases. We selected the most stable parts highlighted by the yellow lines (masks $D$, $E$). 

Figure \ref{fig:slit_time2} right shows the temporal averaged integrated flux of Fe\,{\footnotesize XVII} and Fe\,{\footnotesize XVIII}, which were taken six hours apart. There is a temporal variation mostly in the core of the AR. We adapted our masks to these changes. 
As shown in Figure \ref{fig:dems} there is no thermal variation within at these locations.

\section{EIS radiometric calibration}
\label{app:eis_cal}

Figure \ref{app:calib} shows the correction factor to the pre-flight radiometric calibration of EIS for the two channels, as determined by \cite{delzanna13} (solid line) and \cite{warren14} (dashed line) for the period of our observations.

\bibliography{bib}

\end{document}